\title{Estimating heterogeneous treatment effects in\\
nonstationary time series with state-space models} 
\author{Shu Li
\thanks{
                 Seminar for Statistics, ETH Z\"urich}\hspace{0.15cm}
      \and
    Peter B\"uhlmann
    \footnotemark[1]
  }
\documentclass{article}
\usepackage{graphicx}
\usepackage{amsmath}
\usepackage{amssymb}
\usepackage{amsthm}
\usepackage[longnamesfirst,numbers]{natbib}
\usepackage[top=1in, bottom=1.25in, left=1.25in, right=1.25in]{geometry}

\usepackage{color}
\usepackage{subcaption}
\usepackage{caption}
\usepackage{algorithm}
\usepackage{algorithmic}
\usepackage{url}
\usepackage{enumerate}  

\usepackage{tikz}
\usepackage[framemethod=TikZ]{mdframed}
\usepackage{tikz}
\usetikzlibrary{arrows,automata,decorations.markings,decorations.pathmorphing,backgrounds,fit,petri}
\usetikzlibrary{patterns,snakes}
\usetikzlibrary{positioning}
\usetikzlibrary{calc}
\tikzstyle{vertex}=[circle, draw, fill, inner sep=0pt, minimum size=0.15cm]

\newenvironment{keywords}{}{}

\newcommand{\E}{\mathbb{E}}

\newcommand{\CATE}{\mathrm{CATE}}
\newcommand{\MCATE}{\mathrm{MCATE}}

\newcommand{\MSE}{\mathrm{MSE}}
\newcommand{\Do}{\mathrm{do}}

\begin{document}
\newpage
\maketitle
\begin{abstract}
Randomized trials and observational studies, more often than not, run over a certain period of time. The treatment effect evolves during this period which provides crucial insights into the treatment response and the long-term effects. Many conventional methods for estimating treatment effects are limited to the i.i.d. setting and are not suited for inferring the time dynamics of the
treatment effect. The time series encountered in these settings are highly informative but often nonstationary due to the changing effects of
treatment. This increases the difficulty, since stationarity, a common assumption in time series analysis, cannot be reasonably assumed. Another challenge is the heterogeneity of the treatment effect when the treatment affects units differently. The task of estimating heterogeneous treatment effects from nonstationary and, in particular, interventional time series is highly relevant but has remained unexplored yet.

We propose Causal Transfer, a method which combines regression to adjust for confounding with time series modelling to learn the effect of the treatment and how it evolves over time. Causal Transfer does not assume the data to be stationary and can be applied to randomized trials and observational studies in which treatment is confounded. Causal Transfer adjusts the effect for possible confounders and transfers the learned effect to other time series and, thereby, estimates various forms of treatment effects, such as the average treatment effect (ATE) or the conditional average treatment effect (CATE). By learning the time dynamics of the effect, Causal Transfer can also predict the treatment effect for unobserved future time points and determine the long-term consequences of treatment.
\end{abstract}  


\section{Introduction}
\label{section:introduction}

Many types of experiments run over a period of time. For example, phase 1 clinical trials typically span several months while online experiments for measuring advertising effectiveness last several weeks. The treatment effect continuously evolves during the duration of the experiment. It might strengthen or weaken over time. This provides important information about the response to treatment. The treatment response and the value of the long-term effect, in particular, are critical for whether an experiment is successful. For example, advertising campaigns whose effectiveness last over a duration of time are preferable to campaigns whose effectiveness drop to 0 quickly. In order to predict long-term effects, one has to learn the time dynamics of the treatment effect. This is not an easy task since common time series assumptions like stationarity are not satisfied in the presence of interventions. Common strategies for analysing experimental panel data include analysing the data of every time point in the experiment in separation using an i.i.d. method, e.g., imputation~\cite{imbens2015}, or by aggregating the experimental units and applying a time series method to the aggregated data, e.g., Causal Impact~\cite{brodersen2015}. 
The first approach, while easy to implement, treats every time point as a separate experiment and does not pool information over time. The latter sacrifices information by combining the experimental units into a single unit. We propose novel methodology for estimating (average and heterogeneous) treatment effects from experiments which span multiple time points. Our method, Causal Transfer, makes use of all data from the entire experiment and neither requires data aggregation over time nor over units. We demonstrate this on simulated data and on real data from geo experiments in Section~\ref{section:empirical}.


\paragraph{General setting.} We first describe the general setting before specializing to potentially nonstationary and multivariate time
series. Suppose that we observe $d$ units. For each unit $i$, there exists
a pair of outcomes $(Y_i(0), Y_i(1))$ under treatment and under
control. Let $T$ denote the binary treatment indicator and $Y_i = Y_i(T_i)$
the observed outcome of unit $i$. The observed outcome is equal to $Y_i(1)$
if the unit is assigned to treatment and $Y_i(0)$ otherwise. We assume that
there is no interference between units, i.e., the SUTVA
assumption~\cite{rubin1978}. The measure reported most often from
experiments is the average treatment effect (ATE) defined as
$\mathbb{E}[Y_i(1) -Y_i(0)]$. In randomized experiments, the ATE is
directly identified from $\mathbb{E}[Y_i\,|\,T_i=1] -
\mathbb{E}[Y_i\,|\,T_i=0]$. The ATE implicitly assumes that the units are
drawn from an underlying distribution of a large population. In some
cases, the units might not be representative of such a population, e.g.,
when the units are selected. A more appropriate measure under these
circumstances is the sample average treatment effect (SATE) $\frac{1}{d}\sum_{i=1}^d (Y_i(1) -Y_i(0))$. The SATE is simply the treatment effect on the study units and it avoids assumptions on distributions \cite{balzer2015}. If the treatment effect should be transferable to other units, however, the ATE is arguably more suited than the SATE. The SATE will converge against the ATE in the large sample limit if the units are, in fact, independent, reflective of a population, and the first moments of the potential outcomes exist.\\

When treatment affects units differently, the conditional average treatment effect (CATE) is of special interest:
\begin{equation*}
\mathbb{E}[Y_i(1) -Y_i(0)\,|\, X_i],
\end{equation*}
which captures the heterogeneity of the treatment effect as a function of
some covariates $X_i$. Obviously, the ATE is the expectation of the CATE:
\begin{equation*}
\mathbb{E}[Y_i(1) -Y_i(0)] = \mathbb{E}[\mathbb{E}[Y_i(1) -Y_i(0)\,|\,
X_i]].
\end{equation*}
Moreover, the randomization assumption can be weakened substantially when integrating out the CATE to obtain the ATE,
if the covariates $X_i$ are ``well chosen'' as discussed below.

In many situations, the treatment $T$ cannot be randomized due to ethical reasons and is endogenous. In this case, it is not sufficient to solely compare the difference of the means between the treatment and the control group. In order to identify the treatment effect, we require unconfoundedness, which is weaker than randomization. We assume that the
  treatment is randomized conditional on some covariates $X_i$, that is:
  \begin{equation}\label{eq:condrand}
(Y_i(0), Y_i(1)) \perp T_i \,|\, X_i . 
\end{equation}
We can then infer the ATE as follows:
\begin{eqnarray}
\label{eq:adjustment}
& &\mathbb{E}[Y_i(1) -Y_i(0)] = \mathbb{E}[ \mathbb{E}[Y_i(1) -Y_i(0)\,|\,
    X_i]]\nonumber \\
  &=& \mathbb{E}[ \mathbb{E}[Y_i(1) -Y_i(0)\,|\,
T_i, X_i]]\nonumber \\
 &=& \mathbb{E}[ \mathbb{E}[Y_i|\, T_i = 1, X_i] -
\mathbb{E}[Y_i|\, T_i = 0, X_i] ], 
\end{eqnarray}
where the assumption in \eqref{eq:condrand} is used for the second equality and the last equality follows by the definition of a potential outcome when conditioned on the treatment. The formula~\eqref{eq:adjustment} is also well known in structural equation modelling and often referred to as Pearl's backdoor adjustment formula~\cite{pearl2000}. In this framework, if $X_i$ blocks all backdoor paths from the treatment $T_i$ to the response $Y_i$, the adjustment formula~\eqref{eq:adjustment} is valid. A well-known example is when $X_i$ is the set of parental variables of $T_i$ in the graph corresponding to the underlying structural equation model.
The choice of the set of covariates $X_i$ is in general non-trivial and requires domain knowledge. Nonetheless, time order often helps to simplify this problem since causes precede their effects.
\paragraph{Time series setting.} Equation~\eqref{eq:adjustment} requires the fitting of a regression function. In time series, we only observe one sequence of observations per stochastic process. To make inferences feasible, one either assumes redundancy in terms of stationarity or a parametric model for nonstationary settings. As soon as we intervene on a time series, it changes the distribution of the time series. Therefore, interventions typically break the stationarity of a time series. It goes without saying that many time series are not even stationary before an intervention takes place. For this reason, we focus on the latter approach and adopt the highly-established framework of state-space models for estimating dynamic regression functions~\cite{west2006}. Such state-space models share some properties, which make them well-suited for the task of causal effect estimation as they can deal with nonstationarity and missing observations. Causal inference is a missing data problem at its core.

We develop a novel state-space model and corresponding methodology that estimates heterogeneous causal effects from potentially nonstationary time series. Our method, which we call Causal Transfer, is able to learn the effect of an intervention, transfer this effect to other time series and, thereby, estimate (causal) treatment effects in the form of population, sample, or heterogeneous effects. The state-space model further learns how the treatment effect evolves over time and is able to predict treatment effects for (unobserved) future time points.

\begin{figure}[!htb]
    \centering
        \begin{subfigure}[b]{0.35\textwidth}
        \includegraphics[width=\textwidth]{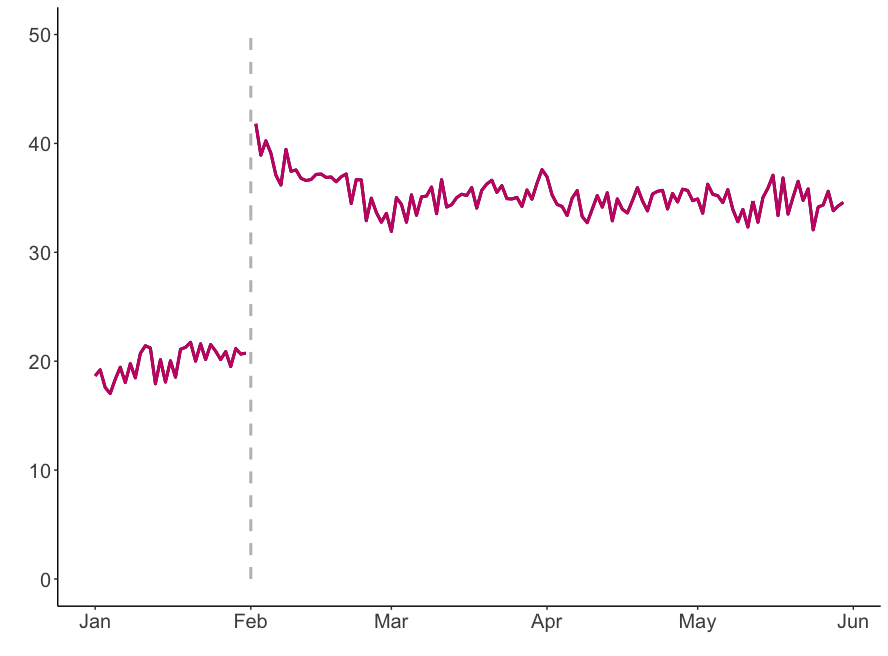}
        \caption{Observed outcomes of unit 1}
        \label{fig:seq1}
    \end{subfigure}
    \hspace{0.5cm}
    ~
       \begin{subfigure}[b]{0.35\textwidth}
        \includegraphics[width=\textwidth]{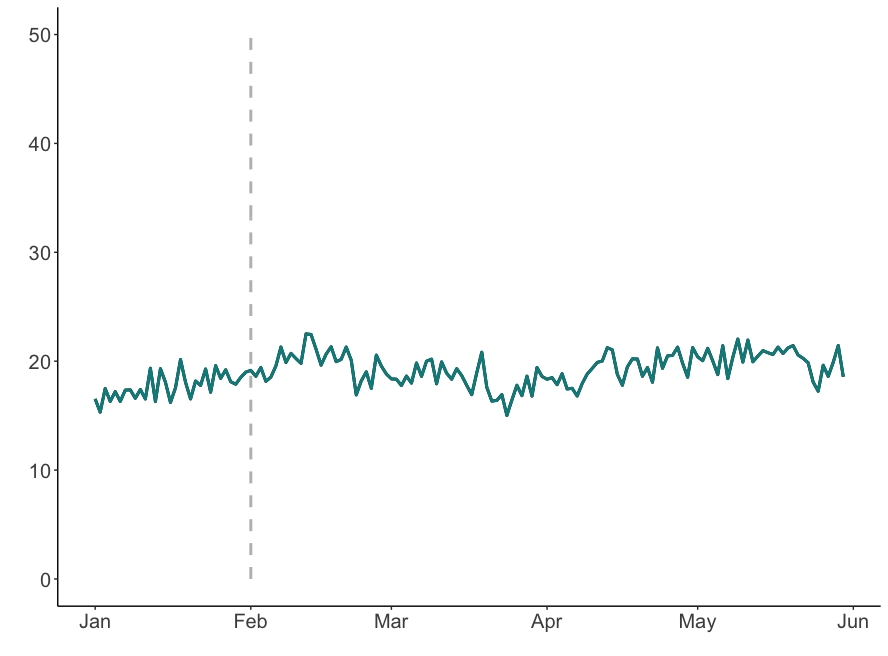}
        \caption{Observed outcomes of unit 2}
        \label{fig:seq2}
    \end{subfigure}
    
        \begin{subfigure}[b]{0.35\textwidth}
        \includegraphics[width=\textwidth]{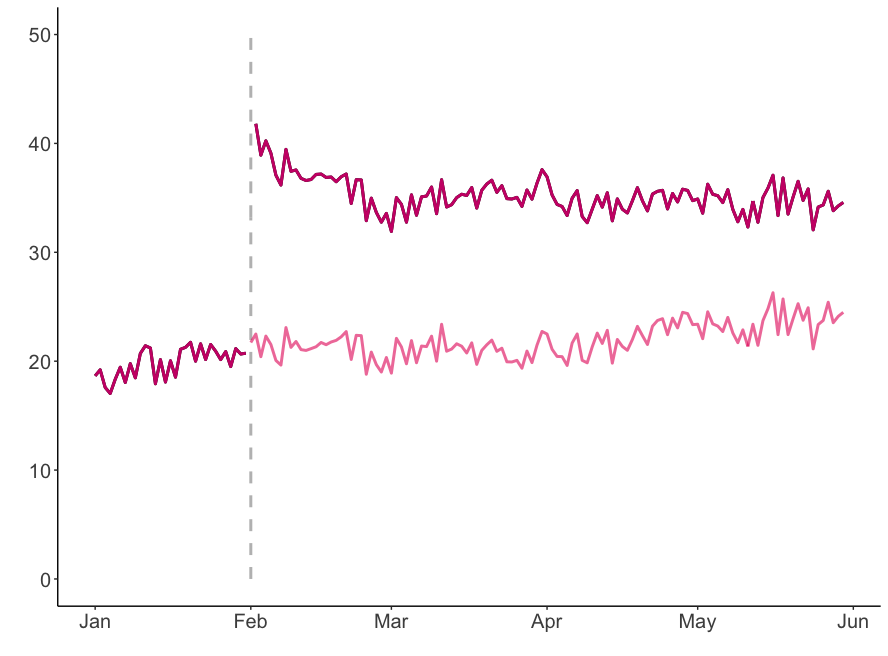}
        \caption{Predicted outcomes of unit 1}
        \label{fig:seq3}
    \end{subfigure}
        \hspace{0.5cm}
    ~
        \begin{subfigure}[b]{0.35\textwidth}
        \includegraphics[width=\textwidth]{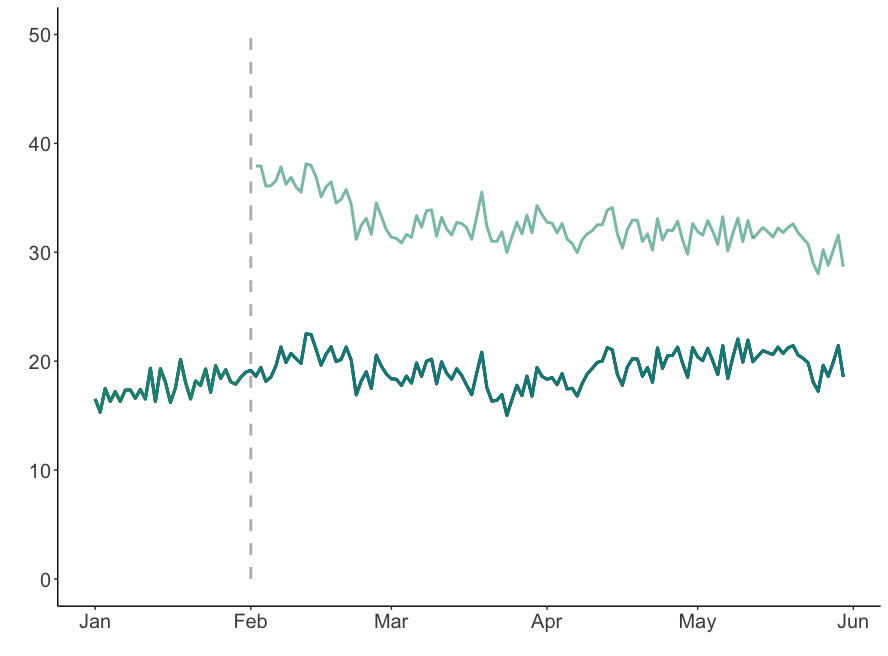}
        \caption{Predicted outcomes of unit 2}
        \label{fig:seq4}
    \end{subfigure}
    
        \begin{subfigure}[b]{0.35\textwidth}
        \includegraphics[width=\textwidth]{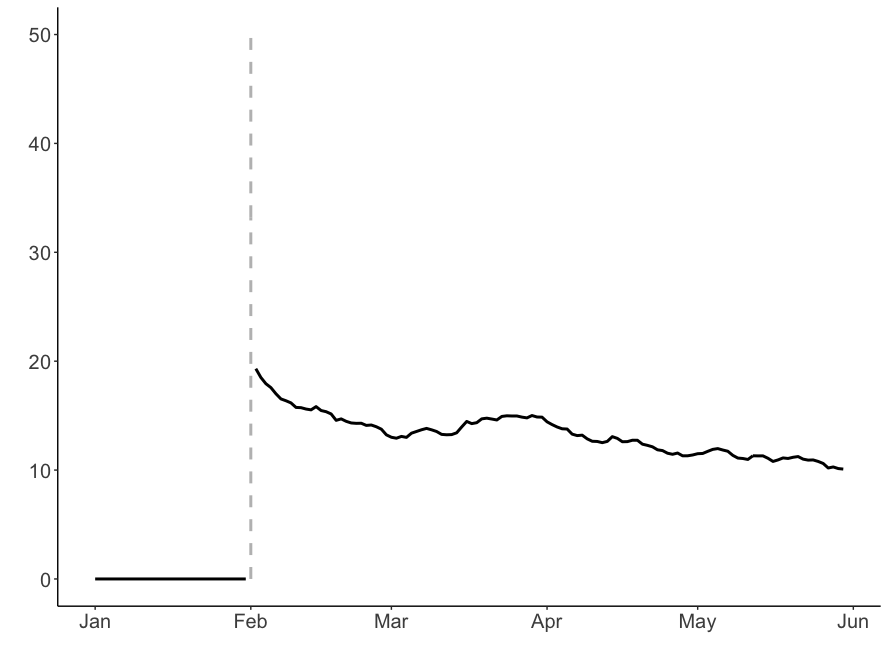}
        \caption{Estimated average treatment effect}
        \label{fig:seq5}
    \end{subfigure}
        \hspace{0.5cm}
    ~
        \begin{subfigure}[b]{0.35\textwidth}
        \includegraphics[width=\textwidth]{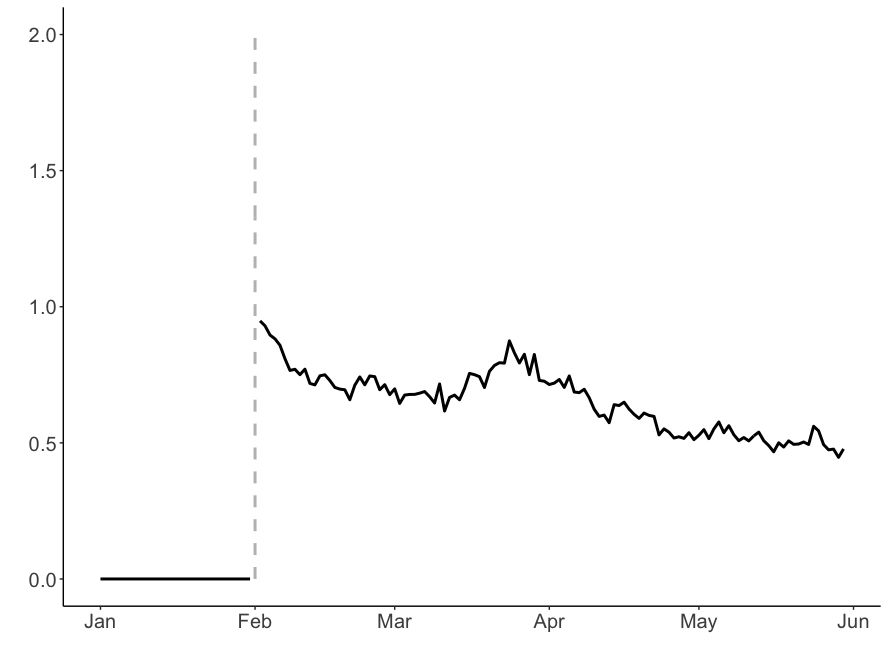}
        \caption{Estimated relative treatment effect}
        \label{fig:seq6}
    \end{subfigure}
\caption{Steps for estimating treatment
  effects: Figure~\ref{fig:seq1} shows the observed outcomes for unit 1 and
  Figure~\ref{fig:seq2} for unit 2. Unit 1 is treated while unit 2 is
  not. The treatment period (or post-intervention period) starts in
  February and is shown after the dashed line. During the pre-period in
  January, neither unit 1 nor unit 2 is treated. We learn the intervention
  from the observed outcomes through a state-space model. Consequently, we
  remove the learned effect from unit 1 in Figure~\ref{fig:seq3} and
  transfer it to unit 2  in Figure~\ref{fig:seq4}. By comparing the
  outcomes in Figures~\ref{fig:seq3} and \ref{fig:seq4}, we are able to
  estimate the average treatment effect (Figure~\ref{fig:seq5}) and the
  average relative treatment effect (Figure~\ref{fig:seq6}). We note that
  in this example the transferred effect is equal to the estimated effect
  on unit 1. Once more treated units become available, the transferred effect will be unit-specific and depend on the units' covariate values.}\label{fig:seq} 
\end{figure}

We illustrate the idea with an example. Suppose we run a simple experiment
on two units (both being a time series). Unit 1 is assigned to treatment and unit 2 to control. We observe the outcomes in Figures~\ref{fig:seq1} and \ref{fig:seq2}. During the treatment or post-intervention period, we are only able to observe the outcomes under treatment for unit 1 and under
control for unit 2.
To predict the counterfactuals, we can learn the intervention from the observed
outcomes through a state-space model.  The learned effect can be transferred from unit 1 (Figure~\ref{fig:seq3}) to unit 2 (Figure~\ref{fig:seq4}).
With both the observed and predicted outcomes, one can estimate the average
treatment effect (Figure~\ref{fig:seq5}), relative treatment effects
(Figure~\ref{fig:seq6}), or any other function of the outcomes. We note
that in this simple example with only one treated unit the transferred
effect from unit 1 to 2 is equal to the estimated effect on unit 1. Once
more treated units are available, one can learn an effect
function and transfer unit-specific treatment effects depending on each
unit's covariate values. Naturally, the estimated average treatment
effects also improve as more treated and control units become
available. Instead of transferring the learned intervention to another
unit, it is also possible to transfer in time to predict the effect of a
hypothetical intervention at another time point under the assumption that
the effect, which is possibly a function of covariates, is invariant under
time shifts. 

\subsection{Related work and our contribution}
A related method which uses state-space models for causal effect estimation is Causal Impact~\cite{brodersen2015}. Causal Impact infers the counterfactual of a treated univariate time series, that is, its outcome under no interventions. For this purpose, it requires a control time series: a covariate which is predictive of the time series of interest but not affected by treatment itself. Causal Impact learns the relationship between the response time series and the control time series by fitting a dynamic regression model during the pre-period. Causal Impact assumes that the learned relationship between the control and the response time series is stable over time. By doing so, Causal Impact is able to predict the counterfactual of the response time series for the treatment period. The treatment effect is estimated by subtracting the observed time series by the predicted counterfactual time series. 

A major difference between Causal Impact and our method, Causal Transfer, is that the former is restricted to counterfactuals of a treated time series. It cannot learn the intervention and predict the counterfactual of control time series. In particular, Causal Transfer has an advantage in that it is able to predict future treatment effects, for 
which no data points have been observed. This is not feasible with Causal Impact since it requires the data of the
treated time series to compute the effect. In many situations, not every
unit can be treated due to ethical or economic reasons. In these cases,
Causal Impact might not be able to estimate the effect or the estimate is
biased since it is based on treated units alone, which might not be
reflective of the control units. In the causal inference literature, this is referred to as the ``treatment effect on the treated''. Causal Transfer, on the other hand, is able to predict the counterfactual of both treated and untreated units as soon as it has seen the intervention on at least one treated unit. It adjusts the treatment effect for confounding. Therefore, it leads to less biased estimates of the treatment effects.

Causal Impact further requires the response time series to be univariate. Multivariate time series have to be aggregated cross-sectionally to estimate average treatment effects. Causal Impact cannot estimate heterogenous effects, such as the CATE, without making further assumptions.\\

Marginal integration~\cite{li2017} is another related method for causal
effect estimation. The main difference is that the regression function in
Equation~\eqref{eq:adjustment} is nonparametric and estimated with kernel
regression before integrating out the adjustment set. Marginal integration
can consistently estimate the ATE with optimal one-dimensional
nonparametric convergence rate $n^{-2/5}$ for continuous treatment variables \cite{li2017}. The price to be paid for such a general result is 
that it requires strict stationarity for the estimation of the smooth regression function and is,
therefore, restricted to observational time series. The theoretical
guarantees hold for estimands which are functions of $\mathbb{E}[Y\,|\,
\Do(T=t)]$ for some $t$ in the support of $T$, such as the ATE. It does not support the estimation of prediction
intervals. In principle, marginal integration can be extended to estimate sample average
treatment effects or heterogenous effects. The theoretical guarantees, however, may not carry over as the estimation of the latter is severely exposed to the curse of dimensionality. Marginal integration is capable of predicting
future effects but only up to the maximum time distance present in the data
and assuming stationarity.\\ 

In econometrics, popular methods for causal inference with panel data include fixed-effects and first-difference models \cite{finkel1995}. These models can control for selection biases due to time-invariant hidden confounders, but they also eliminate the effect of time-invariant observed variables. More recently, hybrid models were proposed that combine fixed with random effects to include the effect of time-invariant observables, for example~\cite{rabe2004}. These models assume that the unobserved time trend in the treatment group is equal to the observed time trend in the control group. \\

Causal Transfer estimates average or
heterogeneous effects (or any other function of the potential outcomes) and
their prediction intervals from potentially nonstationary, interventional
data assuming no latent confounding. It does not differentiate between
time-varying or time-invariant covariates while allowing the units to have
individual time trends (see Section~\ref{section:model2}). Causal Transfer
can be applied to randomized trials or observational studies. This makes
our method widely applicable to many real-world scenarios. Causal Transfer
fits a dynamic regression model to the data. Even though the 
regression model is linear, our method is robust for the ATE against
  misspecification (see Section~\ref{section:robust}). Lastly, Causal Transfer is able to
inform us about unseen future effects, which provides crucial
information when deciding whether a treatment is successful in the long run. 


\section{Causal effect estimation in nonstationary time series}
\label{section:theory}

\subsection{A simple example}
\label{section:example}

We begin with a simple example to help illustrate our goal. Suppose we are given the following table. The column $Y(1)$ lists the outcomes under treatment and the column $Y(0)$ the outcomes under control.

\begin{table}[h!]
\footnotesize
\centering
\begin{tabular}{ c | r | r| r}
 unit & $t$ & $Y_t(1)$ & $Y_t(0)$ \\ \hline
  1 & 1 & 5 & 3 \\
  2 & 1 & 6 & 5 \\
  3 & 1& 4 & 2 \\
  4 & 1 & 6 & 3 \\
  1 & 2 & 7 & 4 \\
  2 & 2 & 4 & 2 \\
  3 & 2 & 6 & 4 \\
  4 & 2 & 6 & 3 
\end{tabular}
\caption{The toy data set}
\end{table}
Given the full data table, it is straightforward to estimate the average
treatment effect (ATE) $\mathbb{E}[Y_t(1)-Y_t(0)]$. We can  estimate the
ATE as the difference in means between the treated outcomes and the control
outcomes, i.e., $\widehat{\mathrm{ATE}} = ((5 + 6 + 4 + 6) - (3 + 5 + 2 +
3))/4 = 2$ for $t=1$ and $\widehat{\mathrm{ATE}} = ((7 + 4 + 6 + 6) - (4 +
2 + 4 + 3))/4 = 2.5$ for $t=2$. It is no more difficult to estimate any
other function of the potential outcomes, e.g., relative treatment effects.  

In practice, we never observe the entire table but only one type of outcome per unit: either $Y(0)$ or $Y(1)$. Therefore, the estimation of causal effects involves predicting the missing outcomes.
\begin{table}[h!]
\footnotesize
\centering
\begin{tabular}{ c | r | r| r}
 unit & $t$ & $Y_t(1)$ & $Y_t(0)$ \\ \hline
  1 & 1 & - & 3 \\
  2 & 1 & - & 5 \\
  3 & 1 & 4 & - \\
  4 & 1 & 6 & - \\
  1 & 2 &- & 4 \\
  2 & 2 & - & 2 \\
  3 & 2 & 6 & - \\
  4 & 2 & 6 & - 
\end{tabular}
\caption{Observed and missing outcomes in the toy data.}
\label{tab:toy_CT1}
\end{table}

A naive approach for imputing the missing outcomes is to fit separate linear regression models of the observed outcomes against the treatment indicator $Y_t \sim T$ to each time point $t$. We then predict the missing outcomes from the fitted regression models. This step is especially easy for the simple linear regression $Y_t \sim T$. It means that, at each time point, we impute the missing treated outcomes by the average over the observed treated units and the missing control outcomes by the average over the observed control units. The resulting table with both observed and imputed outcomes becomes:

\begin{table}[h!]
\footnotesize
\centering
\begin{tabular}{ c | r | r| r}
 unit & $t$ & $Y_t(1)$ & $Y_t(0)$ \\ \hline
 1 & 1 & 5 & 3 \\
 2 & 1 & 5& 5 \\
 3 & 1 & 4 & 4 \\
 4 & 1 & 6 & 4 \\
 1 & 2 & 6& 4 \\
 2 & 2 & 6 & 2 \\
 3 & 2 & 6 & 3 \\
 4 & 2 & 6 & 3 
\end{tabular}
\caption{Predicted outcomes using linear regression.}
\end{table}

With the completed table, we can proceed as before for estimating the ATE or any other function of the outcomes. For example, the estimate of the ATE now equals $((5 + 5 +  4 + 6) - (3 + 5 + 4 + 4))/4 = 1$ for $t=1$ and $((6+6+6+6) - (4+2+3+3))/4  =  3$ for $t=2$. To make the predictions more meaningful, one can include predictive covariates in the regression model.

The above naive approach does not incorporate time dependence and only works well for reasonably large numbers of units. We sketch how to impute the missing outcomes with state-space models instead.
We replace the simple linear regression model by a dynamic regression model:
\begin{align}
Y_t & = \beta_{t} + \mu_{t} T + v_t \nonumber\\
\beta_{t} &= \beta_{t-1} + w_t \nonumber\\
\mu_{t} &= \mu_{t-1} + u_t\,,\nonumber
\end{align}
where $u_t$ denotes the observation noise and $v_t$, $w_t$ denote the
process noise. This is a special case of a linear state-space model where
the Kalman filter can be used. The first equation is the so-called
measurement equation. The second and third equation are the so-called state
equations. The measurement equation contains the regression model, while
the state equations describe the time evolution of the regression
coefficients. In this example, the states follow random walks. We predict the missing outcomes of Table~\ref{tab:toy_CT1}
through the dynamic regression model in Table~\ref{tab:toy_CT2}.

\begin{table}[h!]
\footnotesize
\centering
\begin{tabular}{ c | r | r| r}
 unit & $t$ & $Y_t(1)$ & $Y_t(0)$ \\ \hline
  1 & 1 & 5.5 & 3 \\
  2 & 1 & 5.5 & 5 \\
  3 & 1 & 4 & 3.5 \\
  4 & 1 & 6 & 3.5 \\
  1 & 2 & 5.5 & 4 \\
  2 & 2 & 5.5 & 2 \\
  3 & 2 & 6 & 3.5 \\
  4 & 2 & 6 & 3.5 
\end{tabular}
\caption{Predicted outcomes using dynamic regression.}
\label{tab:toy_CT2}
\end{table}

The estimated ATE using the Kalman filter estimates is $(21 - 15)/4
= 1.5$ for time point 1 and $(23 - 13)/4 = 2.5$ for time
point 2. One advantage of the Kalman filter is that it pools information
over time while learning the time dynamics of the treatment effect.  It makes use of the underlying smoothness between time points in an experiment. For
example, if the study units respond favourably to treatment on one day, it is unlikely that this will change drastically on the following day. The variance of the noise terms in the state equations can serve
as smoothness conditions on the regression coefficients and, thereby, on
the estimated effects. The separate regression models further emerge as a special case of the dynamic regression model by setting the AR coefficients of the state equations to 0. Having learned the time dynamics, the Kalman filter
also predicts unseen future effects. This is not possible with separate
regression models for each time point.
In the following Section, we will outline the imputation scheme with state-space models in general.


\subsection{State-space models for heterogeneous treatment
  effects} \label{section:model1}
  
Causal Transfer combines regression with time series models into a flexible framework for causal effect estimation. Many modelling choices are possible. For example, one can choose linear dynamic regression models (possibly with interactions or fixed/random effects) or nonlinear models.  Linear models can be fitted with the Kalman filter. Instead of the Kalman filter, one can use robust filters~\cite{durovic1999}~\cite{masreliez1977} or nonlinear filters, such as the unscented Kalman filter~\cite{julier2004} or the particle filter~\cite{del1996} to accommodate nonlinear models, non-Gaussian noise distributions, or outliers. A general description of the algorithm is given in Algorithm~\ref{alg:sketch}. 

\begin{algorithm}[!htb]
\small
\begin{flushleft}
       \hspace*{\algorithmicindent} \textbf{Input} Dynamic regression model, data of observed outcomes and covariates.\\
       \hspace*{\algorithmicindent} \textbf{Output} Estimated prediction intervals and treatment effects for every time point.
\end{flushleft}
\begin{algorithmic}[1]
\STATE (Optional) Estimate the unknown parameters of the dynamic regression model with the MLE or MCMC.
\STATE Apply a state-space method, e.g., the Kalman filter to estimate the states.
\STATE Estimate the treatment effect either from the states or by imputing the missing outcomes.
\STATE Estimate prediction intervals from the posterior distribution or through resampling methods.
\end{algorithmic}
\caption{\small Causal Transfer}\label{alg:sketch}
\end{algorithm}

We focus on linear models since they are robust against misspecification for the estimation of the ATE (see Section~\ref{section:robust}) and give examples of such models for the estimation of average and heterogenous effects. We consider a data set with $n \in \mathbb{N}$ time points and $d \in \mathbb{N}$ units.
The Kalman filter is a well-known method for a linear state-space model. It consists of a
measurement and a state equation: 
\begin{align}
X_t&=F_t\theta_{t} + v_t \label{eq:KF1}\\
\theta_{t} &= G_t\theta_{t-1} + w_t \label{eq:KF2}
\end{align}
for $t=1,\ldots, n$. The measurements $X_t = (X_{1,t},\ldots, X_{d,t})$ and the states $\theta_t = (\theta_{1,t},\ldots, \theta_{m,t})$ with $m \in \mathbb{N}$ are multivariate and $F_t$, $G_t$ are $d \times m$ and $m \times m$ matrices respectively. The noise terms are independent of each other and normally distributed according to $v_t \sim \mathcal{N}(0, V_t)$ and $w_t \sim \mathcal{N}(0, W_t)$ with $V_t \in \mathbb{R}^{d\times  d}$ and $W_t \in \mathbb{R}^{m \times m}$.
We assume that only the measurements $(X_t)_{t=1}^n$ of the measurement
equation~\eqref{eq:KF1} are observable but not the states
$(\theta_t)_{t=1}^n$. Therefore, our knowledge of the states comes from the
measurements of $(X_t)_{t=1}^n$ and the state equation~\eqref{eq:KF2}
alone. 
The measurements given the states $(X_t \,|\,\theta_t)_{t=1}^n$ are mutually
independent. The Gaussian assumption for the noise terms is made for simplicity in
  Section~\ref{section:algorithms}. In fact, the Kalman filter is optimal if the noise
  terms are Gaussian and provides consistent estimates of the states while minimizing
the mean-squared error. If the noise is not Gaussian, it is still the
optimal linear estimator (although nonlinear methods might perform
better). 

\paragraph{The working model.} We now switch the notation from $Y_i(0)$ and $Y_i(1)$ in Section~\ref{section:introduction} to a $d$-dim. response variable $X_t = (X_{1,t},\ldots, X_{d,t})$ with potential outcomes $X_{i,t}(0)$ and $X_{i,t}(1)$ for each unit $i=1,\ldots,d$. For the purpose of estimating causal effects, we suggest a dynamic regression model. For example, 
\begin{align}
X_t &= \beta_{0, t} + \beta_{1, t} X_{\text{pre}} + \beta_{2, t} Z_t + T (\mu_{0, t} + \mu_{1, t} X_{\text{pre}} + \mu_{2, t} g) + v_t
\label{eq:measurement} \\
\beta_{t} &= \beta_{t-1} + w_{t} \label{eq:state1} \\
\mu_{t} &= 
  \begin{pmatrix}
    c_1 &  0 & 0 \\
    0 & c_2 & 0 \\
    0 & 0 & c_3
  \end{pmatrix}
 \mu_{t-1} + u_{t} \,.\label{eq:state2}
\end{align} \label{eq:model1}

Here, $T = (T_1,\ldots, T_d)$, $Z_t = (Z_{1,t},\ldots, Z_{d,t})$,
$X_{\text{pre}} = (X_{1, \text{pre}},\ldots, X_{d, \text{pre}})$, and $g =
(g_1,\ldots, g_d)$. The states are $\beta_{t} = (\beta_{0,t}, \beta_{1, t},
\beta_{2, t})$ and $\mu_{t} = (\mu_{0, t}, \mu_{1, t}, \mu_{2, t})$. The
noise $v_t$ is normally distributed with covariance matrix $V_t \equiv V
= \sigma^2 \mathbb{I}_d$. The noise
terms $w_t$, $u_t$ are jointly normally distributed with arbitrary (but
constant) diagonal covariance matrices $W_t \equiv W$. The noise terms
$u_t$, $v_t$, and $w_t$ are independent of each other for all $t =
1,\ldots,n$. 

The variable $T$ is the treatment indicator, i.e., $T_{i} = 1$ for every unit $i$ that is assigned to treatment and $0$ otherwise. The treatment $T$ can also be understood as being continuous dose values, i.e., $T \in \mathbb{R}_0^+$. We assume that $T$ remains fixed over time implying that a whole time series unit $i$ is either treated ($T_i = 1$) or untreated ($T_i = 0$). It is possible to replace $T$ by a time varying indicator $T_t$.  A model with a time-varying $T_t$ will be introduced in Section~\ref{section:model2}.

The covariates in Equation~\eqref{eq:measurement} only serve as
examples. We assume that the covariates that appear in heterogeneous
treatment effects are unaffected by treatment and render $T_i$ and $(X_{i,
  t}(0), X_{i, t}(1))$ independent if conditioned on. The variables
$X_{\text{pre}}, Z_t, g$ represent different types of such
covariates.  The variable $X_{i, \text{pre}}$ is a pre-period covariate,
e.g., the value of unit $i$ before the treatment begins. Intuitively,
it makes sense that the treatment effect is in relation to the levels of
the pre-period covariate. Therefore, an interaction term $TX_{\text{pre}}$
between the treatment indicator and the pre-period covariate can be included in
the model. If the data indicates no such interaction, it can be
omitted. The variable $Z_t$ represents a contemporaneous, time-varying
covariate, which is predictive of the outcome $X_t$ but not affected by
treatment itself. 
The term $g$ is an optional factor with two levels 
indicating group membership. A factor with more than two levels can be
accommodated with multiple dummy variables. Instead of fixed effects, one can also include random effects which are useful for cluster randomized trials.  Of course, there can be more or
less covariates of each type to achieve unconfoundedness as in
\eqref{eq:condrand} or more accurate approximation of heterogeneous
treatment effects. For example, using more than
  one contemporaneous covariate per unit results in matrices for $Z_t$ at each $t$. Alternatively, there could be no covariates at all. In fact, if $T$
is already randomized, no covariates are needed for estimating the ATE. In
the simplest case of no covariates and $c_1 = 1$, the model reduces to the
one in Section~\ref{section:example}. If there is a large set of covariates
$\mathbf{C}_{i, t}=(X_{i, \text{pre}}, Z_{i,t}, g_i, \ldots)$, one can
regress on the propensity score $e(\mathbf{C}_{i, t})$ instead of all
covariates since the propensity score satisfies $(X_{i, t}(0), X_{i,
  t}(1))\perp T_i\,|\,e(\mathbf{C}_{i, t})$ as well. 

In structural equation modelling, causal structures are represented as directed acyclic graphs. Possible graphs that lead to conditional randomization as in \eqref{eq:condrand} are:
\begin{center}
\text{a)}
\begin{tikzpicture}[scale=0.6, line width=0.5pt, minimum size=0.58cm, inner sep=0.3mm, shorten >=1pt, shorten <=1pt, rounded corners=1mm]
     \normalsize
 \draw (0,2) node(1) [ draw ] {$X_{\text{pre}}$};
 \draw (-2,0) node(2) [ draw ] {$T$};
 \draw (2,0) node(3) [ draw ] {$X_t$};
 \draw (0,-2) node(4) [ draw ] {$Z_t$};
 \draw (-5,0) node(5) [ draw ] {$g$};
 
\path [->]   (1)  edge [bend left= 0]    (2);
\path [->]   (2)  edge [bend left= 0]    (3);
\path [->]   (1)  edge [bend left= 0]    (3);
\path  [->]  (4)  edge [bend left= 0]    (3);
\path  [->]  (4)  edge [bend left= 0]    (2);
\path  [->]  (5)  edge [bend left= 0]    (1);
\path  [->]  (5)  edge [bend left= 0]    (4);
\path  [->]  (5)  edge [bend left= 0]    (2);
\path  [->]  (5)  edge [bend left= 30]    (3);
      \end{tikzpicture}
~\hspace{0.75cm}~\text{b)}~\hspace{0.25cm}~
\begin{tikzpicture}[scale=0.6, line width=0.5pt, minimum size=0.58cm, inner sep=0.3mm, shorten >=1pt, shorten <=1pt, rounded corners=1mm]
     \normalsize
 \draw (0,2) node(1) [ draw ] {$X_{\text{pre}}$};
 \draw (-2,0) node(2) [ draw ] {$T$};
 \draw (2,0) node(3) [ draw ] {$X_t$};
 \draw (5,0) node(4) [ draw ] {$Z_t$};
 \draw (-5,0) node(5) [ draw ] {$g$};
 
\path [->]   (1)  edge [bend left= 0]    (2);
\path [->]   (2)  edge [bend left= 0]    (3);
\path [->]   (1)  edge [bend left= 0]    (3);
\path  [->]  (4)  edge [bend left= 0]    (3);
\path  [->]  (5)  edge [bend left= 0]    (2);
      \end{tikzpicture}
\end{center}

The ATE of $T$ on $X_t$ can be identified by regressing on an adjustment
set. When there are no hidden variables, Pearl's backdoor
criterion~\cite{pearl2000} is sufficient for finding a valid adjustment
set. The criterion requires that all backdoor paths between $T$ and $X_t$
are blocked by the adjustment set. For both graphs a) and b), a valid
adjustment set is $\{X_{\text{pre}}, Z_t, g\}$. In graph a),
$\{X_{\text{pre}}, Z_t, g\}$ blocks all backdoor paths. The only backdoor
path in graph b) is $T\leftarrow X_{\text{pre}} \rightarrow X_t$. Although
the adjustment set for b) only needs $X_{\text{pre}}$, including $Z_t$ can
help to improve efficiency and including $g$ does not lead to inconsistency and can be of interest for
heterogeneous effects.  


 \paragraph{An example.} We are interested in the effect of tax penalties on car producers. In the above model, we can choose $X_{i, t}$ as the sales of a car manufacturer $i$ from country $g_i$. The treatment $T$ is an increase in taxation due to tax penalties.  The indicator $T_{i}$ is set to 1 if a company $i$ is subject to the tax increase and $0$ otherwise. Whether or not and how much a company is treated depends on the country $g$ of unit $i$. The pre-period covariate $X_{i, \text{pre}}$ is the sales of company $i$ before the onset of the tax increase. A covariate $Z_{i,t}$, that is predictive of the sales but assumed to be unaffected by treatment could be the inflation rate or the GDP of country $g_i$.\\ 

The evolution of the regression coefficients $(\beta_{j, t}$, $\mu_{j, t})_{j=0,1,2}$ is modelled as random walks or autoregressive processes. The parameters $c_j$ let $\mu_{j, t}$ grow to infinity, or to remain of order $O(1)$ over time, depending on whether the absolute value of $c_j$ is larger or equal to 1, or smaller than 1. One can add optional trend terms $\alpha = (\alpha_1, \alpha_2, \alpha_3)$ and $\delta = (\delta_1, \delta_2, \delta_3)$ to the random walks such that
\begin{align}
\beta_{t} &= \alpha + \beta_{t-1} + w_{t}\nonumber\\
\mu_{t} &= \delta + 
  \begin{pmatrix}
    c_1 &  0 & 0 \\
    0 & c_2 & 0 \\
    0 & 0 & c_3
  \end{pmatrix}
 \mu_{t-1} + u_{t}. \nonumber
\end{align}
The constants $\alpha$ and $\delta$ modify the baselines of the trends of $\beta_t$ and $\mu_t$. The terms $\alpha$ and $\delta$ can also be time-varying $\alpha_t$ and $\delta_t$ as in local linear trend models. Furthermore, the model can be extended to capture the seasonality of the data by including higher-lagged terms in the state equations~\cite{brodersen2015}. The above equations can, of course, be exchanged in favor of other equations if these are better suited.\\

Equation~\eqref{eq:measurement} is the measurement equation. Equations~\eqref{eq:state1} and \eqref{eq:state2} are the state equations. The equations can be easily rewritten into the form needed for the Kalman filter:
\begin{align}
\begin{pmatrix}
X_{1,t}\\
\vdots\\
X_{d,t}
\end{pmatrix}
 &=   
\begin{pmatrix}
    1 &  X_{1, \text{pre}} & Z_{1, t} & T_{1} & T_{1}X_{1, \text{pre}} &T_{1}g_{1}\\
    \vdots & \vdots & \vdots & \vdots & \vdots & \vdots \\
    1 &  X_{d, \text{pre}} & Z_{d, t} & T_{d} & T_{d}X_{d, \text{pre}} &T_{d}g_{d}
  \end{pmatrix} 
  \begin{pmatrix}
    \beta_{0, t} \\
     \beta_{1, t} \\
    \beta_{2, t} \\
    \mu_{0, t}\\
     \mu_{1, t}\\
      \mu_{2, t}
  \end{pmatrix}  + 
   \begin{pmatrix}
v_{1,t}\\
\vdots\\
v_{d,t}
\end{pmatrix} \\
  \begin{pmatrix}
    \beta_{0, t} \\
     \beta_{1, t} \\
    \beta_{2, t} \\
    \mu_{0, t}\\
     \mu_{1, t}\\
      \mu_{2, t}
  \end{pmatrix}   &= 
    \begin{pmatrix}
    1 & 0 & 0 & 0 &  0 & 0 \\
    0 & 1 & 0 & 0 &  0 & 0 \\
    0 & 0 & 1 & 0 &  0 & 0 \\
    0 & 0 & 0 & c_0 & 0 & 0 \\
    0 & 0 & 0 & 0 & c_1 & 0 \\
    0 & 0 & 0 & 0 & 0 & c_2 \\
  \end{pmatrix}
  \begin{pmatrix}
    \beta_{0, t-1} \\
     \beta_{1, t-1} \\
    \beta_{2, t-1} \\
    \mu_{0, t-1}\\
     \mu_{1, t-1}\\
      \mu_{2, t-1}
  \end{pmatrix}  
  + 
    \begin{pmatrix}
    w_{0, t} \\
    w_{1, t} \\
    w_{2, t} \\
    u_{0, t} \\
    u_{1, t} \\
    u_{2, t}
\end{pmatrix}
\end{align}

Regression estimators are known to be sensitive towards differences in covariate distributions between the treatment and the control group. If there is a lack of overlap between the covariate values, it could be beneficial to combine the dynamic regression model~\eqref{eq:measurement} with matching (e.g., \cite{gu1993}), blocking (e.g., \cite{rosenbaum1983}), or weighting (e.g., \cite{robins1997}). These methods can help to balance the covariate values. For example, one can match units with similar covariate values or propensity scores and apply our method to the matched data set.

A further remark concerns the identifiability of model~\eqref{eq:measurement}. If $T \equiv 0$, one cannot possibly say anything about the effect of an intervention in the absence of having seen any, without making further assumptions. If $d=1$ and $T\equiv 1$, the states are not identifiable since some of the predictors are perfectly collinear. The states only become identifiable once there is at least one treated and one control unit or, alternatively, if there is a pre-period in addition to the treatment period as outlined in Section~\ref{section:model2}. Since our goal is to estimate average and heterogeneous treatment effects, we will assume in the remainder of the manuscript to have observed at least one treated and one control unit. 


\subsubsection{Model extension for unit-specific states} \label{section:model2}
One can extend the previous model in a way that allows each unit to have its own set of
regression parameters. The observational part is then individual to each unit while the treatment parameters $\mu_{t}$ are shared across units. In state-space form this can be written as:
\begin{align}
\begin{pmatrix}
X_{1,t}\\
\vdots\\
X_{d,t}
\end{pmatrix}
 =   F_t 
\begin{pmatrix}
    \beta_{0, 1, t} \\
    \vdots \\
    \beta_{0, d, t} \\
    \beta_{1, 1, t} \\
    \vdots \\
    \beta_{1, d, t} \\
    \mu_{0, t}\\
    \mu_{1, t}\\
    \mu_{2, t}\\
\end{pmatrix}  + 
   \begin{pmatrix}
v_{1,t}\\
\vdots\\
v_{d,t}
\end{pmatrix}, \nonumber
\end{align}
where
\begin{align}
F_t = 
\begin{pmatrix}
X_{1, \text{pre}} \ \  0 \ldots  0  \ \ Z_{1, t} \ \  0 \ldots
 0  \ \  T_{1, t}  \ \ T_{1, t}X_{1, \text{pre}}  \ \  T_{1, t}g_{1}\\
 0  \ \  X_{2, \text{pre}} \ldots 0  \ \  0  \ \  Z_{2,t} \ldots 0 \ \ 
T_{2, t}  \ \  T_{2,t}X_{2, \text{pre}} \ \  T_{2, t}g_{2}\\
 \ \  \vdots \\
0 \ \ 0 \ldots X_{d, \text{pre}}  \ \  0 \ \  0 \ldots Z_{d,t}  \ \  T_{d, t}  \ \  T_{d, t}X_{d, \text{pre}}  \ \  T_{d, t}g_{d}
 \end{pmatrix} . \nonumber
\end{align}

For every unit $i$, the corresponding state equations are
\begin{align}
\beta_{0,i, t} &= \beta_{0,i, t-1} + w_{0,i, t}, \nonumber\\
\beta_{1,i, t} &= \beta_{1,i, t-1} + w_{1,i, t},\nonumber\\
\mu_{t} &=
  \begin{pmatrix}
    c_1 &  0 & 0 \\
    0 & c_2 & 0 \\
    0 & 0 & c_3
  \end{pmatrix}
 \mu_{t-1} + u_{t}. \nonumber
\end{align}
We assume that $v_t \sim \mathcal{N} (0, V)$ with $V=\mathrm{diag}(\sigma_1^2, \ldots, \sigma_d^2)$ and that $w_{t} = (w_{0,1,t},\ldots,  w_{1,d,t})$ and $u_t$ are jointly normally distributed with diagonal covariance matrix $W$. The noise terms $u_t$, $v_t$, and $w_t$ are independent of each other for all $t = 1,\ldots,n$.
This model has a time-varying treatment indicator $T_{i, t}$ and requires a pre-period for identifiability. It makes use of data from the pre-period and the treatment period and can also be fitted to a single treated unit (by removing the interaction terms  $T_{t}X_{\text{pre}}$ and $T_{t}g$). Its advantage is that it allows each unit to evolve individually while the treatment effect $X_{i, t}(1) - X_{i, t}(0) = \mu_{0, t} + \mu_{1, t}X_{i, \text{pre}}+\mu_{2, t}g_i$ is still a transferable function of the covariate values of all units.  This model extension, however, increases the number of unknown variance parameters ($d$ variance parameters for $v_t$) which need to be estimated and potentially the risk of overfitting. In the following empirical sections, we will use the simpler dynamic regression model~\eqref{eq:measurement} instead. 

\subsection{Robustness against model misspecification} \label{section:robust}
We argue that a misspecified (dynamic) regression model is not problematic when aiming to estimate the ATE, while it has a (negative) effect for the CATE. For simplicity, we present the argument in the notation of Section~\ref{section:introduction} without time series.

The potential outcomes are denoted by $Y(1)$ and $Y(0)$: we dropped the
index $i$ of the units here. We assume that the treatment assignment is randomized conditional on the covariates $X$ as
in \eqref{eq:condrand}. From formula~\eqref{eq:adjustment}, we know that the ATE can be inferred by 
$$\E[m_1(X)] - \E[m_0(X)]$$
with $m_1(x) = \E[Y|T=1,X=X]$ and $m_0(x) = \E[Y|T=0,X=X]$.
Assume now that we have misspecified the regression functions $m_T(\cdot)$
for $T \in \{0,1\}$. The misspecified function is assumed to be of the form 
\begin{eqnarray*}
\mu_T + g_{\theta_T}(\cdot),
\end{eqnarray*}
with intercept $\mu_T$. The best $L_2$-approximation is
defined by 
\begin{eqnarray*}
\mu_T^*,\theta_T^* =\ \mbox{argmin}_{\mu_T,\theta_T} \E[(m_T(X) - \mu_T -
  g_{\theta_T}(X))^2].
\end{eqnarray*}
The partial derivative with respect to $\mu_T$ then leads to 
\begin{eqnarray*}
\E[m_T(X) - \mu_T^* - g_{\theta^*_T}(X)] = 0. 
\end{eqnarray*}
This implies that if we integrate out $X$ in the misspecified model,
we obtain
\begin{eqnarray*}
\E[\mu^*_T + g_{\theta^*_T}(X)] = \E[m_T(X)] = \E[Y(T)],
\end{eqnarray*}
where the last inequality follows by \eqref{eq:condrand}. Thus, even when
we have misspecified the regression functions $m_T(\cdot)$ with a
parametric model, the ATE can be consistently inferred. The empirical 
results in Section~\ref{section:empirical} confirm this.

Obviously, when $m_T(\cdot)$ is misspecified, the heterogeneous treatment
effect will be misspecified: it is only when averaging out $X$ as in
the ATE that the misspecification has no effect on the inference.

\subsection{Robustness against outliers}

Real data sets often contain outliers. The Kalman filter is sensitive towards outliers. Manual inspection of the data and removal of outliers is cumbersome. To guard against the influence of outliers, one can replace the Kalman filter with a robust filter in Causal Transfer. Robust filters in the literature include filters that minimize a robust loss function, e.g., \cite{durovic1999}, \cite{chan2005} and \cite{kovavcevic1992} and filters that learn weights for each observation and assign less weights to outliers, e.g., \cite{agamennoni2011} and \cite{ting2007}. One way to implement a filter that minimizes a robust loss is to use the equivalence between the Kalman filter and least squares.  Suppose the model is linear and of the general form:
\begin{align}
X_t&=F_t\theta_{t} + v_t, \hspace{0.1cm} v_t \sim \mathcal{N}(0, V_t)\nonumber\\
\theta_{t} &= G_t\theta_{t-1} + w_t, \hspace{0.1cm} w_t \sim \mathcal{N}(0, W_t) \text{ for } t=1,\ldots, n\nonumber
\end{align}

The estimating equations of the Kalman filter can then be rewritten into a sequence of equivalent least squares problems: 
\begin{align}
[I \hspace{0.1cm} F_t]^T \theta_{t} &= [G_t m_{t-1} \hspace{0.1cm} X_t]^T+ e_t, \hspace{0.1cm} e_t \sim \mathcal{N}(0, \Sigma_t) \text{ for } t=1,\ldots, n\nonumber.
\end{align}

The covariance $\Sigma_t$ is blockdiagonal with diagonal elements $(R_t, V_t)$.  The matrix $R_t$ is equal to $R_t = G_tC_{t-1}G_t^T+ W_t$ and the vector $m_t$ is the solution to the above least squares problem at time point $t-1$. M-estimation can now be carried out on the equivalent least square problems. In practice, this is achieved by applying iterated reweighted least squares.

Instead of robust filters, one can combine nonlinear filters with Causal Transfer, such as the particle filter (also known sequential Monte Carlo methods). Nonlinear filters are beneficial if the state or the measurement equation contain nonlinearities or the noise is non-Gaussian, albeit at the price of higher computational cost.

\subsection{Imputation with state-space models} \label{section:algorithms}

We now discuss the Kalman filter version of the algorithm in greater detail. Before describing the steps, we will briefly outline how inferences are done with the Kalman filter. This is by no means exhaustive and we refer to \cite{petris2009} for more details. 

\paragraph{Kalman filter.} As before, we consider the model:
\begin{align}
X_t&=F_t\theta_{t} + v_t \nonumber\\
\theta_{t} &= G_t\theta_{t-1} + w_t \text{ for } t=1,\ldots, n\nonumber.
\end{align}
The precise form of each element can be customized. Two proposals are given in Section~\ref{section:model1}. At time point $0$, we assume that the distribution of the states are known with $\theta_0 \sim \mathcal{N}(m_0, C_0)$. This is our starting point. At each time point $t$, one can propagate the distribution of the states according to the state equation~\eqref{eq:KF2}. The resulting distribution is the one-step-ahead predictive distribution of $\theta_t\,|\,X_{1,\ldots,t-1}$. If $\theta_{t-1}\,|\,X_{1,\ldots,t-1}\sim \mathcal{N}(m_{t-1}, C_{t-1})$, the one-step-ahead predictive distribution is $\theta_t\,|\,X_{1,\ldots,t-1} \sim \mathcal{N}(a_t, R_t)$ with $a_t = G_t m_{t-1}$ and $R_t = G_t C_{t-1}G_t' + W_t$. Similarly, one can compute the one-step-ahead predictive distribution of the measurements $X_t\,|\, X_{1,\ldots,t-1}$ according to the measurement equation~\eqref{eq:KF1}, which is $\mathcal{N}(f_t, Q_t)$ with $f_t = F_t a_{t}$ and $Q_t = F_t R_{t}F_t' + V_t$. The one-step-ahead predictive distribution can be combined with the likelihood of the next observation $X_t\,|\,\theta_t$ for obtaining the filtering distribution of $\theta_t\,|\,X_{1,\ldots,t}$ using Bayes theorem:
$$p(\theta_t\,|\,X_{1,\ldots,t}) = p(\theta_t\,|\,X_{1,\ldots,t-1}) p(X_t\,|\,\theta_t) / p(X_t\,|\, X_{1,\ldots,t-1})\,.$$
The filtering distribution of $\theta_t\,|\,X_{1,\ldots,t}$ is $\mathcal{N}(m_t, C_t)$ with $m_t = a_{t} + R_tF_t'Q_t^{-1}e_t$ and $C_t = R_{t} - R_tF_t'Q_t^{-1}F_tR_t$. The forecast error is $e_t = X_t-f_t$. Intuitively, this step can be understood as a correction. The predicted distributions according to the state-space model are corrected for the latest observation. The weight of the correction term $R_tF_t'Q_t^{-1}$ is referred to as the Kalman gain matrix. It depends on $W_t$ and $V_t$ through $R_t$ and $Q_t$. If the error variance of the predicted states $R_t$ is small, the Kalman gain is small giving the correction term little weight and $m_t \approx a_t$. If, on the other hand, the error variance of the measurement equation $V_t$ is small, $K_t=R_tF_t'Q_t^{-1} \approx R_tF_t'(F_t R_{t}F_t' )^{-1} = F_t^{-1}$ and $m_t \approx F_t^{-1} X_t$ and more weight is given to the most recent observation $X_t$. The choice of the Kalman gain as $R_tF_t'Q_t^{-1}$ is optimal in terms of minimizing the mean-squared error of the forecast $m_t$ for $\theta_t$.\\

The filtering distribution of $\theta_t\,|\,X_{1,\ldots,t}$ only make use of the observations up to time point $t$. In retrospect, it is often desirable to estimate the distributions with all available data, i.e., the smoothing distribution of  $\theta_t\,|\,X_{1,\ldots,n}$. The smoothing distributions can be estimated recursively and backwards in time starting from the filtering distribution of $\theta_n\,|\,X_{1,\ldots,n}$. If $\theta_{t+1}\,|\,X_{1,\ldots,n} \sim \mathcal{N}(s_{t+1}, S_{t+1})$, the smoothing distribution of $\theta_{t}\,|\,X_{1,\ldots,n}$ for $t=1,\ldots, n-1$ is $\mathcal{N}(s_{t}, S_{t})$ with 
\begin{align}
\label{eq:smoothrec}
s_t&=m_t+ C_t G_{t+1}' R_{t+1}^{-1} (s_{t+1}-a_{t+1})\\
S_t&=C_t-C_t G_{t+1}'R_{t+1}^{-1}(R_{t+1}-S_{t+1})R_{t+1}^{-1}G_{t+1}C_t\,.
\end{align} 
The smoothing recursions can be derived from $p(\theta_{t}\,|\,X_{1,\ldots,n}) = \int p(\theta_{t+1}\,|\,X_{1,\ldots,n}) p(\theta_{t}\,|\, \theta_{t+1}, X_{1,\ldots,n}) d\theta_{t+1}$. Intuitively, they can be understood as adding the information that is gained from the  time points $t+1,\ldots, n$, which is not included in the filtering distributions.\\

In many situations, the parameters in the state-space model are not fully known. For example, the noise covariance matrices $V_t$ and $W_t$ or the parameters in $G_t$ may be unknown and have to be estimated from data. In these cases, one can use the maximum-likelihood estimation (MLE) for inferring the unknown parameters. Let $\psi$ denote the unknown parameters. The distribution of $X_{1,\ldots,n}\,|\,\psi$ can be factorized into $p(X_{1,\ldots,n}\,|\,\psi) = \prod_{t=1}^n p(X_t\,|\, X_{1,\ldots, t-1},\psi)$. Each term in this product can be recognized as the one-step-ahead predictive distribution of $X_t\,|\, X_{1,\ldots,t-1},\psi$. From before, we know that the terms are equal to  $\mathcal{N}(f_t, Q_t)$. Therefore, one can rewrite the log-likelihood as 
$$l(\psi) = -\frac{1}{2} \sum_{t=1}^{n} \log|Q_t| -\frac{1}{2} \sum_{t=1}^{n} (x_t-f_t)' Q_t^{-1} (x_t-f_t)\,.$$
This expression depends on $\psi$ implicitly through the terms $f_t$ and $Q_t$. Maximizing the log-likelihood or minimizing the negative log-likelihood with respect to $\psi$ yields the MLE estimate of $\psi$.
For computing the MLE, one can utilize the function \texttt{dlmMLE} from the R-package dlm~\cite{petris2010}. This function builds the likelihood of the Kalman filter as a function of $\psi$ and parses this expression to the function \texttt{optim} for optimization. A word of caution is necessary here. The likelihood function of the Kalman filter can exhibit many local minima. Therefore, it is advised to try different starting values for the MLE routine and to compare the resulting estimates before proceeding with further steps.
The MLE in conjunction with the Kalman filter has been shown to be consistent and asymptotically normal for stationary and nonstationary but asymptotically identifiable models \cite{tommaso2012}. \citet{chang2009} have proven its consistency and asymptotic mixed normality for nonstationary models that involve integrated time series. Although, to the best of our knowledge, there is no asymptotic theory for the MLE in general nonstationary models, the Kalman filter with ML estimated parameters is still regarded to work well. Some examples are reviewed by \citet{kim1999}.

\paragraph{Causal Transfer.} 
We now turn to two versions of Causal Transfer for predicting population and sample treatment effects. For illustration purposes, we include examples for model~\eqref{eq:measurement}.

\paragraph{Sample version} This version consists of the following steps. We first estimate the unknown parameters with MLE if needed, such as the covariance terms and the constants $c_0$, $c_1$, $c_2$. We then plug in the estimated parameters into the state-space model. For all time points, we estimate the smoothing distribution $\mathcal{N}(s_t, S_t)$ of the states by iterating the smoothing recursions~\eqref{eq:smoothrec}. Given the smoothing distribution, we estimate the distribution $\mathcal{N}(\tilde{f}_t, \tilde{Q}_t)$ of the missing outcomes through $\tilde{f}_t = \tilde{F}_t s_t$ and $\tilde{Q}_t = \tilde{F}_t S_t \tilde{F}_t'+V_t$. The matrix $\tilde{F}_t$ contains the covariates of the counterfactuals. This implies that $\tilde{F}_t$ is the same as $F_t$ except for $T$ being replaced by $\tilde{T} = |T-1|$. We draw $B$ set of samples $(\tilde{x}_{t}^{(b)})_{b=1}^B$ from $\mathcal{N}(\tilde{f}_t, \tilde{Q}_t)$ to impute the missing outcomes. We estimate the estimand of interest $\hat{\tau}_{t}^{(b)}$ from the data $x_t$ and the sample $\tilde{x}_{t}^{(b)}$. For the SATE, the effect sample is given by $\hat{\tau}_t^{(b)} = (\tilde{x}_{t}^{(b)}-x_{t}) \cdot (\tilde{T} - T)$ for every $t$. Lastly, we summarize the effect samples $(\hat{\tau}_t^{(b)})_{b=1}^B$ into prediction intervals and point estimates, e.g., by taking the percentiles and the average. A pseudo-code version is shown below in algorithm~\ref{alg:CT}. One can also estimate other estimands than the SATE, e.g., relative treatment effects. \\

\begin{algorithm}[!htb]
\small
\begin{algorithmic}[1]
\vspace{0.4cm}
\STATE (Optional) Estimate the unknown coefficients with the MLE.
\STATE Estimate the smoothing distribution of the time series $\mathcal{N}(s_t, S_t)$ at every time point $t = 1,\ldots, n$.
\STATE Estimate the distribution of the counterfactual series $\mathcal{N}(\tilde{f}_t, \tilde{Q}_t)$ at every time point $t = 1,\ldots, n$.
\FOR{$b = 1,\ldots, B$}
\STATE (i)\hspace{0.11cm} Draw one sample $\tilde{x}_{t}^{(b)}$ from $\mathcal{N}(\tilde{f}_t, \tilde{Q}_t)$.
\STATE (ii)\hspace{0.03cm} Estimate effect sample $\hat{\tau}_{t}^{(b)}$ from the observed $x_t$ and imputed $\tilde{x}_{t}^{(b)}$.
\ENDFOR
\STATE Summarize the effect samples $(\hat{\tau}_{t}^{(b)})_{b=1}^B$ into prediction intervals.
\STATE Estimate sample treatment effect as  $\hat{\tau}_{t} = \frac{1}{B}\sum_{b=1}^B\hat{\tau}_{t}^{(b)}$.
\RETURN Estimated prediction intervals and treatment effects for time points $t=1,\ldots,n$.
\end{algorithmic}
\caption{\small Causal Transfer for sample treatment effects}\label{alg:CT}
\end{algorithm}

Causal Transfer naturally extends to the prediction of unseen future treatment effects.  We denote the future time points by $t = n+1,\ldots, n_{\text{ahead}}$. We predict the distribution of the states $\mathcal{N}(m_t, C_t)$ for the time period of interest recursively starting from  $\mathcal{N}(m_n, C_n)$. $\mathcal{N}(m_n, C_n)$ is the filtering distribution at $t = n$, which coincides with the smoothing distribution at this time point. The parameters of the distribution are updated according to $m_t = G_t m_{t-1}$ and $C_t = G_t C_{t-1} G_t' + W_t$ for $t = n+1,\ldots, n_{\text{ahead}}$. No correction step with the Kalman gain is needed since no observations have been made. We draw the predicted outcomes of the time series from $\mathcal{N}(f_t, Q_t)$ with $f_t = F_t m_t$ and $Q_t = F_t C_t F_t'+V_t$ and the predicted outcomes of the counterfactual series from $\mathcal{N}(\tilde{f}_t, \tilde{Q}_t)$ with $\tilde{f}_t = \tilde{F}_t m_t$ and $\tilde{Q}_t = \tilde{F}_t C_t \tilde{F}_t'+V_t$. We then proceed with the same steps as before for obtaining the prediction intervals and the point estimates. Algorithm~\ref{alg:CT_pred} contains the pseudo-code version.

\begin{algorithm}[!htb]
\small
\begin{algorithmic}[1]
\vspace{0.4cm}
\STATE Predict the distribution $\mathcal{N}(m_t, C_t)$ from the previous $\mathcal{N}(m_{t-1}, C_{t-1})$ for $t = n + 1,\ldots, n_{\text{ahead}}$.
\STATE Predict the distribution of the time series $\mathcal{N}(f_t, Q_t)$.
\STATE Predict the distribution of the counterfactual series $\mathcal{N}(\tilde{f}_t, \tilde{Q}_t)$.
\FOR{$b = 1,\ldots, B$}
\STATE (i)\hspace{0.11cm} Draw one sample $\tilde{x}_{t}^{(b)}$ and $x_{t}^{(b)}$ from $\mathcal{N}(\tilde{f}_t, \tilde{Q}_t)$ and $\mathcal{N}(f_t, Q_t)$.
\STATE (ii)\hspace{0.03cm} Predict effect sample $\hat{\tau}_{t}^{(b)}$ from $x_{t}^{(b)}$  and $\tilde{x}_{t}^{(b)}$.
\ENDFOR
\STATE Summarize the effect samples $(\hat{\tau}_{t}^{(b)})_{b=1}^B$ into prediction intervals.
\STATE Predict sample treatment effect as $\hat{\tau}_{t} = \frac{1}{B}\sum_{b=1}^B \hat{\tau}_{t}^{(b)}$. 
\RETURN Prediction intervals and predicted treatment effects for $t=n + 1,\ldots, n_{\text{ahead}}$.
\end{algorithmic}
\caption{\small Causal Transfer for (unseen) future sample treatment effects}\label{alg:CT_pred}
\end{algorithm}

\paragraph{Population version}
In linear models, the ATE is especially simple and can be read off directly from the model coefficients: $\mu_{0,t} + \mu_{1,t} \mathbb{E}[X_{\text{pre}}] + \mu_{2,t} \mathbb{E}[g]$. This example illustrates that the ATE and other population treatment effects can be estimated directly from $\mu_{t}$ without having to impute the missing potential outcomes. Hence, one can skip the steps 4-6 for population treatment effects, such as the ATE. Instead, one draws samples $(s_{t}^{(b)})_{b=1}^B$ directly from the smoothing distribution of the states $\mathcal{N}(s_t, S_t)$ to estimate the effect samples $(\hat{\tau}_{t}^{(b)})_{b=1}^B$. For the ATE, $\hat{\tau}_{t}^{(b)}= \frac{1}{d}\sum_{i=1}^d ( s_{0,t}^{(b)} + s_{1,t}^{(b)}  X_{i, \text{pre}} + s_{2,t}^{(b)} g_i)$. This is equivalent to replacing all outcomes by the expected outcomes. Suppose we sample the states $s_{t}^{(b)}$ from $\mathcal{N}(s_t, S_t)$. We replace the observed outcomes by the expected outcomes $f_{t}^{(b)} = F_t s_{t}^{(b)}$ and the missing outcomes by the expected outcomes $\tilde{f}_{t}^{(b)} = \tilde{F}_t s_{t}^{(b)}$. The effect sample $\hat{\tau}_{t}^{(b)} = (\tilde{f}_{t}^{(b)}-f_{t}^{(b)}) \cdot (\tilde{T} - T)$ then reduces to $\frac{1}{d}\sum_{i=1}^d (s_{0,t}^{(b)} + s_{1,t}^{(b)} X_{i, \text{pre}} + s_{2,t}^{(b)} g_i)$. The samples $(\hat{\tau}_{t}^{(b)})_{b=1}^B$ can be summarized into prediction intervals as before. The ATE point estimate $\hat{\tau}_{t}$ can be estimated directly from $s_t$, e.g., $\hat{\tau}_{t}=\frac{1}{d}\sum_{i=1}^d (s_{0,t} + s_{1,t}  X_{i, \text{pre}} + s_{2,t} g_i)$. The population version of Causal Transfer is shown in algorithm~\ref{alg:CT_sp}. Algorithm~\ref{alg:CT_sp} is computationally faster than algorithm~\ref{alg:CT} and accelerates computation time by roughly a factor 3. Both versions can be accelerated by parallelization of the sampling process or by sequential processing of the data points~\cite{durbin2012}.

\begin{algorithm}[!htb]
\small
\begin{algorithmic}[1]
\vspace{0.4cm}
\STATE (Optional) Estimate the unknown coefficients with the MLE.
\STATE Estimate the smoothing distribution of the time series $\mathcal{N}(s_t, S_t)$ at every time point $t = 1,\ldots, n$.
\FOR{$b = 1,\ldots, B$}
\STATE (i)\hspace{0.11cm} Draw one sample $s_{t}^{(b)}$ from $\mathcal{N}(s_t, S_t)$.
\STATE (ii)\hspace{0.03cm} Estimate effect sample $\hat{\tau}_{t}^{(b)}$ from $s_{t}^{(b)}$.
\ENDFOR
\STATE Summarize the effect samples $(\hat{\tau}_{t}^{(b)})_{b=1}^B$ into prediction intervals.
\STATE Estimate population treatment effect $\hat{\tau}_{t}$ from $s_t$.
\RETURN Estimated prediction intervals and treatment effects for time points $t=1,\ldots,n$.
\end{algorithmic}
\caption{\small Causal Transfer for population treatment effects}\label{alg:CT_sp}
\end{algorithm}

\paragraph{Heterogeneous version}
Population treatment effects implicitly assume that the units in the study are reflective of a larger population with an underlying distribution function. Naturally, population treatment effects can be transferred to new units that did not take part in the study as long as they arise from the same population. The individual effect for a new unit with covariates $X_{\text{pre}}$ and $g$ is the CATE $\mathbb{E}[X_t(1)-X_t(0)\,|\, X_{\text{pre}}, g]$. To estimate the CATE with algorithm~\ref{alg:CT_sp}, one estimates the effect sample in step 6 of algorithm~\ref{alg:CT_sp} as $\hat{\tau}_{t}^{(b)} = s_{0,t}^{(b)} + s_{1,t}^{(b)}  X_{\text{pre}} + s_{2,t}^{(b)}g$ and the effect in step 9 of algorithm~\ref{alg:CT_sp} as $\hat{\tau}_{t}=s_{0,t} + s_{1,t}  X_{\text{pre}} + s_{2,t}g$. 

We can integrate $X_{\text{pre}}$ out if we are interested solely in the effects as a function of $g$ since $\mathbb{E}[\mathbb{E}[X_t(1)-X_t(0)\,|\, g, X_{\text{pre}}]\,|\, g] = \mathbb{E}[X_t(1)-X_t(0)\,|\, g]$. The expression $\mathbb{E}[X_t(1)-X_t(0)\,|\, g]$ is the so-called marginal conditional average treatment effect (MCATE)~\cite{grimmer2017}. The MCATE is useful for comparing treatment effects between groups of units with different values of $g$. The effect samples for the MCATE in step 6 can be estimated as $\hat{\tau}_{t, g=0}^{(b)} = \frac{1}{d_0}\sum_{i; g_i = 0} (s_{0,t}^{(b)} + s_{1,t}^{(b)}  X_{i, \text{pre}})$ for group 1 with $g=0$ and $\hat{\tau}_{t,g=1}^{(b)} = \frac{1}{d_1}\sum_{i; g_i = 1} (s_{0,t}^{(b)} + s_{1,t}^{(b)}  X_{i, \text{pre}} + s_{2,t}^{(b)})$ for group 2 with $g=1$. Here, $d_0$ denotes the number of units in the group with $g=0$ and $d_1$ the number of units in the group $g=1$. The effect samples can be summarized into prediction intervals within each group. In the same way, the effect in step 9 can be estimated as  $ \hat{\tau}_{t, g=0}=\frac{1}{d_0}\sum_{i; g_i = 0} (s_{0,t} + s_{1,t}  X_{i, \text{pre}}) $ and $ \hat{\tau}_{t, g=1}=\frac{1}{d_1}\sum_{i; g_i = 1}  (s_{0,t} + s_{1,t}  X_{i, \text{pre}} + s_{2,t} )$  for each group respectively. The samples of the effect differences between the groups $\mathbb{E}[X_t(1)-X_t(0)\,|\, g=0] -\mathbb{E}[X_t(1)-X_t(0)\,|\, g=1]$ can be estimated conveniently from the effect samples $\hat{\tau}_{t,\Delta_g}^{(b)} = \hat{\tau}_{t,g=0}^{(b)}-\hat{\tau}_{t,g=1}^{(b)}$. The estimated average effect difference is equal to $\hat{\tau}_{t, \Delta_g} = \hat{\tau}_{t, g=0} - \hat{\tau}_{t,g=1}$.

The prediction of unseen future population treatment effects (Algorithm~\ref{alg:CT_pred_sp}) is as simple as for sample treatment effects.

\begin{algorithm}[!htb]
\small
\begin{algorithmic}[1]
\vspace{0.4cm}
\STATE Predict the distribution $\mathcal{N}(m_t, C_t)$ from the previous $\mathcal{N}(m_{t-1}, C_{t-1})$ for $t = n + 1,\ldots, n_{\text{ahead}}$..
\FOR{$b = 1,\ldots, B$}
\STATE (i)\hspace{0.11cm} Draw one sample $s_{t}^{(b)}$ from $\mathcal{N}(m_t, C_t)$
\STATE (ii)\hspace{0.03cm} Predict effect sample $\hat{\tau}_{t}^{(b)}$ from $s_{t}^{(b)}$.
\ENDFOR
\STATE Summarize the effect samples $(\hat{\tau}_{t}^{(b)})_{b=1}^B$ into prediction intervals.
\STATE Predict population treatment effect $\hat{\tau}_{t}$ from $m_t$.
\RETURN Prediction intervals and predicted treatment effects for $t=n + 1,\ldots, n_{\text{ahead}}$.
\end{algorithmic}
\caption{\small Causal Transfer for (unseen) future population treatment effects}\label{alg:CT_pred_sp}
\end{algorithm}


\section{Empirical results}
\label{section:empirical}

We assessed the empirical properties of Causal Transfer in simulations. We further compared its performance with two reference methods: Bayesian imputation and Causal Impact. Bayesian imputation is considered to be a standard method for inferring causal effects from cross-sectional data \cite{imbens2015}. Causal impact, like Causal Transfer, uses state-space models for causal effect estimation. The methods represent two common strategies for estimating causal effects in panel data: One approach is to analyse the data of every time point in the experiment separately with an i.i.d method, such as Bayesian imputation. Another is to aggregate the data over all experimental units and to analyse the resulting univariate time series instead. More details on the reference method are provided in Section~\ref{section:ref_methods}.

Here, we provide the empirical results for a variety of simulation set-ups. Each of the following sections contains the results for different kinds of estimands. We begin with a section on sample treatment effects on synthetic data. We then proceed to population and heterogenous effects. Thereafter, we consider treatment assignments with confounding. Experimental findings on real data are presented last.

\subsection{Sample treatment effects}\label{section:simulations}

We simulated data points according to the following true and misspecified models:

\begin{itemize}
\item Model 1 (correctly specified): 
$$X_{i, t} = \beta_{0, t} + \beta_{1, t} X_{i, \text{pre}} + \beta_{2, t} Z_{i, t} + T_i (\mu_{0, t} + \mu_{1, t} X_{i, \text{pre}} + \mu_{2, t} g_i) + v_{i, t}.$$
The pre-period covariate $X_{i, \text{pre}}$ is drawn from the uniform distribution $\mathcal{U}(0, 1)$ for each unit $i = 1,\ldots, d$. 
We partition the units into two groups and let the variable $g$ indicate the group membership, i.e., $g_i = 0$ for units of group 1 and $g_i = 1$ for the units of group 2.  For example, half of the units could originate from country $A$ and half of the units from country $B$. It seems natural to use one covariate for the units from group 1, e.g., GDP of country A and another covariate for units from group 2, e.g., GDP from country B. Therefore, the time-varying covariate $Z_t = (Z_{1,t},Z_{1,t}\ldots,Z_{2,t},Z_{2,t})$ is chosen to be identical for half of the units: $Z_{1,t}\sim\mathcal{N}(m_1, 0.1^2)$ and $Z_{2,t}\sim\mathcal{N}(m_2, 0.1^2)$, where $m_1$ is drawn from $\mathcal{U}(0, 1)$ and $m_2$ from $\mathcal{U}(-1, 0)$. The indicator $T_{i}$ is equal to 1 for treated units and 0 otherwise. The states are modelled as random walks
\begin{align}
\beta_{t} &= \beta_{t-1} + w_{t} \\
\mu_{t} &= 
  \begin{pmatrix}
    0.8 &  0 & 0 \\
    0 & 0.9 & 0 \\
    0 & 0 & 1
  \end{pmatrix}
 \mu_{t-1} + u_{t} 
\end{align}
with $\beta_{0} = (0.2, 0.6, 0.3)$ and $\mu_{0} = (1, 0.5, 0.3)$. The states $\mu_{0,t}$, $\mu_{1,t}$, and $\mu_{2,t}$ do not decay at the same rate. In fact, $\mu_{2,t}$ does not decay implying that the long-term treatment effect is not equal to 0. The noise terms $v_{i, t}, w_{j, t}, u_{j, t}$ are independent of each other, centered, and normally distributed with standard deviations 0.1, 0.01, and 0.01 respectively.

\item Model 2 (multiplicative effect): this model replaces the additive effect in model 1 by a multiplicative one. As before, we generate the observational part as 
$X_{i, t}^{\text{obs}} = \beta_{0, t} + \beta_{1, t} X_{i, \text{pre}} + \beta_{2, t} Z_{i, t} + v_{i, t} $. Now instead of adding the treatment effect, we multiply the observational part by the treatment effect, i.e., $X_{i, t} = T_i \mu_t  X_{i, t}^{\text{obs}}$. The multiplicative effect is modelled as $\mu_t = 1.5 + (\mu_{t-1}-1.5) 0.9 + u_t = 0.15 + 0.9\mu_{t-1}  + u_t$ with $\mu_0 = 2$ and $u_t$ being normally distributed with standard deviation 0.01. This equation implies that the treatment effect gradually decays from 2 to 1.5 over time. 

\item Model 3 (AR model): we use an AR model instead of a regression model for the observational part of the time series. 
\begin{align}
X_{i, t} &= X_{i, t}^{\text{obs}} + X_{i, t}^{\text{int}}\nonumber\\
X_{i, t}^{\text{obs}} & = \beta_{0, t} + \beta_{1} X_{i, t-1}^{\text{obs}} + \beta_{2, t} Z_{i, t} + v_{i, t}\nonumber\\
X_{i, t}^{\text{int}} & = T_i (\mu_{0, t} + \mu_{1, t} X_{i, \text{pre}} + \mu_{2, t} g_i) + \nu_{i, t}\nonumber
\end{align}
The variable $X_{\text{pre}}$ is replaced by $X_{t-1}^{\text{obs}}$ to add the AR term. The term
$\beta_1$ is constant in time because, otherwise, there would be an interaction between two states. The states $\beta_{0, t}$, $\beta_{2, t}$, $\mu_{0, t}$, $\mu_{1, t}$, and $\mu_{2, t}$  are generated as in model 1.  The noise variables $v_{i, t}$ and $\nu_{i, t}$ are independent of each other, centered, and normally distributed with standard deviation 0.1.

\item Model 4 (unit-specific parameters): the observational part of the time series is individual to each unit:
\begin{align}
X_{i, t} &= X_{i, t}^{\text{obs}} + X_{i, t}^{\text{int}}\nonumber\\
X_{i, t}^{\text{obs}} & = \beta_{0, i} + \beta_{1, i} X_{i, \text{pre}}^2 + \beta_{2,i} Z_{i, t} + v_{i, t}\nonumber\\
X_{i, t}^{\text{int}} & = T_i (\mu_{0, t} + \mu_{1, t} X_{i, \text{pre}} + \mu_{2, t} g_i) + \nu_{i, t}\nonumber
\end{align}
The coefficient $\beta_{0, i}$ is sampled with replacement from $\{0.1, 0.11, 0.12 \ldots, 0.3\}$, $\beta_{1, i}$ from \\$\{0.5, 0.51, 0.52, \ldots, 0.7\}$, and $\beta_{2, i}$ from $\{0.2, 0.21, 0.22, \ldots, 0.4\}$ for each unit. The noise distribution $v_{i,t}\sim\mathcal{N}(0, \sigma_i^2)$ is unit-specific as well. The standard deviations $\sigma_i$ are sampled with replacement from $\{0.09, 0.091, 0.092, \ldots, 0.11\}$. The noise term $\nu_{i, t}$ is simulated from $\mathcal{N}(0, 0.1^2)$.
Furthermore, we add a nonlinearity by replacing $X_{\text{pre}}$ with $X_{\text{pre}}^2$ in the equation for $X_t^{\text{obs}}$. The remaining states $\mu_{0, t}$, $\mu_{1, t}$, and $\mu_{2, t}$  are generated as in model 1. 

\item Model 5 (nonlinear effect): we misspecify the treatment effect by replacing $X_{\text{pre}}$ with $\cos(X_{\text{pre}})$ in
$$X_{i, t}= \beta_{0, t} + \beta_{1, t} X_{i, \text{pre}} + \beta_{2,t} Z_{i, t} + T_i \mu_{i, t} \cos(X_{i, \text{pre}}) + v_{i, t} \,.$$
Furthermore, the starting value $\mu_{i, 0}$ is drawn with replacement from $\{0.9, 0.91, 0.92 \ldots, 1.1\}$ for each unit. This implies that the effect is unit-specific. Each state evolves according to
\begin{equation}
\mu_{i,t} = 1.002 \mu_{i,t-1} + u_{i,t} \nonumber
\end{equation}
with $u_{i, t} \sim  \mathcal{N}(0, 0.01^2)$. Hence, the effect grows slightly over time. The states $\beta_{0, t}$, $\beta_{1, t}$, $\beta_{2, t}$ and the noise $v_{i, t}$ are generated as in model 1.

\item Model 6 (deterministic effect): the treatment effect is misspecified by replacing $X_{\text{pre}}$ with $X_{\text{pre}}^2$ in
$$X_{i, t} = \beta_{0, t} + \beta_{1, t} X_{i, \text{pre}} + \beta_{2,t} Z_{i, t} + T_i (\mu_{0,i, t} + \mu_{1,i, t} X_{i, \text{pre}}^2) + v_{i, t} \,.$$
Furthermore, the starting value $\mu_{0, i, 0}$ is sampled with replacement from $\{0.9, 0.91, 0.92 \ldots, 1.1\}$ and $\mu_{1, i, 0}$ from $\{0.4, 0.41, 0.42, \ldots, 0.6\}$ for each unit. The states evolve deterministically as 
\begin{equation}
\mu_{i,t} = 
 \begin{pmatrix}
   0.9 &  0 \\
   0 & 1 \\
 \end{pmatrix}
\mu_{i,t-1}\nonumber.
\end{equation}
The states $\beta_{0, t}$, $\beta_{1, t}$, $\beta_{2, t}$ and the noise $v_{i, t}$ are generated as in model 1.
\end{itemize}

\begin{figure}[!htb]
    \centering
        \begin{subfigure}[b]{0.485\textwidth}
        \includegraphics[width=\textwidth]{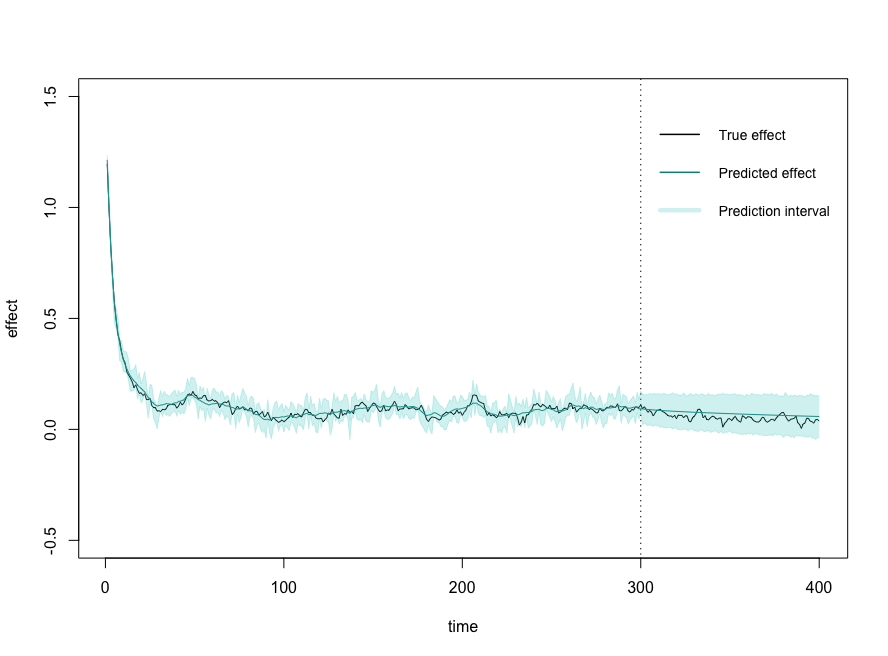}
        \caption{Model 1 (CT)}
        \label{fig:model1_CT}
    \end{subfigure}
    ~
       \begin{subfigure}[b]{0.485\textwidth}
        \includegraphics[width=\textwidth]{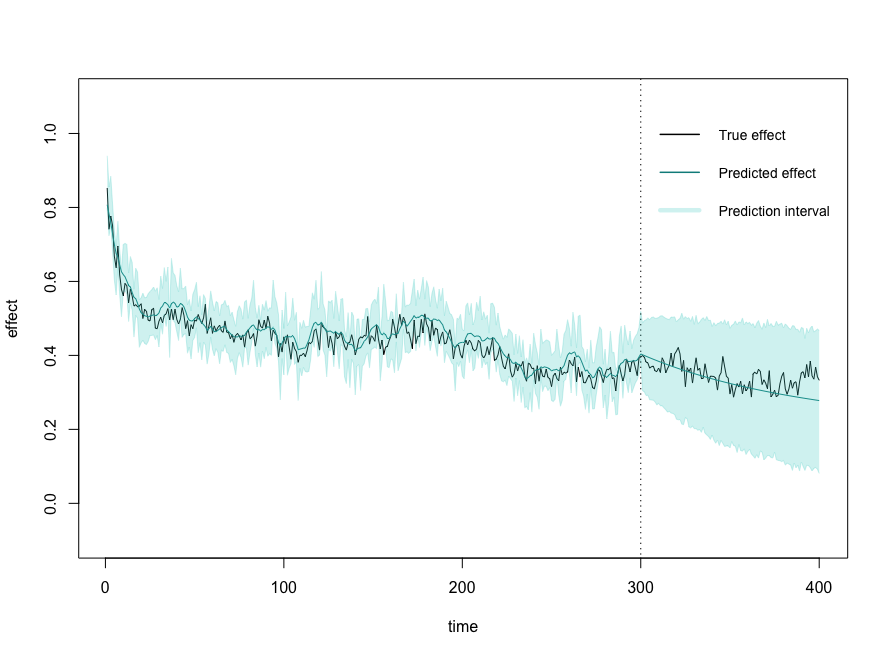}
        \caption{Model 2 (CT)}
        \label{fig:model2_CT}
    \end{subfigure}
    
        \begin{subfigure}[b]{0.485\textwidth}
        \includegraphics[width=\textwidth]{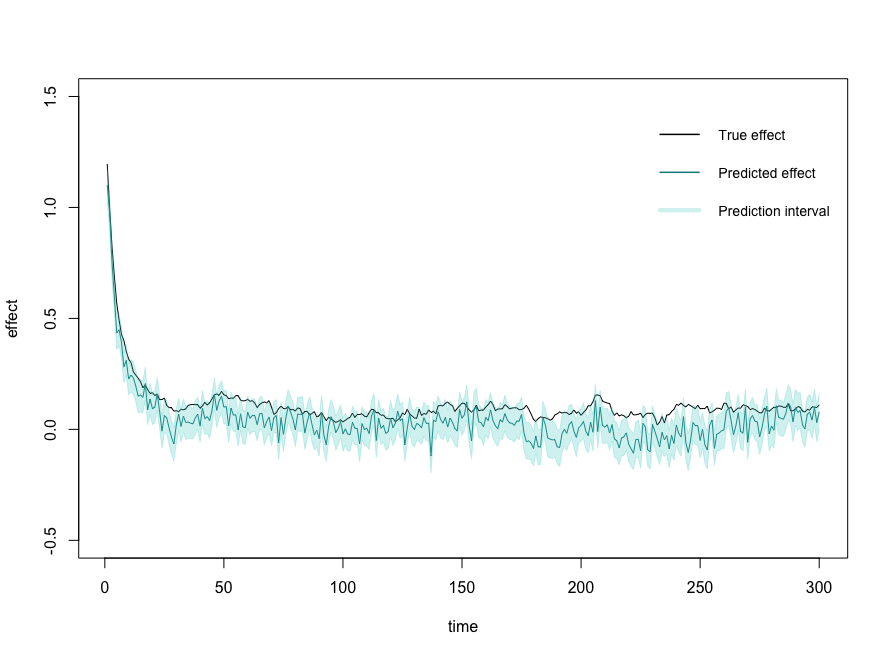}
        \caption{Model 1 (CI)}
        \label{fig:model1_CI}
    \end{subfigure}
    ~
        \begin{subfigure}[b]{0.485\textwidth}
        \includegraphics[width=\textwidth]{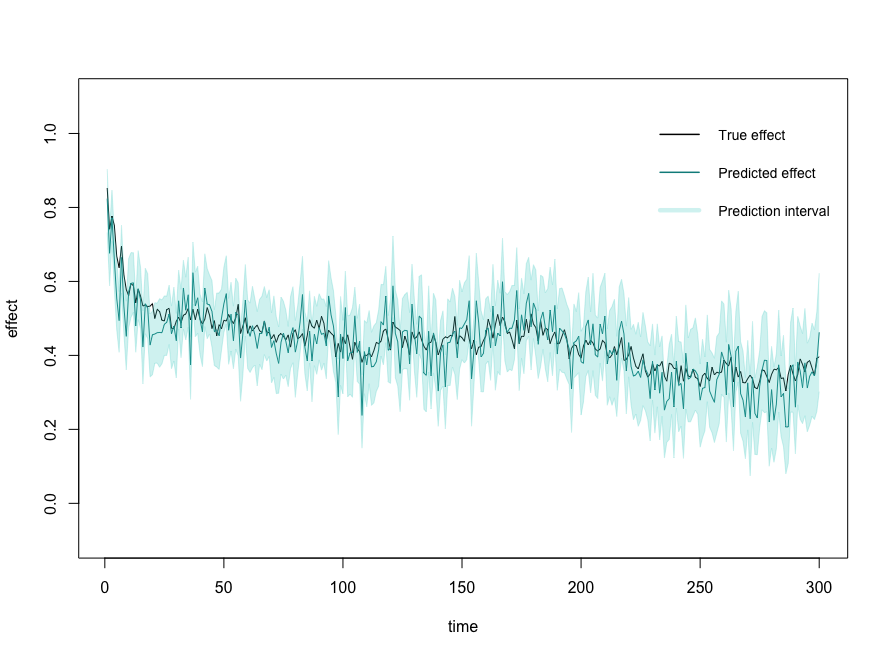}
        \caption{Model 2 (CI)}
        \label{fig:model2_CI}
    \end{subfigure}
    
        \begin{subfigure}[b]{0.485\textwidth}
        \includegraphics[width=\textwidth]{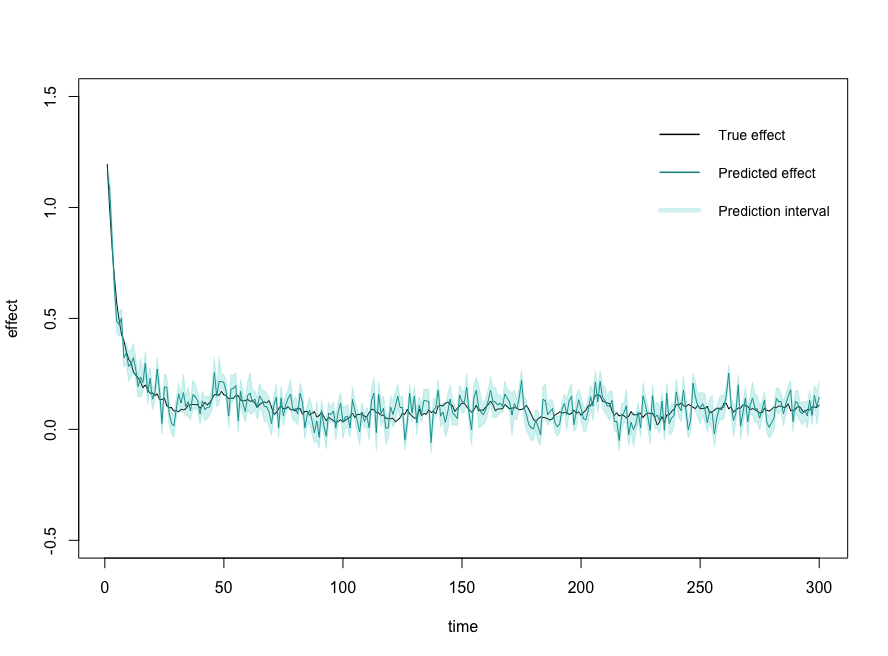}
        \caption{Model 1 (BI)}
        \label{fig:model1_BI}
    \end{subfigure}
    ~
       \begin{subfigure}[b]{0.485\textwidth}
        \includegraphics[width=\textwidth]{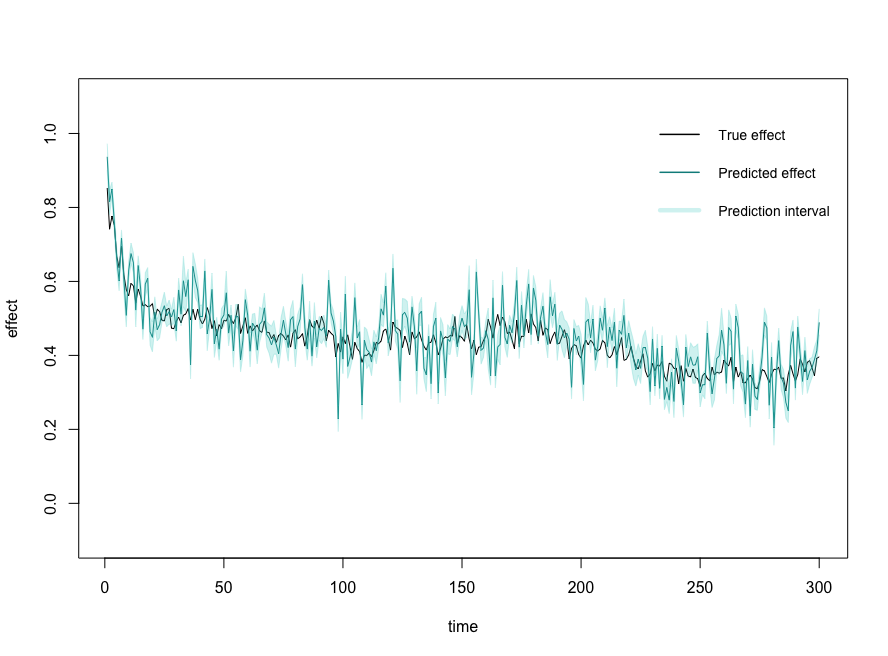}
        \caption{Model 2 (BI)}
        \label{fig:model2_BI}
    \end{subfigure}
\caption{Example analysis plots for the SATE on models 1 and 2. Causal Transfer (Figure~\ref{fig:model1_CT} and \ref{fig:model2_CT}) is able to predict ``future'' effects, which are shown after the dashed line. The prediction intervals for ``future'' effects are wider than for the past effects. The prediction intervals are also wider for model 2 than for model 1. Bayesian imputation (Figure~\ref{fig:model1_BI} and \ref{fig:model2_BI}) has the narrowest and Causal Impact (Figure~\ref{fig:model1_CI} and \ref{fig:model2_CI}) the widest intervals. }\label{fig:models1_2} 
\end{figure}

\begin{figure}[!htb]
    \centering
        \begin{subfigure}[b]{0.485\textwidth}
        \includegraphics[width=\textwidth]{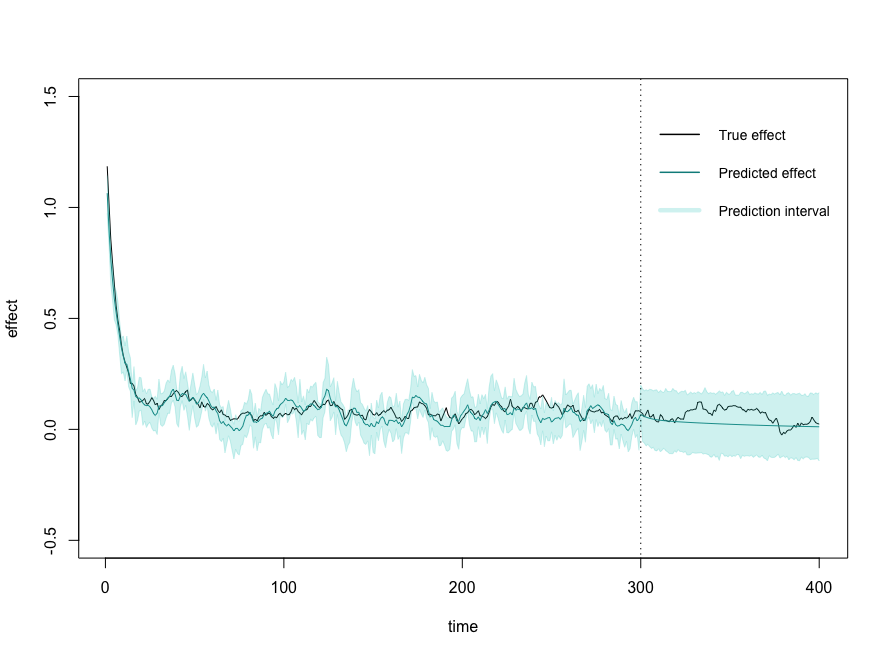}
        \caption{Model 3 (CT)}
        \label{fig:model3_CT}
    \end{subfigure}
    ~
       \begin{subfigure}[b]{0.485\textwidth}
        \includegraphics[width=\textwidth]{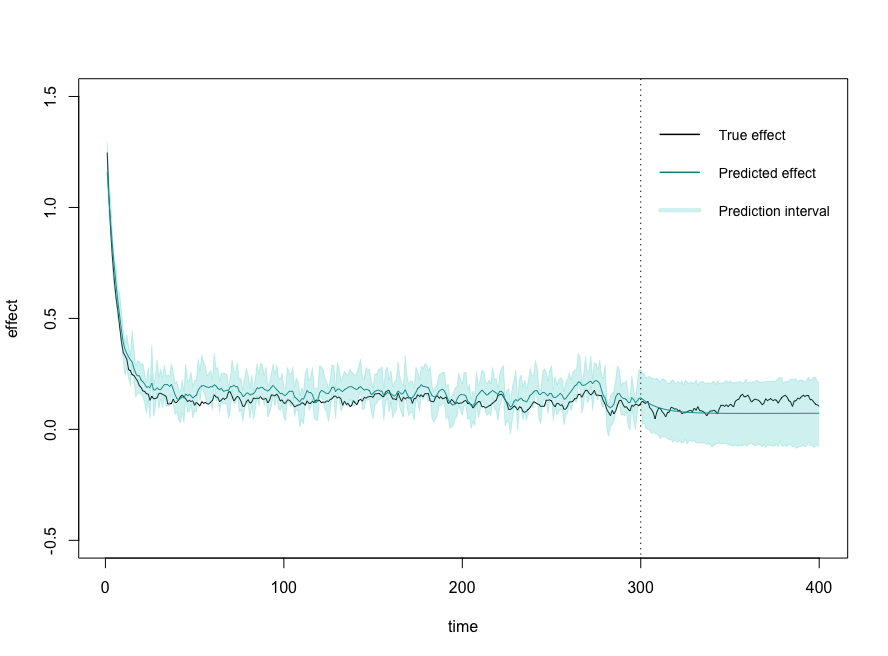}
        \caption{Model 4 (CT)}
        \label{fig:model4_CT}
    \end{subfigure}
    
        \begin{subfigure}[b]{0.485\textwidth}
        \includegraphics[width=\textwidth]{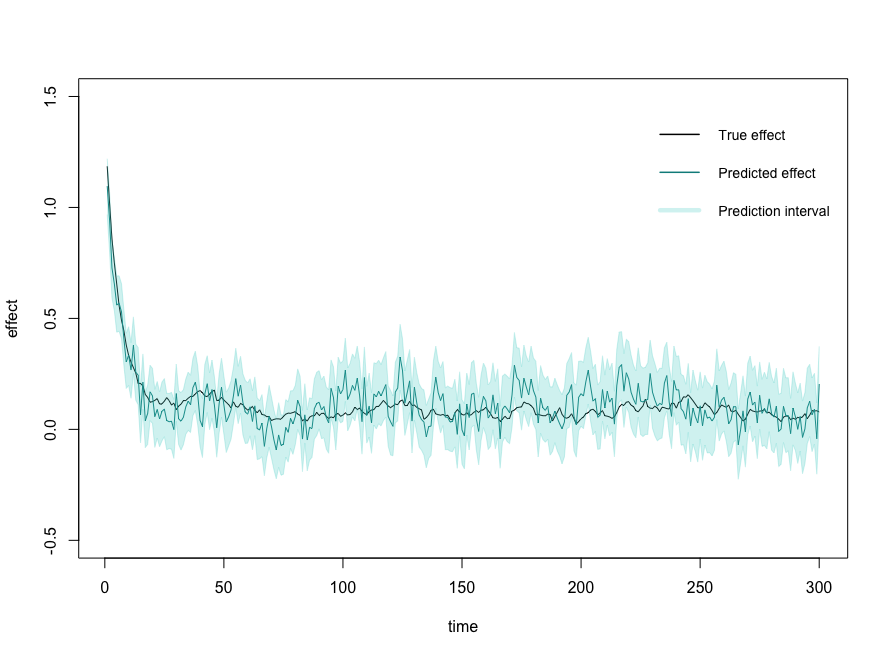}
        \caption{Model 3 (CI)}
        \label{fig:model3_CI}
    \end{subfigure}
    ~
        \begin{subfigure}[b]{0.485\textwidth}
        \includegraphics[width=\textwidth]{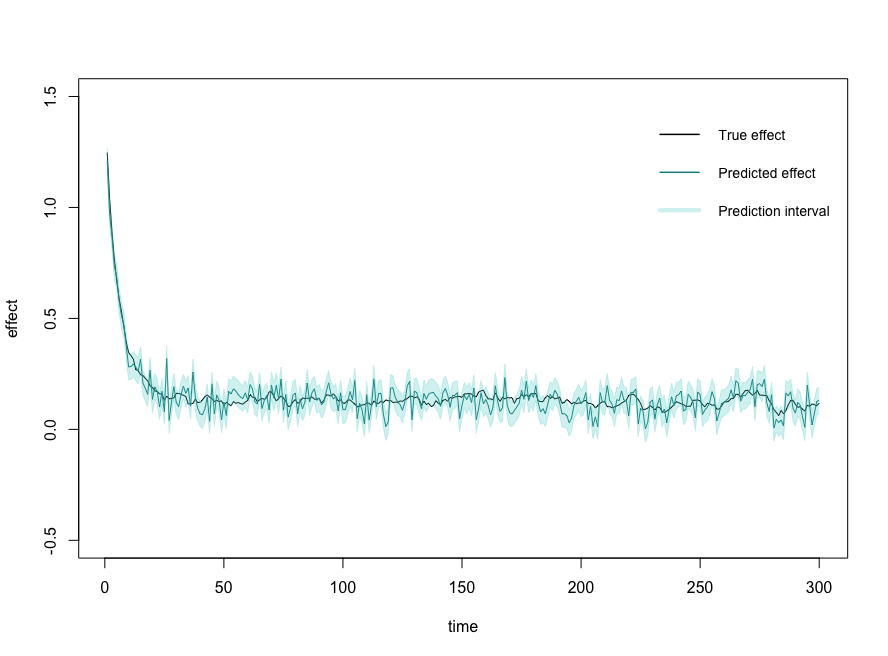}
        \caption{Model 4 (CI)}
        \label{fig:model4_CI}
    \end{subfigure}
    
        \begin{subfigure}[b]{0.485\textwidth}
        \includegraphics[width=\textwidth]{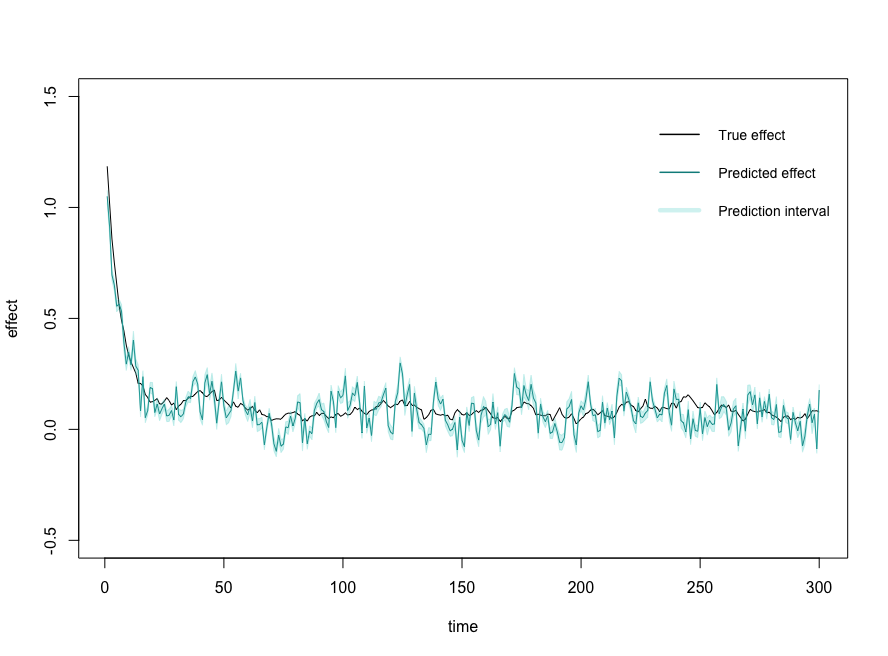}
        \caption{Model 3 (BI)}
        \label{fig:model3_BI}
    \end{subfigure}
    ~
        \begin{subfigure}[b]{0.485\textwidth}
        \includegraphics[width=\textwidth]{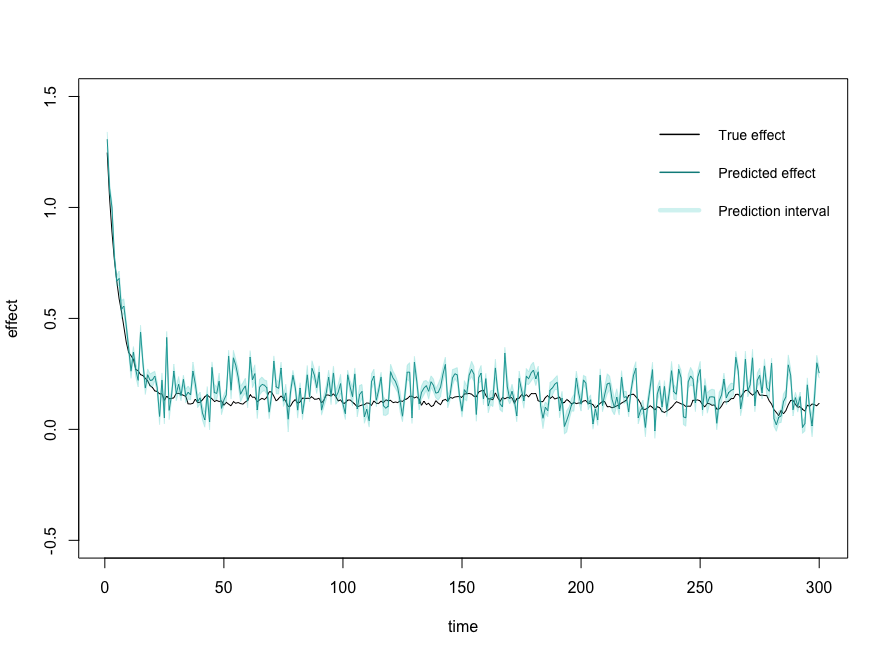}
        \caption{Model 4 (BI)}
        \label{fig:model4_BI}
    \end{subfigure}
\caption{Example analysis plots for the SATE on models 3 and 4. Causal Transfer (Figure~\ref{fig:model3_CT} and \ref{fig:model4_CT}) is able to predict ``future'' effects, which are shown after the dashed line. Causal Impact performs noticeably better on model 4 (Figure~\ref{fig:model4_CI}) than on model 3 (Figure~\ref{fig:model4_CI}). Bayesian imputation undercovers on both models (Figure~\ref{fig:model3_BI} and \ref{fig:model4_BI}).}\label{fig:models3_4} 
\end{figure}

\begin{figure}[!htb]
    \centering
        \begin{subfigure}[b]{0.485\textwidth}
        \includegraphics[width=\textwidth]{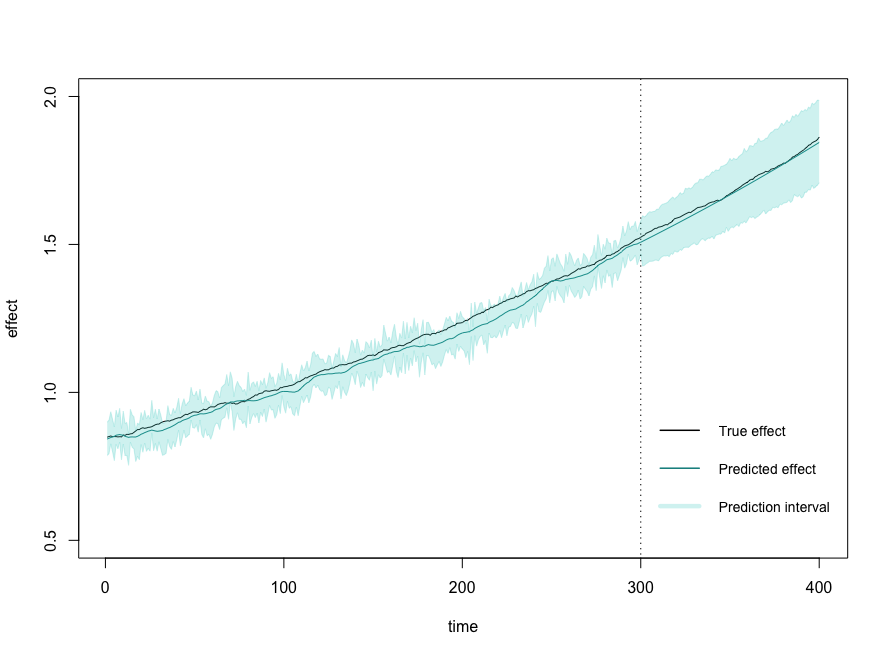}
        \caption{Model 5 (CT)}
        \label{fig:model5_CT}
    \end{subfigure}
    ~
       \begin{subfigure}[b]{0.485\textwidth}
        \includegraphics[width=\textwidth]{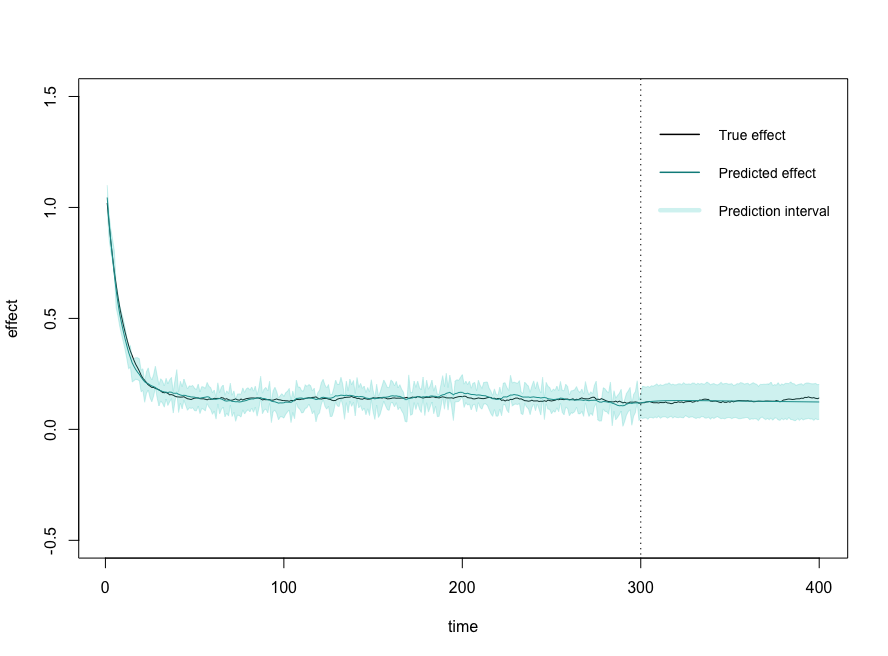}
        \caption{Model 6 (CT)}
        \label{fig:model6_CT}
    \end{subfigure}
    
        \begin{subfigure}[b]{0.485\textwidth}
        \includegraphics[width=\textwidth]{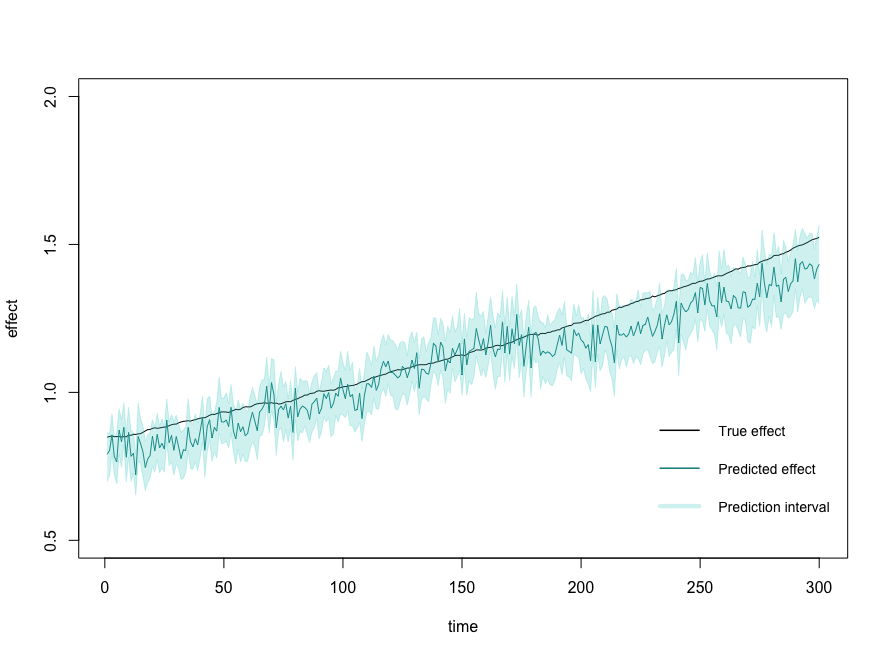}
        \caption{Model 5 (CI)}
        \label{fig:model5_CI}
    \end{subfigure}
    ~
        \begin{subfigure}[b]{0.485\textwidth}
        \includegraphics[width=\textwidth]{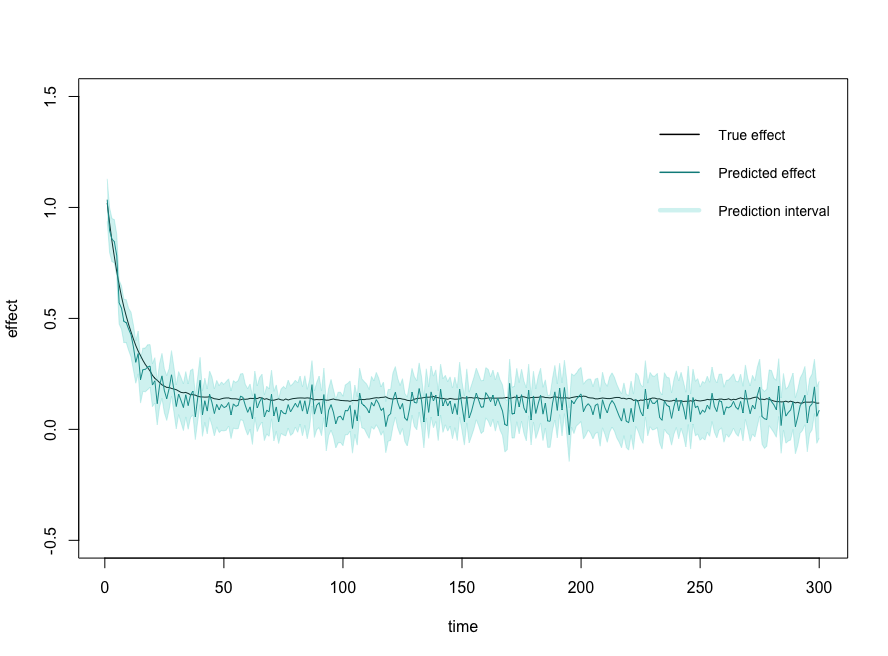}
        \caption{Model 6 (CI)}
        \label{fig:model6_CI}
    \end{subfigure}
    
        \begin{subfigure}[b]{0.485\textwidth}
        \includegraphics[width=\textwidth]{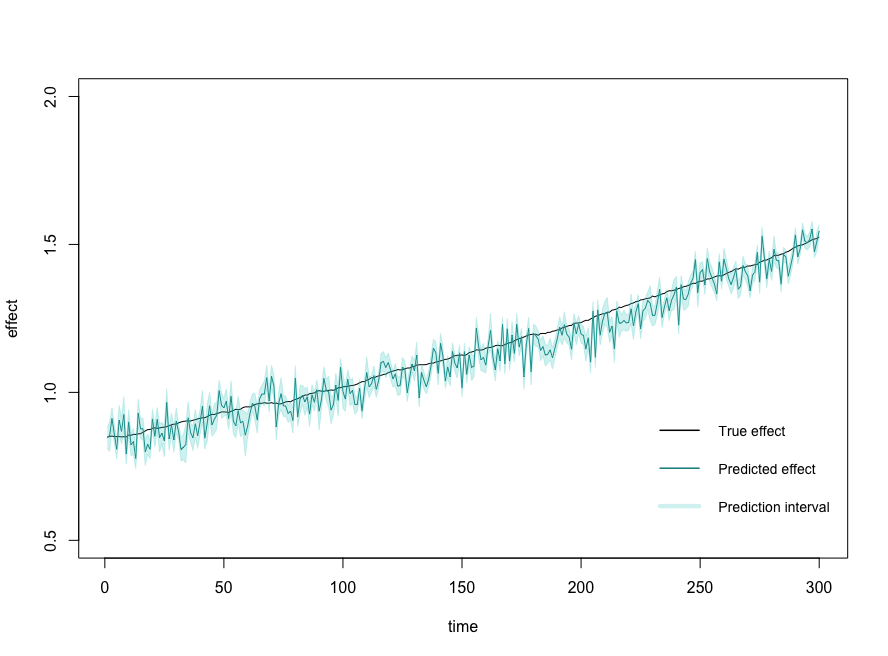}
        \caption{Model 5 (BI)}
        \label{fig:model5_BI}
    \end{subfigure}
    ~
        \begin{subfigure}[b]{0.485\textwidth}
        \includegraphics[width=\textwidth]{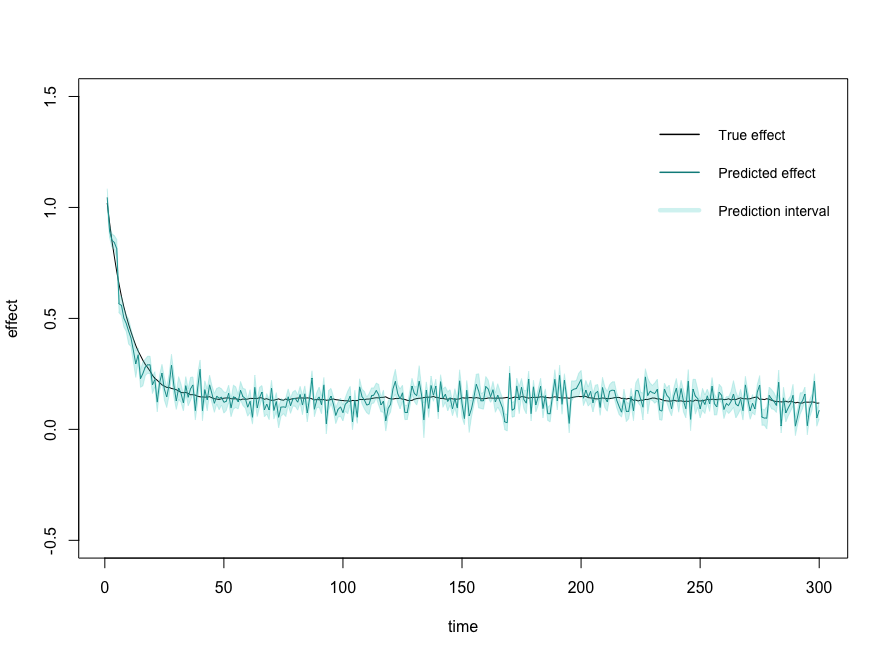}
        \caption{Model 6 (BI)}
        \label{fig:model6_BI}
    \end{subfigure}
\caption{Example analysis plots for the SATE on models 5 and 6. Causal Transfer (Figure~\ref{fig:model5_CT} and \ref{fig:model6_CT}) is able to predict ``future'' effects, which are shown after the dashed line. Bayesian imputation (Figure~\ref{fig:model5_BI} and \ref{fig:model6_BI}) undercovers while Causal Impact (Figure~\ref{fig:model5_CI} and \ref{fig:model6_CI}) estimates the widest intervals. The ground truth in these models is smoother than in models 1-4. Causal Transfer is able to adapt its smoothness to the data.}\label{fig:models5_6} 
\end{figure}

Model 1 is correctly specified, while the remaining models are misspecified. For all models, we generated $n = 300$ data points for $d=20$ units. 10 units were assigned to the treatment group (5 units with $g=0$ and 5 units with $g=1$) and the remaining 10 units to the control group. We ran Causal Transfer with the dynamic regression model $X_t\sim Z_t +  X_{\text{pre}} * T + T g$ on models 1, 3, and 4 and $X_t\sim Z_t + X_{\text{pre}} * T $ on models 2, 5, and 6 and estimated the SATE $\frac{1}{d}\sum_{i=1}^d (X_{i, t}(1) - X_{i, t}(0))$. We summarized the MSE, the coverage and width of the 95$\%$-prediction intervals over the 100 simulation runs in Table~\ref{tab:sim_CT}. Our method is further capable of predicting unseen ``future'' treatment effects, for which no data point has been observed yet. We predicted ``future'' effects for time points $t= n+1, n+2,\ldots, n+100$ and compared the ``future'' to the ``past'' effects for $t = 1, \ldots, n$ in Table~\ref{tab:sim_CT}. Example plots for one simulation run are shown in Figures~\ref{fig:model1_CT} - \ref{fig:model6_CT}.

\begin{table}[h!]
\small
\centering
\setlength\tabcolsep{4pt}
\begin{tabular}{lrrrrrrr}
\hline
 &\multicolumn{2}{c}{MSE$\cdot10^3$}&\multicolumn{2}{c}{Coverage}&\multicolumn{2}{c}{Width} \\
 &Past&Future&Past&Future&Past&Future\\
\hline
Model 1&0.3&2.1&0.92&0.94&0.09&0.17\\
Model 2&0.5&5.3&0.93&0.95&0.13&0.26\\
Model 3&1.1&3.7&0.88&0.99&0.15&0.30\\
Model 4&2.1&6.3&0.81&0.92&0.14&0.26\\
Model 5&1.3&4.9&0.86&0.85& 0.12&0.21\\
Model 6&0.4&4.7&0.90&0.94&0.10&0.19\\
\hline
\end{tabular}
\caption{Comparison between ``past'' and ``future'' SATE effects in terms of MSE, coverage, and width of the 95$\%$-prediction intervals. The values were computed separately for the ``past'' for which data is available and the ``future'', for which no outcomes have been observed yet. The results were averaged over 100 simulation runs and the respective time period, e.g., $\MSE_{\text{past}} = \frac{1}{100}\sum_{b=1}^{100} \frac{1}{n}\sum_{t=1}^n (\tau_t^{(b)}-\hat{\tau}_t^{(b)})^2$ and $\MSE_{\text{future}} = \frac{1}{100}\sum_{b=1}^{100} \frac{1}{100}\sum_{t=n+1}^{n+100} (\tau_t^{(b)}-\hat{\tau}_t^{(b)})^2$ with $\tau_t$ being the true effect, $\hat{\tau}_t$ the estimated effect at time $t$, and $n=300$. The ``future'' effects are predicted from the model trained on the data points from ``past''.}\label{tab:sim_CT}
\end{table}

The MSE of the estimated ``past'' effects and the width of its prediction intervals are smallest for the correctly specified model (model 1) although it is hard to compare between the different models. The coverage for the ``past'' effects is close to the desired $95\%$  on model 1. 
  The coverage remains close to 95$\%$ on the multiplicative model 2 irrespective of misspecification. The coverage on the remaining models varies between 81 and 90 $\%$. Some undercoverage has to be expected, since these models contain misspecifications and more variability than what is  assumed by our method.

When comparing ``past'' to ``future'' effects, it is noticeable that the MSE and the width of the prediction intervals increase. Due to the width increase, the coverage is closer to the desired rate of 95$\%$ for the ``future'' effects than the ``past'' effects. Figure~\ref{fig:model1_CT}, for example, shows how the prediction intervals increase in width over time. Intuitively it makes sense, as the last observation lies more distant in the past the estimates become increasingly uncertain. Since no outcomes are available for the estimation of ``future'' effects, all outcomes need to be estimated. This adds a source of variability. The parameters of the distributions are estimated by iterating the state equation~\eqref{eq:KF2} and the measurement equation~\eqref{eq:KF1} without the subsequent correction step with the Kalman gain. The MSE of the predicted SATE increases for ``future'' effects due to the lack of the correction step, which would have corrected the predicted towards the observed values. As a result, the predicted ``future'' effects are smoother than the ``past'' effects. The overall trend of the predicted ``future'' effects in Figures~\ref{fig:model1_CT} - \ref{fig:model6_CT} remains close to the truth.\\

Next, we compare Causal Transfer to Causal Impact and Bayesian imputation. Predictions of ``future'' effects cannot be made with Bayesian imputation nor Causal Impact. Therefore, we restrict the comparison to the ``past'' effects for $t=1,\ldots,n$.  Causal Impact only accepts univariate time series. Therefore, we aggregated the time series cross-sectionally and used the aggregated series as input for Causal Impact. Causal Impact further requires a pre-period to learn the relationship between the response and the control time series in absence of any interventions. This relationship is then exploited to predict the counterfactuals. Therefore, we generated $2n$ data points for Causal Impact: $n$ pre-period points, on which Causal Impact is trained, and $n$ treatment points, for which predictions are made. We used the same predictors for Bayesian imputation and Causal Transfer. 


We estimated the SATE with all methods. The results averaged over 100 simulations are shown in Table~\ref{tab:sim_comp}. The estimates are shown for one simulation run in Figures~\ref{fig:models1_2} -  \ref{fig:models5_6}.

Causal Transfer achieves the lowest MSE among the methods compared except on model 4. In addition, the coverage of Causal Transfer is closest to the desired rate of 95$\%$ except on model 4. Causal Impact produces the widest prediction intervals on all models but model 4 while its coverage rates are too low. Causal Impact performs best on model 4 but the results are not substantially better than for Causal Transfer. Bayesian imputation has, overall, the narrowest intervals but also the least coverage. It is likely that the coverage of Bayesian imputation will improve for a larger number of units than $d = 20$. 


The computation time reflects the complexity of the methods. Causal Impact estimates a single univariate dynamic regression model. Bayesian imputation fits one regression model per time point, and Causal Transfer a multivariate dynamic regression model which accounts for time dependence. As a result, Causal Impact is faster than Causal Transfer and Bayesian imputation in our simulations. Causal Transfer can be accelerated by parallelization of the sampling process or by sequential processing of the data. \\

\begin{table}[h!]
\small
\centering
\setlength\tabcolsep{4pt}
\begin{tabular}{lrrrrrrrrrrrr}
\hline
 &\multicolumn{3}{c}{MSE$\cdot10^3$}&\multicolumn{3}{c}{Coverage}&\multicolumn{3}{c}{Width}&\multicolumn{3}{c}{Time$[s]$} \\
 &CT&CI&BI&CT&CI&BI& CT&CI&BI& CT&CI&BI\\
\hline
Model 1&0.3&5.6&2.2&0.92&0.79&0.69&0.09&0.18&0.10&210&4&367\\
Model 2&0.5&7.0&3.5&0.93&0.72&0.44&0.13&0.18&0.07&184&3&367\\
Model 3&1.1&8.1&5.5&0.88&0.84&0.31&0.15&0.26&0.06&221&3&373\\
Model 4&2.1&2&5.5&0.81&0.82&0.30&0.14&0.12&0.06&219&3&366\\
Model 5&1.3&8.6&3.3&0.86&0.71&0.44&0.12&0.18&0.07&183&3&363\\
Model 6&0.4&6.1&2.3&0.90&0.78&0.64&0.10&0.18&0.09&199&3&359\\
\hline
\end{tabular}
\caption{Comparison between Causal Transfer (CT), Causal Impact (CI), and Bayesian imputation (BI) in terms of MSE, coverage and width of the 95$\%$-prediction intervals, and CPU time consumption for the estimation of the SATE. The comparison is restricted to the ``past'' effects. The results were averaged over $n=300$ time points and 100 simulation runs, e.g., $\MSE = \frac{1}{100}\sum_{b=1}^{100} \frac{1}{n}\sum_{t=1}^n (\tau_t^{(b)}-\hat{\tau}_t^{(b)})^2$ with $\tau_t$ being the true and $\hat{\tau}_t$ the estimated effect at time $t$. }\label{tab:sim_comp}
\end{table}

\clearpage
\subsection{Population and heterogenous treatment effects}\label{section:heteffects}

We now turn to the estimation of population and heterogenous effects. In many situations, we are interested in how the treatment effect varies across individuals. One could imagine that a new unit appears after a study concludes. The new unit is similar to the study units. What would the treatment effect for the new unit be if it would have taken part in the experiment? The treatment effect is individual to each unit through the dependence on its covariate values. To transfer the effect to the new unit, we learn the $\CATE = \mathbb{E}[X_t(1)-X_t(0)\,|\, X_{\text{pre}},\,g]$ from the experiment and plug in the covariate values of the new unit. 

A related topic is the difference in average treatment effects between groups of individuals. For example, how does the efficacy of a drug differ for young and old patients or how do policy changes affect big and small companies? These questions also raise the issue of heterogeneous treatment effects. In the following, we restrict ourselves to a setting with a single factor $g$ with levels 0 and 1. But more groups can be handled in the same manner. The average treatment effect
for units with $g=0$ is $\mathbb{E}[X_t(1)-X_t(0)\,|\,g = 0]$ and analogously  $\mathbb{E}[X_t(1)-X_t(0)\,|\,g = 1]$ for $g=1$. The expression $\mathbb{E}[X_t(1)-X_t(0)\,|\,g]$ is the marginal conditional average treatment effect (MCATE), since it can be derived by taking the iterated (conditional) expectation of the CATE.  
The difference in the average treatment effects between the groups is equal to $\MCATE_{\Delta_g} = \MCATE_{g=1} - \MCATE_{g=0}$. \\

\begin{figure}[!htb]
    \centering
        \begin{subfigure}[b]{0.485\textwidth}
        \includegraphics[width=\textwidth]{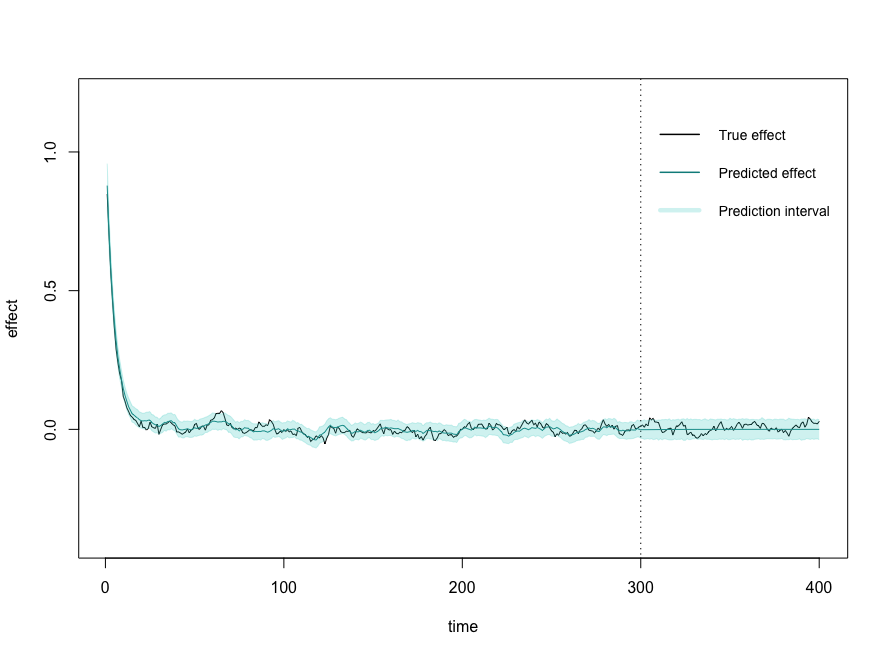}
        \caption{CATE (CT)}
        \label{fig:CATE_CT}
    \end{subfigure}
    ~
       \begin{subfigure}[b]{0.485\textwidth}
        \includegraphics[width=\textwidth]{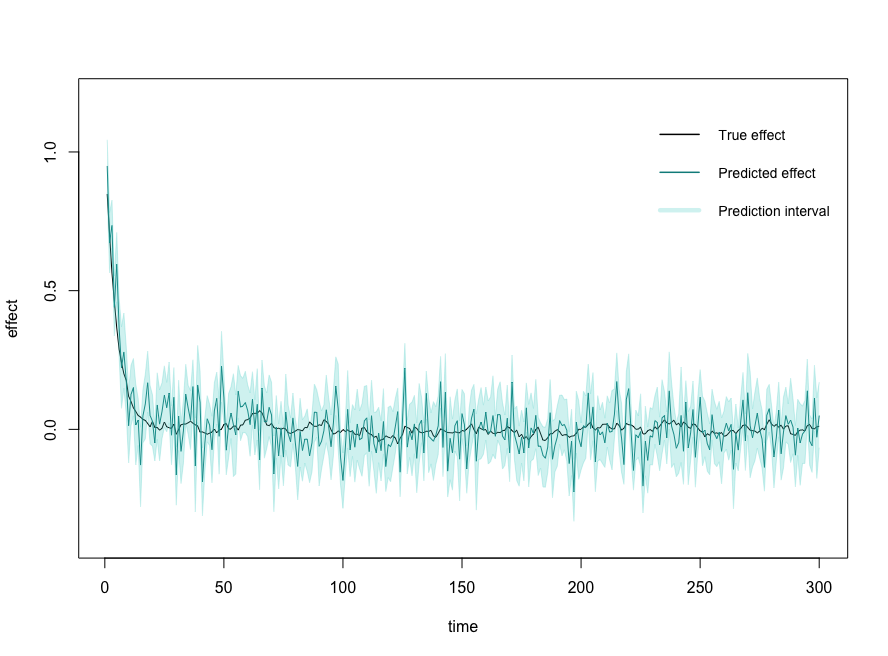}
        \caption{CATE (BI)}
        \label{fig:CATE_BI}
    \end{subfigure}
    
        \begin{subfigure}[b]{0.485\textwidth}
        \includegraphics[width=\textwidth]{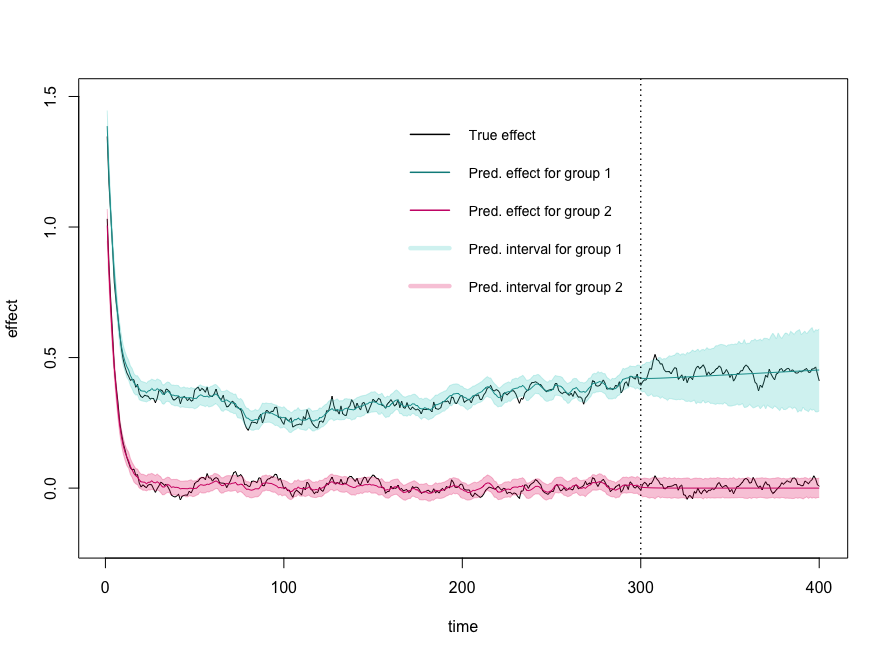}
        \caption{MCATE  (CT)}
        \label{fig:MCATE_CT}
    \end{subfigure}
    ~
        \begin{subfigure}[b]{0.485\textwidth}
        \includegraphics[width=\textwidth]{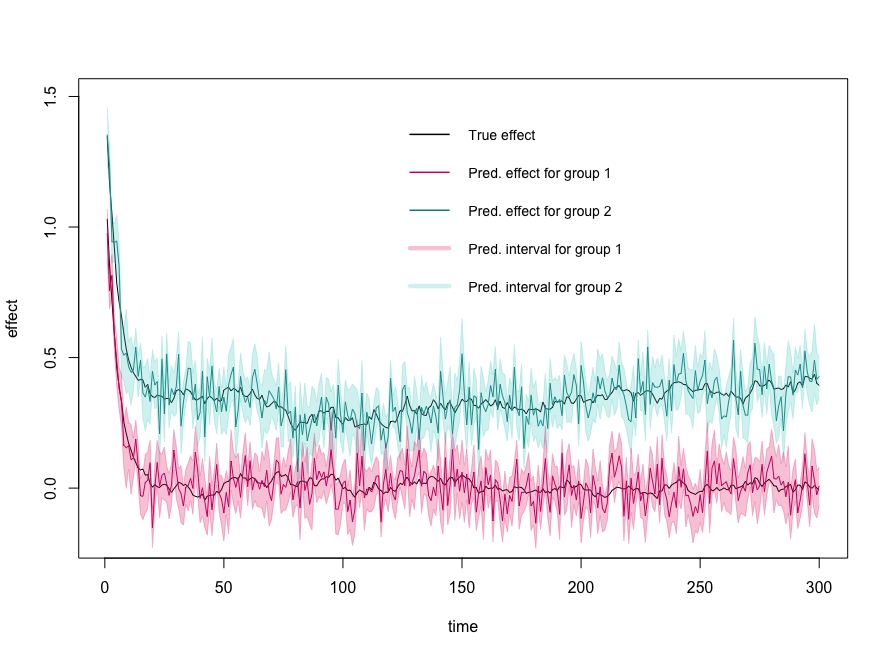}
        \caption{MCATE (BI)}
        \label{fig:MCATE_BI}
    \end{subfigure}
    
        \begin{subfigure}[b]{0.485\textwidth}
        \includegraphics[width=\textwidth]{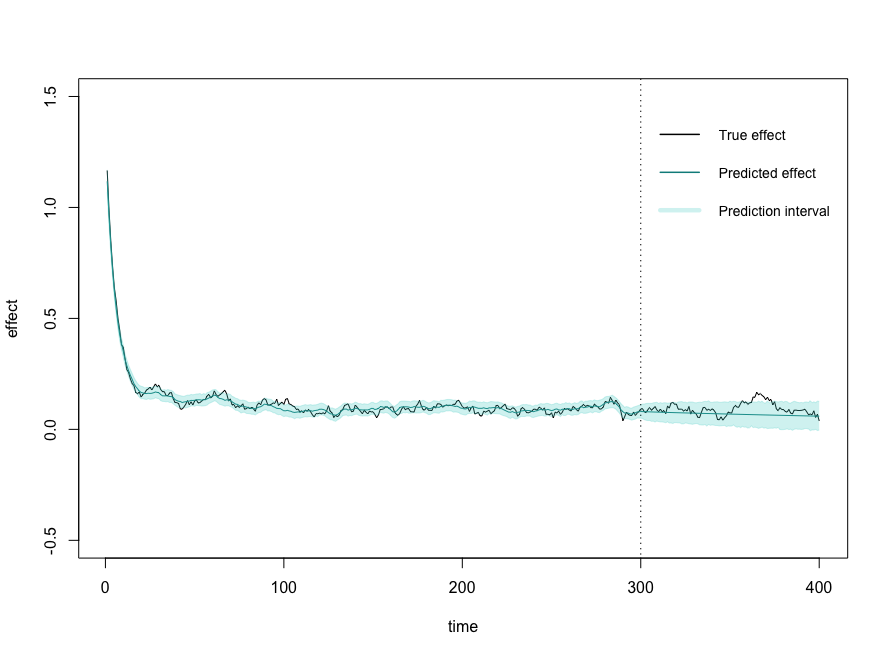}
        \caption{ATE (CT)}
        \label{fig:ATE_CT}
    \end{subfigure}
    ~
        \begin{subfigure}[b]{0.485\textwidth}
        \includegraphics[width=\textwidth]{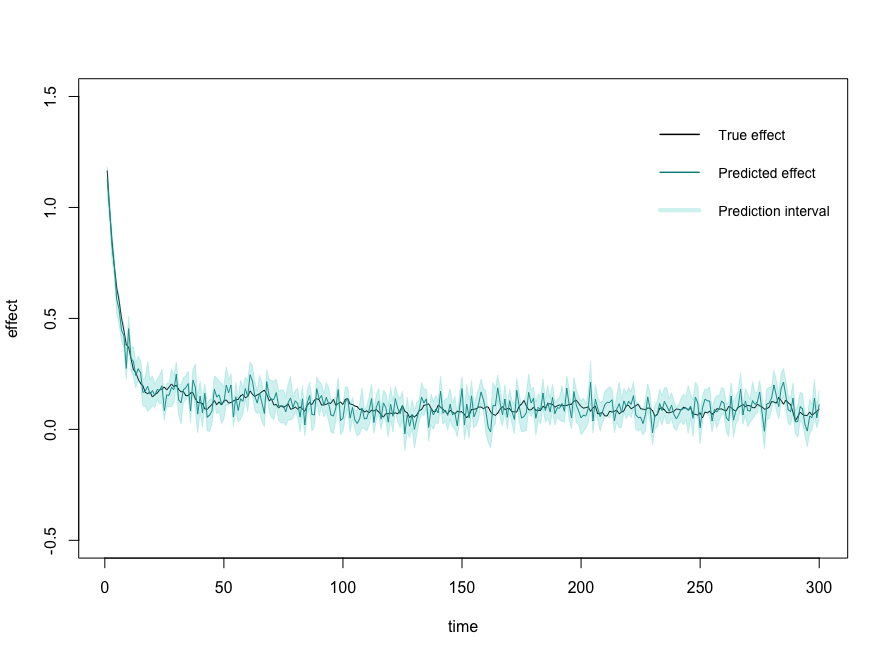}
        \caption{ATE (BI)}
        \label{fig:ATE_BI}
    \end{subfigure}
\caption{Example analysis plots for heterogeneous treatment effects in model 1. Causal Transfer (Figures~\ref{fig:CATE_CT}, \ref{fig:MCATE_CT}, and \ref{fig:ATE_CT}) is able to predict ``future'' effects, which are shown after the dashed line. Causal Transfer produces smoother estimates and narrower intervals than Bayesian imputation (Figures~\ref{fig:CATE_BI}, \ref{fig:MCATE_BI}, and \ref{fig:ATE_BI}) by pooling information over time.}
\label{fig:heteffects} 
\end{figure}

We test Causal Transfer for the task of inferring and transferring heterogeneous effects. We simulated data from model 1 and estimated the CATE, the MCATE, and the ATE. For the CATE, we drew the covariate $X_{\text{pre}}$ of the new unit from the uniform distribution and set $g$ to 0. The averaged results are shown in Table~\ref{tab:sim_het_CT}. The estimated heterogeneous effects are plotted for one simulation run in Figure~\ref{fig:heteffects}. In addition, we compare Causal Transfer to Bayesian imputation in Table~\ref{tab:sim_het}. Causal Transfer outperforms Bayesian imputation in all aspects compared. Causal Impact cannot estimate heterogenous effects since it requires univariate time series. It can, however, estimate the ATE. Causal Impact estimates for the ATE are shown in Section~\ref{section:confeffects}. 

The prediction of ``future'' effects is more uncertain than the prediction of the``past'' effects since outcomes have yet to be observed. This uncertainty is reflected in wider intervals in Figures~\ref{fig:CATE_CT}-\ref{fig:ATE_CT}. The width and the MSE increase the most for the MCATE with $g=1$ or $\Delta_g$. Surprisingly, the width and the MSE for the prediction of the MCATE for $g=0$ are less affected by the additional uncertainty. This could be because for $g=0$, algorithm~\ref{alg:CT_sp} does not need to draw samples of the state $\mu_{2, t}$. 

\begin{table}[h!]
\small
\centering
\setlength\tabcolsep{4pt}
\begin{tabular}{lrrrrrrrr}
\hline
 &\multicolumn{2}{c}{MSE$\cdot10^3$}&\multicolumn{2}{c}{Coverage}&\multicolumn{2}{c}{Width} \\
 &CT&BI&CT&BI& CT&BI\\
\hline
CATE&0.3&8.0&0.91&0.82&0.06&0.24\\
$\MCATE_{g=0}$&0.3&4.4&0.91&0.82&0.06&0.18\\
$\MCATE_{g=1}$&0.5&4.3&0.93&0.82&0.08&0.18\\
$\MCATE_{\Delta_g}$&0.5&8.7&0.92&0.82&0.07&0.26\\
ATE&0.3&2.2&0.93&0.83&0.06&0.13\\
\hline
\end{tabular}
\caption{Comparison between Causal Transfer (CT)  and Bayesian imputation (BI) for the estimation of heterogeneous effects in model 1. The results were averaged over 300 time points and 100 simulation runs. The desired coverage is 95$\%$. The comparison is restricted to ``past'' effects.}\label{tab:sim_het}
\end{table}

\begin{table}[h!]
\small
\centering
\setlength\tabcolsep{4pt}
\begin{tabular}{lrrrrrrr}
\hline
 &\multicolumn{2}{c}{MSE$\cdot10^3$}&\multicolumn{2}{c}{Coverage}&\multicolumn{2}{c}{Width} \\
 &Past&Future&Past&Future&Past&Future\\
\hline
CATE&0.3&0.4&0.91&0.90&0.06&0.08\\
$\MCATE_{g=0}$&0.3&0.4&0.91&0.91&0.06&0.08\\
$\MCATE_{g=1}$&0.5&7.0&0.93&0.90&0.08&0.27\\
$\MCATE_{\Delta_g}$&0.5&6.6&0.92&0.90&0.07&0.26\\
ATE&0.3&2.1&0.93&0.91&0.06&0.15\\
\hline
\end{tabular}
\caption{Comparison between ``past'' and ``future'' heterogeneous effects in model 1. The results were averaged over 100 simulation runs and the respective time period. The desired coverage is 95$\%$. The ``future'' effects are predicted from the model trained on the data points from the ``past''.}\label{tab:sim_het_CT}
\end{table}

\newpage
\subsection{Confounded treatment assignment}\label{section:confeffects}

In the previous simulations, every unit had the same probability for being allocated to treatment. We now switch to confounded treatment assignments in which the probability of receiving the treatment depends on the units' covariate values:

\begin{itemize}
\item Assignment 1: five units with $g=1$ and five units with $g=0$ are randomly assigned to treatment. The remaining units are assigned to the control group. 
\item Assignment 2: nine units with $g=1$ and one unit with $g=0$ are randomly assigned to treatment. The remaining units are assigned to the control group. 
\item Assignment 3: one unit with $g=1$ and nine units with $g=0$ are randomly assigned to treatment. The remaining units are assigned to the control group. 
\end{itemize}

The units are divided into two strata with $g=0$ and $g=1$. Stratified sampling is proportionally allocated in the first set-up and disproportionately allocated in the second and third set-up.\\

We estimated the ATE and the SATE with Causal Transfer and the reference methods under the different assignment mechanisms. The results are shown in Table~\ref{tab:sim_conf}. Causal Transfer is clearly the most stable across the assignment mechanisms for both the ATE and the SATE. The MSE, coverage and width of the prediction intervals are not affected much by the confounding. When moving from proportional allocation (assignment 1) to disproportionate allocation (assignments 2 and 3), the MSE and the width increase by approximately 2-3 times for Bayesian imputation while the coverage is too low. The coverage drops by up to 50$\%$ when switching from assignment 1 to assignments 2 and 3 for Causal Impact. 
The reason for the undercoverage seems to be that Causal Impact does not adjust the treatment effect for possible confounders. It simply takes the difference between the observed treated time series and the predicted control time series while neglecting the possible dependence on confounding variables. Bayesian imputation adjusts the treatment for confounding. However, it is not sufficiently powerful when only a few units per strata is assigned to treatment. For example, only one treated observation from the strata with $g=1$ is available per time point in assignment 3. Causal Transfer is much less affected by this issue since it pools the data over time and makes the observations from other time points available instead.

\begin{table}[h!]
\small
\centering
\setlength\tabcolsep{4pt}
\begin{tabular}{lrrrrrrrrrrrr}
\hline
 &\multicolumn{3}{c}{MSE$\cdot10^3$}&\multicolumn{3}{c}{Coverage}&\multicolumn{3}{c}{Width}&\multicolumn{3}{c}{Time$[s]$} \\
 &CT&CI&BI&CT&CI&BI& CT&CI&BI& CT&CI&BI\\
\hline
Assign. 1 (SATE)&0.3&5.6&2.2&0.92&0.79&0.69&0.09&0.18&0.10&210&4&367\\
Assign. 2 (SATE)&0.3&5.7&6.2&0.94&0.36&0.79&0.10&0.29&0.20&222&3&378\\
Assign. 3 (SATE)&0.3&3.1&6.1&0.94&0.51&0.79&0.10&0.28&0.20&226&3&465\\
Assign. 1 (ATE)&0.3&5.6&2.2&0.93&0.78&0.83&0.06&0.18&0.13&90&3&261\\
Assign. 2 (ATE)&0.3&4.2&6.2&0.93&0.41&0.82&0.07&0.27&0.22& 86&3&269\\
Assign. 3 (ATE)&0.4&3.1&6.1&0.92&0.51&0.82&0.07&0.28&0.22&89&3&271\\
\hline
\end{tabular}
\caption{Comparison between Causal Transfer (CT), Causal Impact (CI), and Bayesian imputation (BI) under confounded treatment assignments in model 1. The results were averaged over 300 time points and 100 simulation runs. The desired coverage is 95$\%$. The comparison is restricted to ``past'' effects.}\label{tab:sim_conf}
\end{table}

Furthermore, we predicted ``future'' effects with Causal Transfer under the different treatment assignments and compared the results to the ``past'' effects in Table~\ref{tab:sim_conf_pred}. As before, the MSE and the width of the prediction intervals increase when predicting ``future'' as opposed to ``past'' effects since no outcomes are observed yet. The results are, as a whole, stable across different assignment mechanisms.



\begin{table}[h!]
\small
\centering
\setlength\tabcolsep{4pt}
\begin{tabular}{lrrrrrrr}
\hline
 &\multicolumn{2}{c}{MSE$\cdot10^3$}&\multicolumn{2}{c}{Coverage}&\multicolumn{2}{c}{Width} \\
 &Past&Future&Past&Future&Past&Future\\
\hline
Assign. 1 (SATE)&0.3&2.1&0.92&0.94&0.09&0.17\\
Assign. 2 (SATE)&0.3&2.0&0.94&0.98&0.10&0.25\\
Assign. 3 (SATE)&0.3&2.4&0.94&0.98& 0.10&0.25\\
Assign. 1 (ATE)&0.3&2.1&0.93&0.91&0.06&0.15\\
Assign. 2 (ATE)&0.3&2.0&0.93&0.91&0.07&0.15\\
Assign. 3 (ATE)&0.4&2.4&0.92&0.89&0.07&0.15\\
\hline
\end{tabular}
\caption{Comparison between ``past'' and ``future'' effects under confounded treatment assignments in model 1. The results were averaged over 100 simulation runs and the respective time period. The desired coverage is 95$\%$. The ``future'' effects are predicted from the model trained on the data points from ``past''.}\label{tab:sim_conf_pred}
\end{table}

\clearpage
\subsection{Real data}

Geo experiments divide markets into non-overlapping regions, so called geos, and use these geos as the experimental units. Treatment is randomly applied across the geos to minimize contamination. In the context of online marketing, the treatment can be, e.g., an ad campaign. The advantages of geo experiments are that they protect the privacy of the users and that they are straightforward to set up. Geo experiments are challenging to analyse, though, because of the relatively small number of available geos. The geos also tend to be highly heterogeneous which increases the difficulty further.\\

\begin{figure}[!htb]
    \centering
        \begin{subfigure}[b]{0.485\textwidth}
        \includegraphics[width=\textwidth]{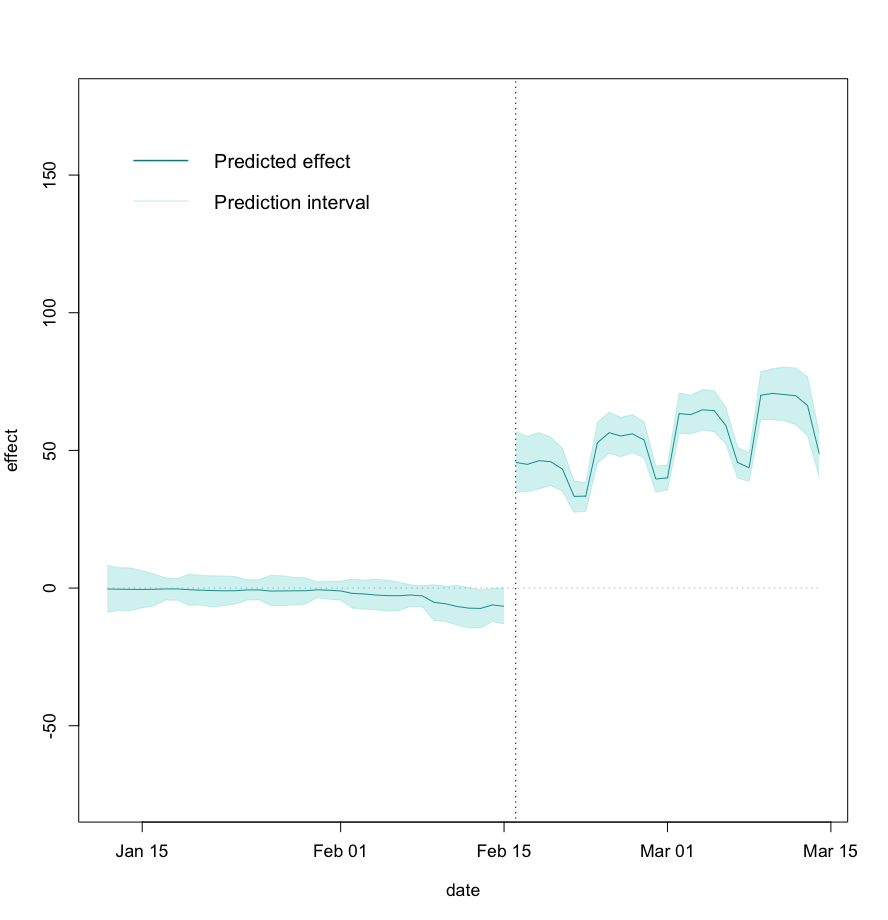}
        \caption{Causal Transfer}
        \label{fig:CT_GeoX}
    \end{subfigure}
   ~        
       \begin{subfigure}[b]{0.485\textwidth}
        \includegraphics[width=\textwidth]{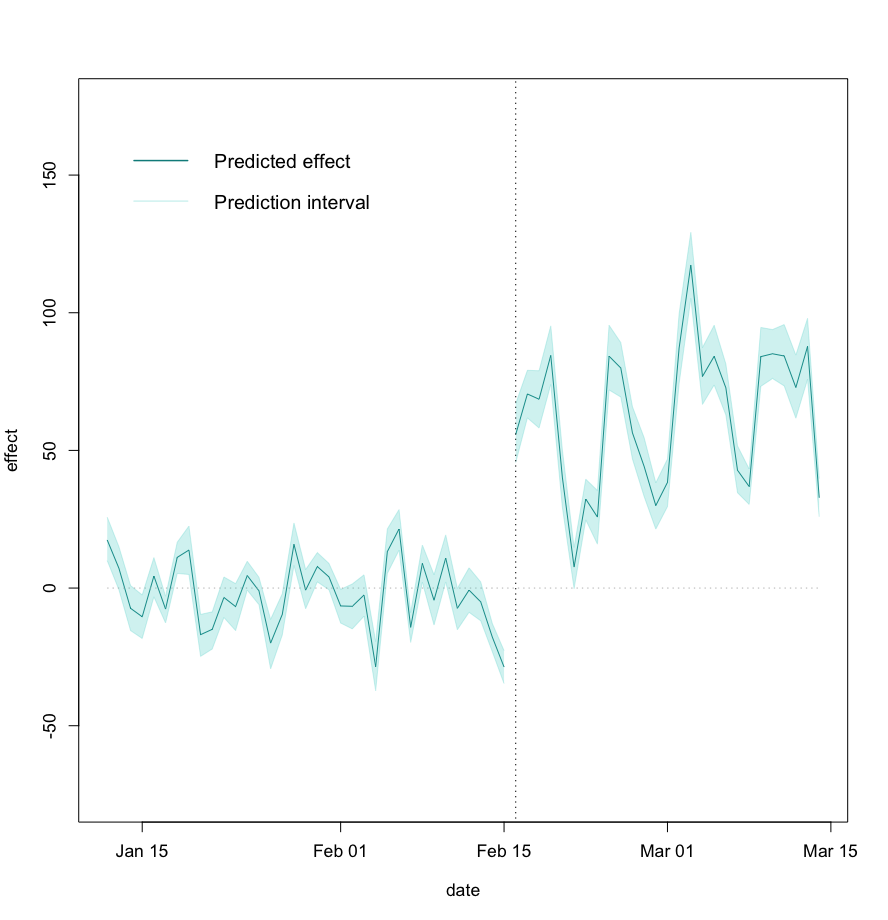}
        \caption{Bayesian imputation}
        \label{fig:BI_GeoX}
    \end{subfigure}
~
        \begin{subfigure}[b]{0.485\textwidth}
        \includegraphics[width=\textwidth]{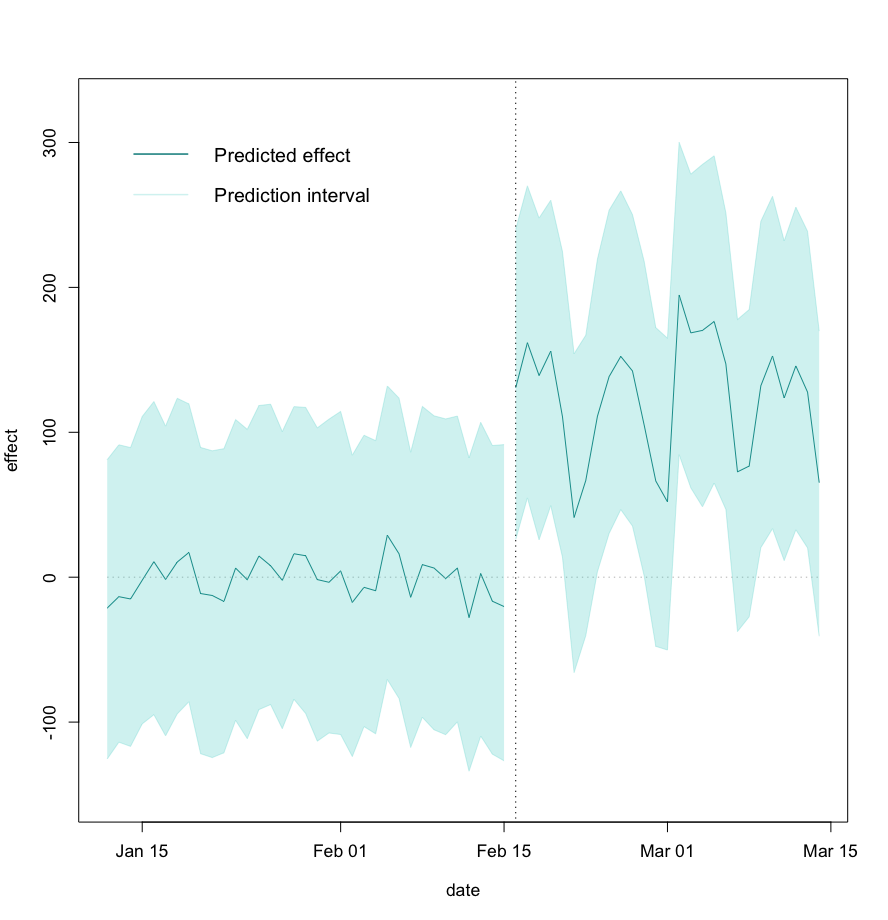}
        \caption{Causal Impact}
        \label{fig:CI_GeoX}
    \end{subfigure}
    ~
            \begin{subfigure}[b]{0.485\textwidth}
        \includegraphics[width=\textwidth]{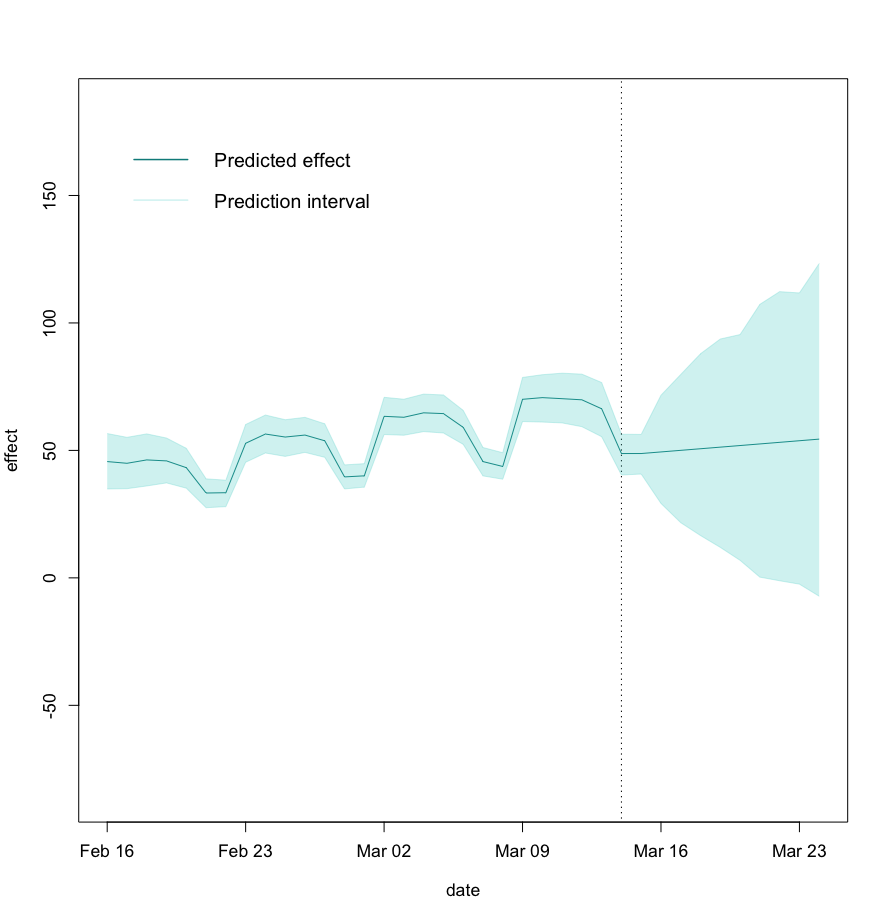}
        \caption{Future effects predicted by Causal Transfer}
        \label{fig:CT_GeoX_pred}
            \end{subfigure}
\caption{The estimated ATE and 90$\%$-prediction intervals for a geo experiment~\cite{kerman2017}. The treatment period started on February 16 and ended on March 15 2015. The true effect should be close to 0 in the pre-period between January 12 and February 15 since no treatment is applied. The prediction intervals of Causal Transfer and Causal Impact correctly cover 0 during the pre-period. However, the prediction intervals of Causal Impact also cover 0 on some days of the treatment period. Bayesian imputation, on the other hand, gives tight intervals but is not neutral in the pre-period indicating undercoverage. The effect estimates all exhibit weakly patterns. Causal Transfer can predict future effects which are shown after the dashed line in Figure~\ref{fig:CT_GeoX_pred}. The prediction quickly grows uncertain over time. This is not surprising because the pre-period covariate loses predictive power over time. The pre-period covariate does not help to capture seasonality either. If time-varying or contemporaneous covariates are available, the predictions are likely to improve. Note, that the axis of the Causal Impact plots are wider in comparison to the other plots.}
\label{fig:GeoX} 
\end{figure}

We analysed publicly available data from a geo experiment~\cite{kerman2017} \cite{GeoX}. The time series consists of sales from 100 geos from January 5 to March 15 2015. The treatment, an ad campaign, is applied between February 16 and March 15. The experiment randomly assigned 50$\%$ of the geos to treatment. We preprocessed the data with the square root transform to correct for skewness. We further excluded 16 geos which had missing values from the analysis. 
We applied Causal Transfer with the dynamic regression model $X_t  = \beta_{0, t} + \beta_{1, t} X_{\text{pre}} + \mu_t T$ to the experimental data. The response variable $X_t$ denotes the sales at time point $t$. The covariate $X_{\text{pre}}$ is constructed from the average sales over the week from January 5 to January 11 for the period before the treatment starts and from February 9 to February 15 for the treatment period. The states $\beta_{0, t}$ and $\beta_{1, t}$ were modelled as random walks and $\mu_t $ as a local linear trend. The observations were given weights of $1/\sqrt{X_{\text{pre}}}$ to adjust for heteroskedasticity. The resulting estimates are shown in Figure ~\ref{fig:CT_GeoX}. The prediction intervals of Causal Transfer include zero in the pre-period, as expected, since no treatment is applied. It also seems plausible that the treatment effect grows slightly during the treatment period since it takes some time for users to see the ad and to decide to respond to it.
We also applied the reference methods, Bayesian Imputation and Causal Impact, under the same conditions, i.e., the same information is given to both methods (in aggregated form for Causal Impact since it requires univariate time series). Results are shown in Figure~\ref{fig:BI_GeoX} and \ref{fig:CT_GeoX}. The point estimates of Bayesian Imputation and Causal Transfer are more variable (also in the pre-period where the effect is known to be neutral). Compared to Causal Impact, the prediction intervals of Causal Transfer are tighter. The prediction intervals of Causal Impact even include 0 on certain days during the treatment period. This implies that we cannot detect a significant effect on these days. Bayesian imputation returns tight intervals but they are not neutral in the pre-period which indicates undercoverage.\\

In addition to the treatment effect, Causal Transfer can be used to estimate the so-called "incremental return on ad spend" (iROAS) if the marketing cost is available. The iROAS is of interest in marketing. It is defined as the incremental sales divided by the incremental ad spend. To estimate the iROAS, one can regress the sales against the marketing cost, i.e., $X_t  = \beta_{0, t} + \beta_{1, t} X_{\text{pre}, t} + \eta_t \text{cost}_t$. The regression coefficient $\eta_t$ corresponds to the iROAS at time point $t$. The iROAS cannot be estimated with Causal Impact directly. Causal Impact requires fitting two separate models, one for the sales and one for the cost, and computes the ratio between the estimated incremental sales and the estimated incremental cost to obtain the iROAS.

On another note, geos are typically made up of smaller constituents, e.g., stores. If finer data than geos is available, one can analyse these finer units instead of geos to increase precision. This corresponds to a cluster randomized trial in which the geos are equivalent to the clusters. To account for the additional correlations between the units within a geo, one can combine a hierarchical model or the cluster bootstrap~\cite{cameron2015} with Causal Transfer.

\clearpage
\section{Conclusions}

We suggest new methodology and a new estimation algorithm for heterogeneous treatment effects in time series. Our method, which we call Causal Transfer, applies state-space models to infer the treatment effect from experiments and observational studies. We note that especially interventional time series exhibit nonstationarities. The underlying model of Causal Transfer can be customized to describe many structures, making it more appropriate for real-world applications than methods which require stationarity or complete randomization. 


Causal Transfer can estimate a variety of estimands in both forms of the population or sample version, e.g., the ATE, the SATE, the CATE, or the MCATE. Each of these estimands is useful for different purposes. The ATE is well-recognized in many fields, but the SATE is better suited when units are selected and not reflective of a population. The CATE, MCATE are helpful when treatment effects are heterogenous. Importantly, Causal Impact also provides prediction intervals for these effects.

Causal Transfer assumes a linear state-space model for fitting the intervention. In simulation studies, the model has proven to be quite robust against misspecifications. Theoretically, this can be justified for average treatment effects. Causal Transfer can further be combined with robust or nonlinear filters  in the presence of outliers or non-Gaussian distributions.\\
Causal Transfer outperformed the reference methods, Causal Impact and Bayesian imputation, in most of our simulations and leads to more meaningful results in a real data application. Our method complements the existing methods for a number of reasons. Methods for the i.i.d. setting, such as Bayesian imputation or difference in difference, performs well when many units are available. For  smaller number of units, they are not sufficiently powered. Causal Transfer is able to increase power by pooling information over time. Time series methods, such as Causal Impact, on the other hand, are limited to predicting the counterfactual of a single treated time series, whereas our method predicts the counterfactual of treated and control time series and, thereby, estimates heterogeneous treatment effects. Lastly, Causal Transfer is able to provide information on future effects: long-term effects, in particular, are crucial for deciding whether a treatment was successful. 

\subsection*{Acknowledgements}
The research of P. B\"uhlmann was supported in part by the European Research Council under the Grant Agreement No. 786461 (CausalStats - ERC-2017-ADG). We would like to thank Nicolas Remy and Jouni Kerman for helpful discussions, insights, and comments.

\newpage
\bibliography{mybib}
\bibliographystyle{plainnat}





\newpage
\appendix
\section{Reference methods}
\label{section:ref_methods}
\subsection{Bayesian imputation}

We can fit separate regression models to the data at each time point as in the example in Section~\ref{section:example}. For instance, $X_t\sim Z_t + T * X_{\text{pre}} + T g$, which is the static version of model~\eqref{eq:measurement}. This approach requires that there are enough units at each time point to fit an informative model. 

More precisely, one puts prior distributions on the regression coefficients $\beta_t$ and the error variance $\sigma_t^2$. Typically, one would choose the same prior for each $\beta_t$ and $\sigma_t^2$ for different time points $t$. For example, $P(\beta_t) \propto 1$ and $P(\sigma_t^2) \propto 1/\sigma_t^2$. These priors are non-informative as they correspond to flat priors on $\beta_t$ and $\log(\sigma_t^2)$. They are also improper as they do not integrate to 1. 
For inference, one samples $\sigma_{t}^{2(b)}$ from the marginal posterior distribution $IG((n-p)/2, ((n-p)/2)\hat{\sigma}_t^2)$. The distribution $IG$ denotes the inverse-gamma distribution and $\hat{\sigma}_t= \sqrt{\frac{1}{n-p} (X_t-F_t\hat{\beta}_t)' (X_t-F_t\hat{\beta}_t)}$ the residual standard error. Note that $F_t$ is the design matrix, $p$ the number of columns of $F_t$, and $\hat{\beta}_t = (F_t' F_t)^{-1} F_t' X_t$ the estimated coefficients (solution of the normal equation). Subsequently, one is able to draw $\beta_{t}^{(b)}$ from the conditional posterior distribution $\mathcal{N}(\hat{\beta}_t, (F_t' F_t)^{-1} \sigma_{t}^{2(b)})$. The pair ($\sigma_{t}^{2(b)}$, $\beta_{t}^{(b)}$) now constitutes a draw from the joint posterior distribution of $\sigma_t^2$ and $\beta_t$. We then draw the missing outcomes from the posterior predictive distribution $\mathcal{N}(\tilde{F}_t\beta_{t}^{(b)}, \mathbb{I}_d\sigma_{t}^{2(b)})$, where the matrix $\tilde{F}_t$ contains the covariates of the counterfactuals. We estimate the effect sample $\hat{\tau}_{t}^{(b)}$ from the observed and the imputed outcomes. Finally, we can summarize the effect samples $(\hat{\tau}_{t}^{(b)})_{b=1}^B$ as prediction intervals and point estimates, e.g., by taking the percentiles and the average. Algorithm~\ref{alg:BI} contains the pseudo-code version. This reference method is known as Bayesian imputation in the causal inference literature. More information on Bayesian imputation can be found in \cite{imbens2015}.\\

\begin{algorithm}[!htb]
\small
\begin{algorithmic}[1]
\vspace{0.4cm}
\FOR{$t=1,\ldots,n$}
\STATE Fit linear regression model to data from time point $t$.
\FOR{$b=1,\ldots,B$}
\STATE (i) \hspace{0.2cm} Sample $\sigma_{t}^{2(b)}$ from $IG((n-p)/2, ((n-p)/2)\hat{\sigma}_t^2)$.
\STATE (ii) \hspace{0.12cm} Sample $\beta_{t}^{(b)}$ from $\mathcal{N}(\hat{\beta}_t, (F' F)^{-1} \sigma_{t}^{2(b)})$.
\STATE (iii) \hspace{0.03cm} Sample $\tilde{X}_{t}^{(b)}$ from $\mathcal{N}(\tilde{F}\beta_{t}^{(b)}, \mathbb{I}_d\sigma_{t}^{2(b)})$.
\STATE (iv) \hspace{0.07cm} Estimate effect sample $\hat{\tau}_{t}^{(b)}$ from $X_t$ and $\tilde{X}_{t}^{(b)}$.
\ENDFOR
\STATE Summarize the effect samples $(\hat{\tau}_{t}^{(b)})_{b=1}^B$ into prediction intervals.
\STATE Estimate sample treatment effect as $\hat{\tau}_t = \frac{1}{B} \sum_{b=1}^B \hat{\tau}_{t}^{(b)}$. 
\ENDFOR
\RETURN Estimated prediction intervals and treatment effects for $t=1,\ldots,n$.
\end{algorithmic}
\caption{\small Bayesian imputation for sample treatment effects}\label{alg:BI}
\end{algorithm}

In the above form, Bayesian imputation estimates sample treatment effects. Algorithm~\ref{alg:BI} can be modified for estimating population treatment effects. The modified version is shown in algorithm~\ref{alg:BI_sp}. The main difference is that we skipped step 6 in favor of estimating the effect samples $\hat{\tau}_{t}^{(b)}$ directly from $\beta_{t}^{(b)}$ and the effect $\hat{\tau}_t$ directly from $\hat{\beta}_t$.

\begin{algorithm}[!htb]
\small
\begin{algorithmic}[1]
\FOR{$t=1,\ldots,n$}
\STATE Fit linear regression model to data from time point $t$.
\FOR{$b=1,\ldots,B$}
\STATE (i) \hspace{0.2cm} Sample $\sigma_{t}^{2(b)}$ from $IG((n-p)/2, ((n-p)/2)\hat{\sigma}_t^2)$.
\STATE (ii) \hspace{0.12cm} Sample $\beta_{t}^{(b)}$ from $\mathcal{N}(\hat{\beta}_t, (F' F)^{-1}\sigma_{t}^{2(b)})$.
\STATE (iii) \hspace{0.03cm} Estimate effect sample $\hat{\tau}_{t}^{(b)}$ from $\beta_{t}^{(b)}$.
\ENDFOR
\STATE Summarize the effect samples $(\hat{\tau}_{t}^{(b)})_{b=1}^B$ into prediction intervals.
\STATE Estimate population treatment effect $\hat{\tau}_t$ from $\hat{\beta}_t$.
\ENDFOR
\RETURN Estimated prediction intervals and treatment effects for  $t=1,\ldots,n$.
\end{algorithmic}
\caption{\small Bayesian imputation for population treatment effects}\label{alg:BI_sp}
\end{algorithm}

\subsection{Causal Impact}

Causal Impact aims to predict the counterfactual of a treated time series $X_t$. It predicts what would have happened to $X_t$ in absence of treatment. 
To estimate the counterfactual, it requires at least one control time series $Z_t$ that is predictive of the response time series but not affected by treatment itself (similar to $Z_t$ 
in model~\eqref{eq:measurement}). It learns the relationship between the response and the control time series during the pre-period. It then assumes that the learned relationship is not changed by treatment. Thereby, it is able to predict the counterfactual series during the treatment period. Causal Impact relies on a dynamic regression model. For a single control time series $Z_t$ the model reads:
\begin{align}
X_t &= \mu_t + \beta Z_t + v_t\nonumber\\
\mu_t &= \mu_{t-1} + \alpha_{t-1} + w_t\nonumber\\
\alpha_t &= \alpha_{t-1} + u_t \,,
\end{align}
for $t =1,\ldots, n$. The noise variables $u_t$, $v_t$, and $w_t$ are mutually independent, centered, and normally distributed with constant variances. The assumptions that the relationship between control $Z_t$ and response time series $X_t$ remains constant before and during treatment can be relaxed by letting the regression coefficient $\beta$ evolve according to a random walk. The local trend and level terms $\mu_t$ and $\alpha_t$ can be adjusted for seasonality by adding higher order lagged terms in the state equations. For a large number of control time series, a spike and slab prior is placed on the regression coefficients for model selection. All inferences are done within a Bayesian framework using MCMC.

Causal Impact requires the response time series $X_t$ to be univariate.  To run the analysis for multivariate response time series, we  aggregate the multivariate time series (cross-sectional) and input the aggregated time series.

\newpage
\section{Supplementary plots}

\begin{figure}[!htb]
    \centering
        \begin{subfigure}[b]{0.485\textwidth}
        \includegraphics[width=\textwidth]{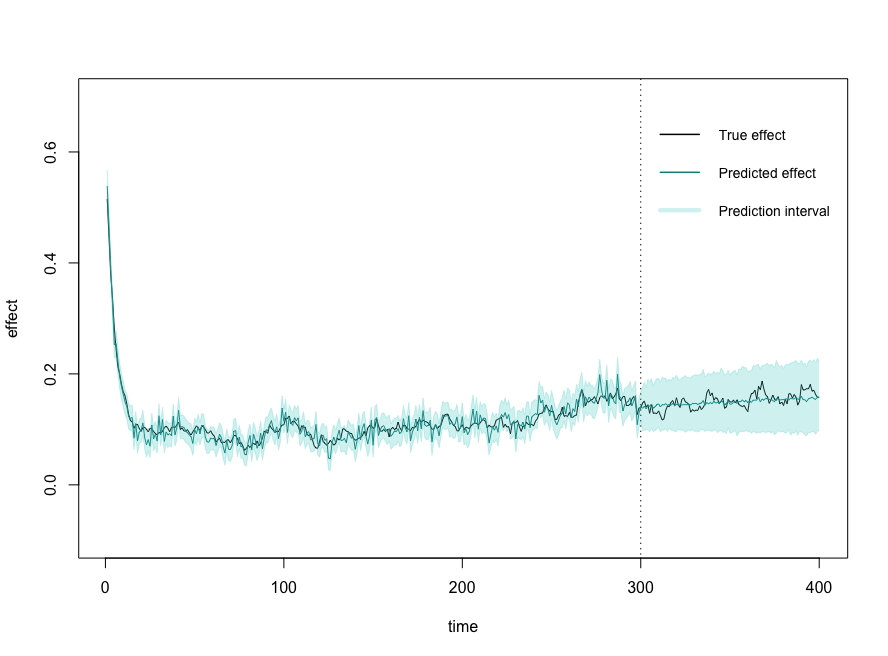}
        \caption{Causal Transfer}
        \label{fig:SARTE_CT}
    \end{subfigure}
    ~
       \begin{subfigure}[b]{0.485\textwidth}
        \includegraphics[width=\textwidth]{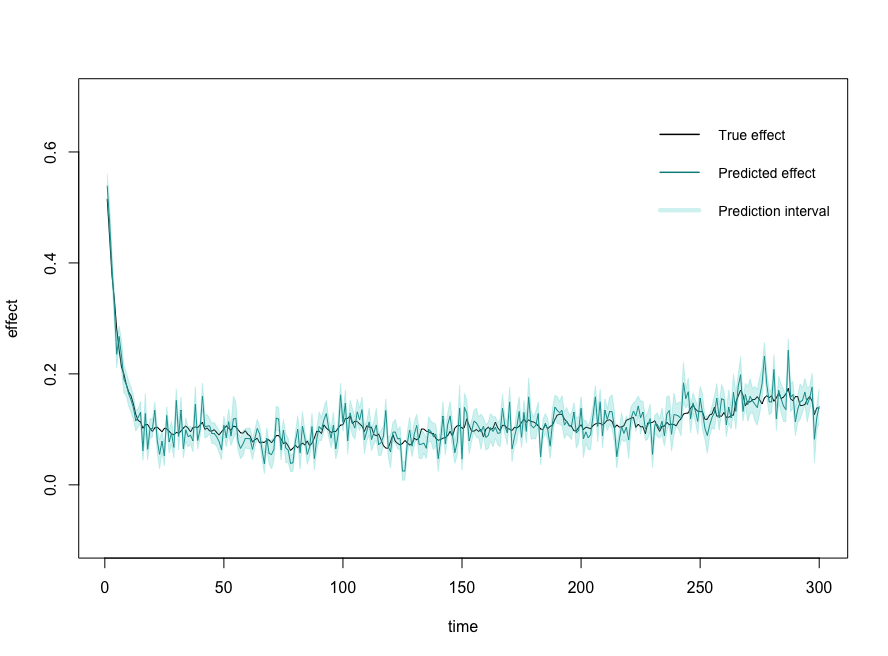}
        \caption{Bayesian imputation}
        \label{fig:SARTE_BI}
    \end{subfigure}
\caption{Estimated SARTE effects of Causal Transfer and Bayesian imputation for one simulation run in model 1. Figure~\ref{fig:SARTE_CT} shows the predicted ``future'' effects after the dashed line.}\label{fig:SARTE} 
\end{figure}

\begin{figure}[!htb]
    \centering
        \begin{subfigure}[b]{0.485\textwidth}
        \includegraphics[width=\textwidth]{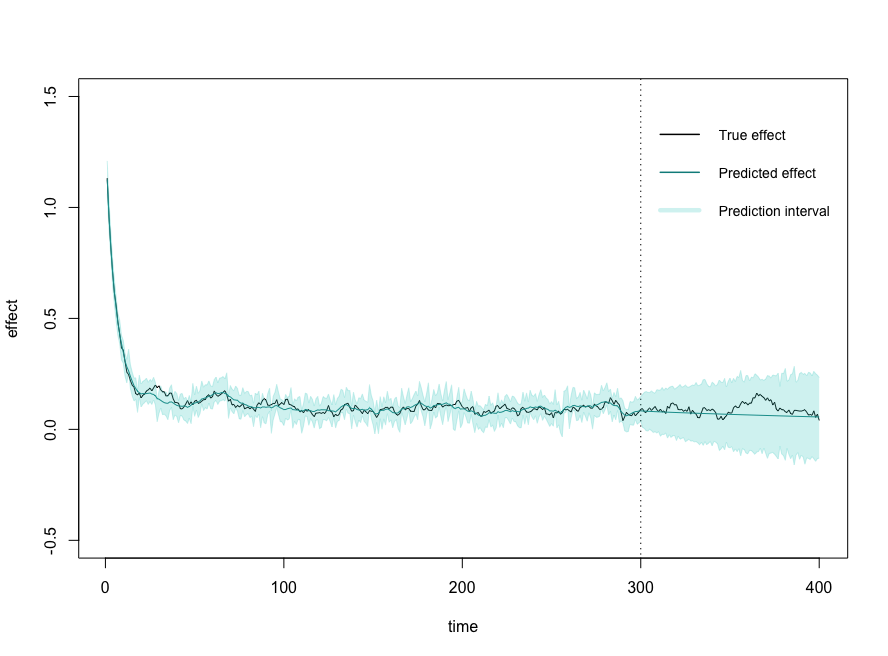}
        \caption{SATE (CT)}
        \label{fig:SATE_CT_conf}
    \end{subfigure}
    ~
       \begin{subfigure}[b]{0.485\textwidth}
        \includegraphics[width=\textwidth]{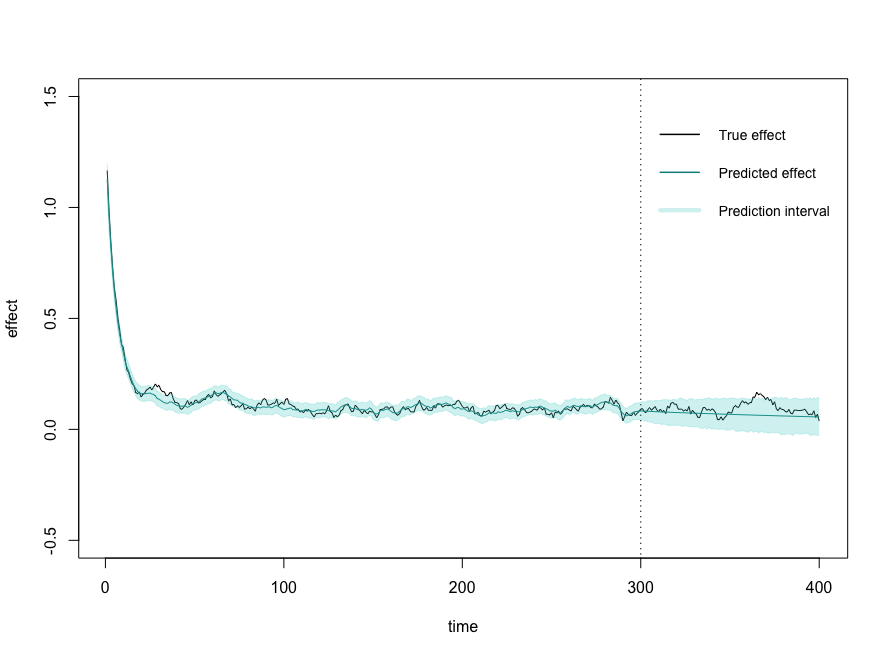}
        \caption{ATE (CT)}
        \label{fig:ATE_CT_conf}
    \end{subfigure}
    
        \begin{subfigure}[b]{0.485\textwidth}
        \includegraphics[width=\textwidth]{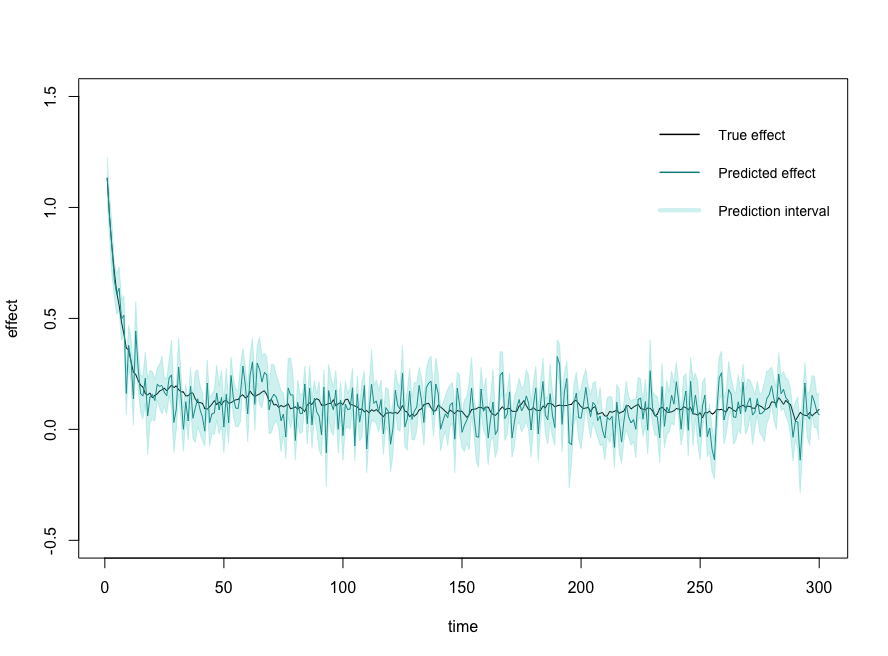}
        \caption{SATE (BI)}
        \label{fig:SATE_BI_conf}
    \end{subfigure}
    ~
        \begin{subfigure}[b]{0.485\textwidth}
        \includegraphics[width=\textwidth]{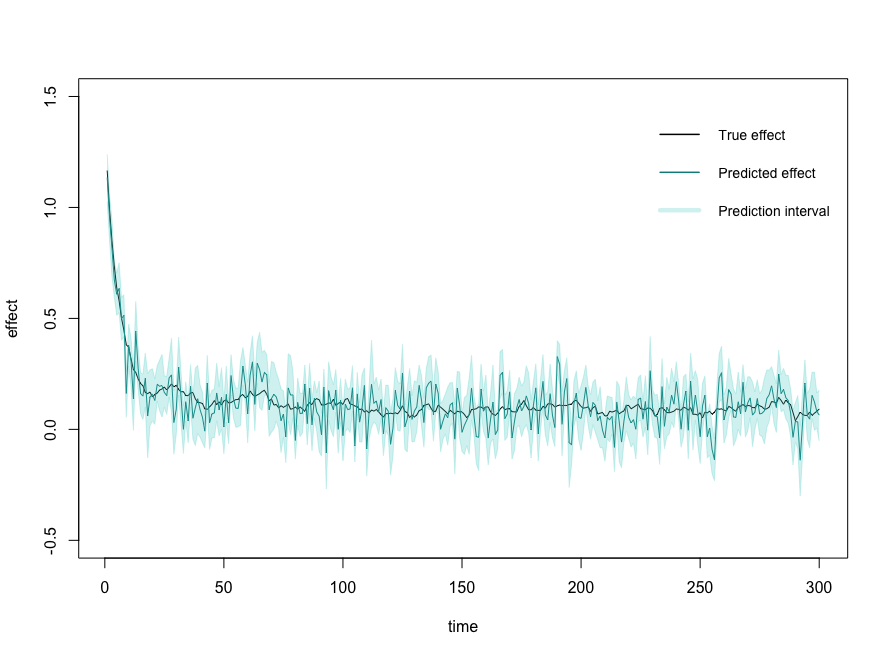}
        \caption{ATE (BI)}
        \label{fig:ATE_BI_conf}
    \end{subfigure}
    
        \begin{subfigure}[b]{0.485\textwidth}
        \includegraphics[width=\textwidth]{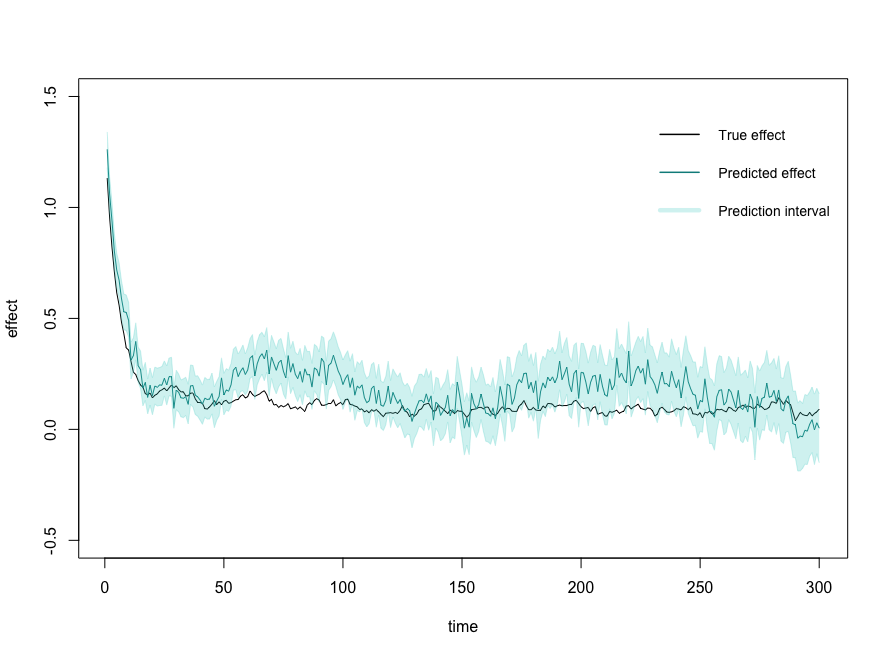}
        \caption{SATE (CI)}
        \label{fig:SATE_CI_conf}
       \end{subfigure}
     ~
        \begin{subfigure}[b]{0.485\textwidth}
        \includegraphics[width=\textwidth]{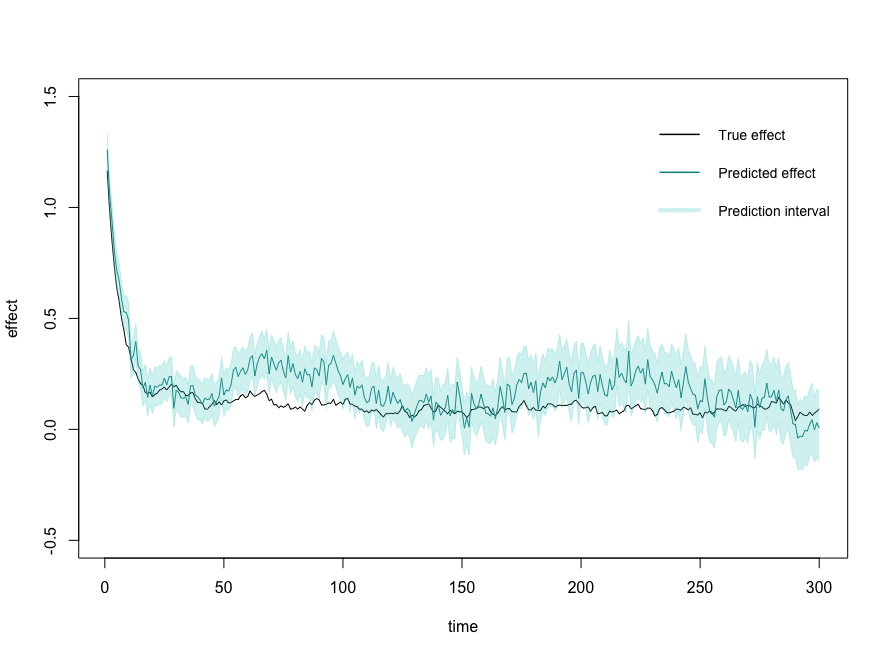}
        \caption{ATE (CI)}
        \label{fig:ATE_CI_conf}
    \end{subfigure}
\caption{Example analysis plots for the ATE and the SATE under (confounded) treatment assignment 2 in model 1. The true ATE and SATE effects are very similar since the units were drawn from a distribution. Causal Transfer (Figure~\ref{fig:SATE_CT_conf} and \ref{fig:ATE_CT_conf}) is able to predict ``future'' effects, which are shown after the dashed line. Bayesian imputation (Figure~\ref{fig:SATE_BI_conf} and \ref{fig:ATE_BI_conf}) and Causal Impact (Figure~\ref{fig:SATE_CI_conf} and Figure \ref{fig:ATE_CI_conf}) both undercover and have wider intervals than Causal Transfer. }\label{fig:confounded} 
\end{figure}

\begin{figure}[!htb]
    \centering
        \begin{subfigure}[b]{0.485\textwidth}
        \includegraphics[width=\textwidth]{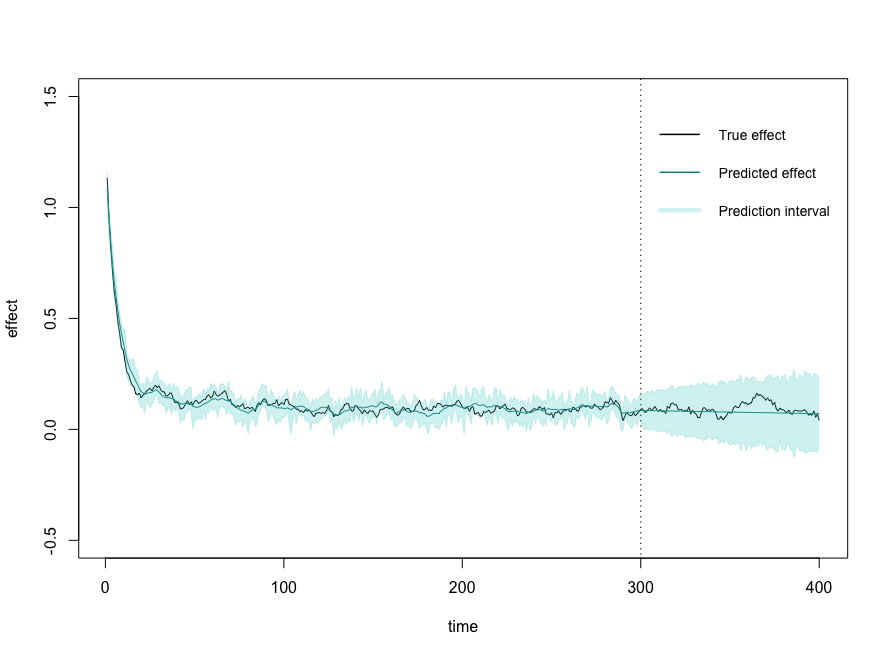}
        \caption{SATE (CT)}
        \label{fig:SATE_CT_conf2}
    \end{subfigure}
    ~
       \begin{subfigure}[b]{0.485\textwidth}
        \includegraphics[width=\textwidth]{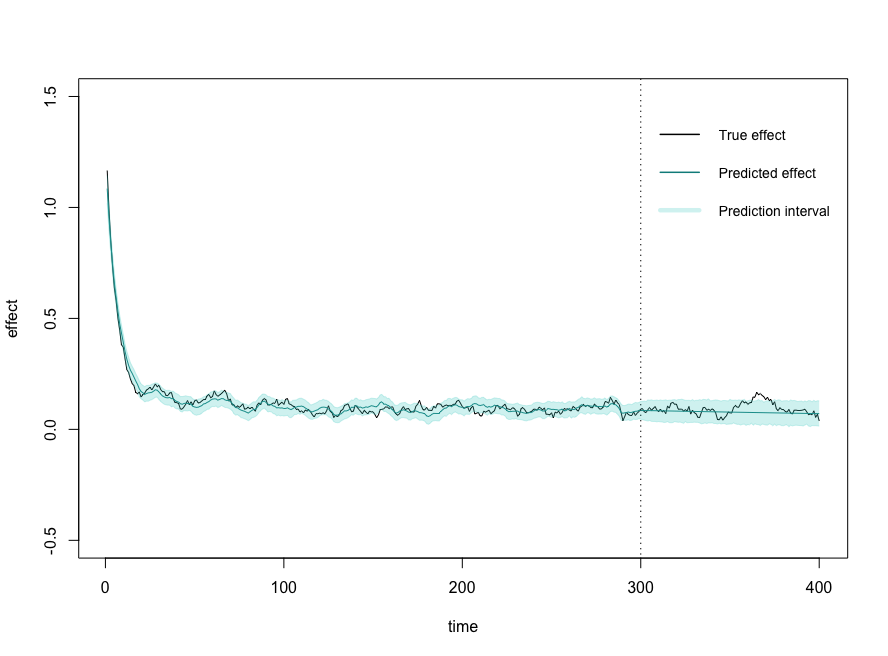}
        \caption{ATE (CT)}
        \label{fig:ATE_CT_conf2}
    \end{subfigure}
    
        \begin{subfigure}[b]{0.485\textwidth}
        \includegraphics[width=\textwidth]{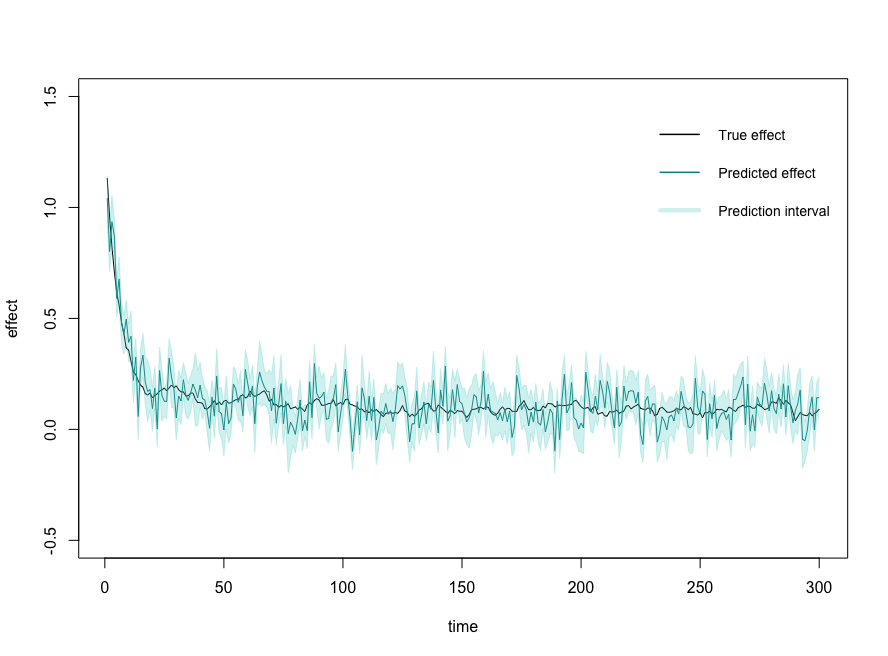}
        \caption{SATE (BI)}
        \label{fig:SATE_BI_conf2}
    \end{subfigure}
    ~
        \begin{subfigure}[b]{0.485\textwidth}
        \includegraphics[width=\textwidth]{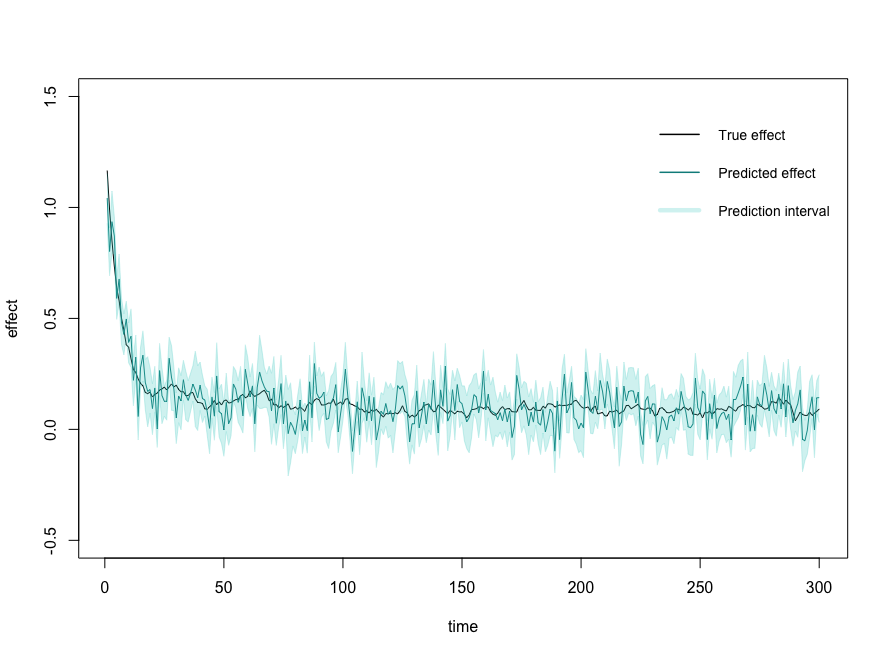}
        \caption{ATE (BI)}
        \label{fig:ATE_BI_conf2}
    \end{subfigure}
    
        \begin{subfigure}[b]{0.485\textwidth}
        \includegraphics[width=\textwidth]{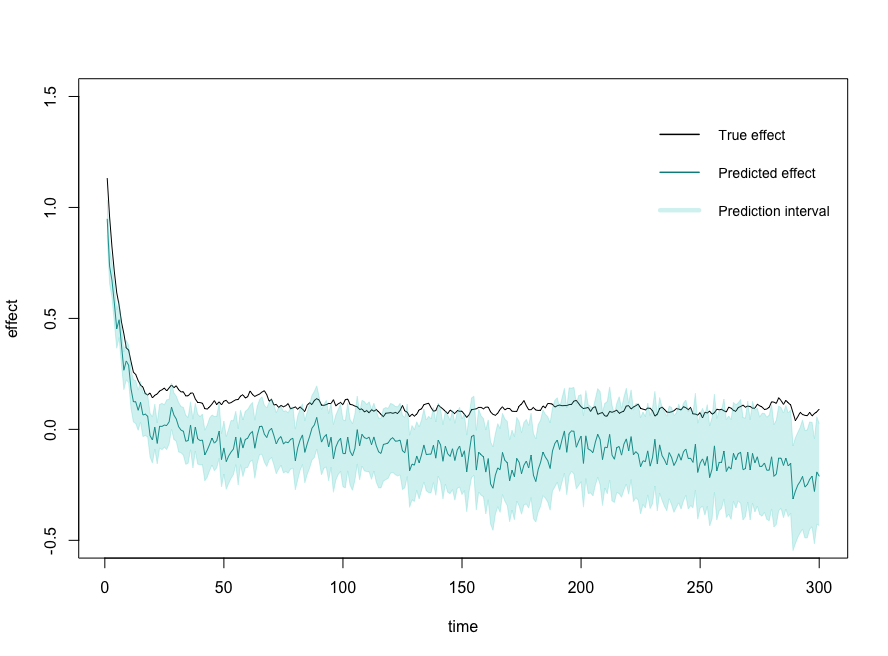}
        \caption{SATE (CI)}
        \label{fig:SATE_CI_conf2}
        \end{subfigure}
     ~
        \begin{subfigure}[b]{0.485\textwidth}
        \includegraphics[width=\textwidth]{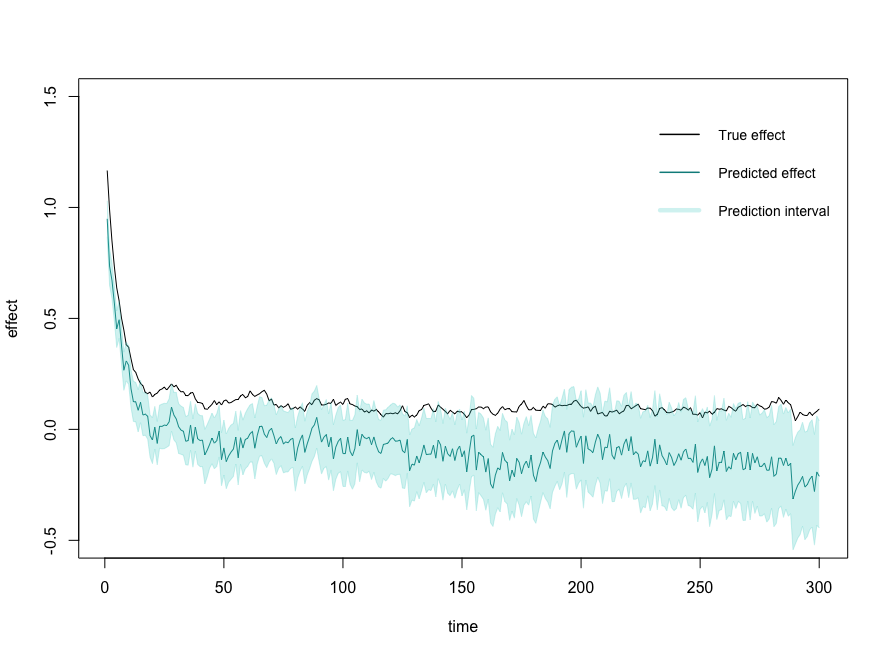}
        \caption{ATE (CI)}
        \label{fig:ATE_CI_conf2}
    \end{subfigure}
\caption{Example analysis plots for the ATE and the SATE under (confounded) treatment assignment 3 in model 1. The true ATE and SATE effects are very similar since the units were drawn from a distribution. Causal Transfer (Figure~\ref{fig:SATE_CT_conf2} and \ref{fig:ATE_CT_conf2}) is able to predict ``future'' effects, which are shown after the dashed line. Bayesian imputation (Figure~\ref{fig:SATE_BI_conf2} and \ref{fig:ATE_BI_conf2}) and Causal Impact (Figure~\ref{fig:SATE_CI_conf2} and Figure \ref{fig:ATE_CI_conf2}) both undercover and have wider intervals than Causal Transfer. }\label{fig:confounded2} 
\end{figure}

\end{document}